\documentclass[twocolumn]{aastex61}

\usepackage{threeparttable}
\usepackage{amsmath}
\usepackage{xspace}

\newcommand*{\rom}[1]{\expandafter\@slowromancap\romannumeral #1@}

\newcommand{\msun}{M$_{\odot}$ }
\newcommand{\myr}{M$_\odot$~yr$^{-1}$ }

\newcommand{\ha}{H$\alpha$\xspace}
\newcommand{\hb}{H$\beta$\xspace}
\newcommand{\nii}{[N{\sc II}]\xspace}
\newcommand{\oiii}{[O{\sc III}]\xspace}
\newcommand{\oii}{[O{\sc II}]\xspace}
\newcommand{\sii}{[S{\sc II}]\xspace}

\newcommand{\kms}{km\,s$^{-1}$ } 

\newcommand{\ferg}{erg s$^{-1}$ cm$^{-2}$ }
\newcommand{\ferga}{erg s$^{-1}$ cm$^{-2}\rm\AA^{-1}~$ }
\newcommand{\ergs}{erg s$^{-1}$ }

\newcommand{\momfluxsfr}{$\dot{P}_{SFR}$ }
\newcommand{\momfluxagn}{$\dot{P}_{AGN}$ }
\newcommand{\momfluxout}{$\dot{P}_{outflow}$ }
\newcommand{\kineticlum}{$\dot{E}_{outflow}$ }
\newcommand{\energydepsne}{$\dot{E}_{SNe}$ }
\newcommand{\eden}{cm$^{-3}$ }

\newcommand{\momfluxratio}{$\frac{\dot{P}_{outflow}}{\dot{P}_{AGN}}$}

\newcommand{\lbolqso}{$L_{Bol}$}
\newcommand{\msigma}{$M_{\bullet}-\sigma~$}
\newcommand{\mstellar}{$M_{\bullet}-M_{*}~$}

\received{July 1, 2017}
\revised{September 27, 2017}
\accepted{\today}

\submitjournal{ApJ}

\shorttitle{Kinematics and Energetics of Ionization Gas}
\shortauthors{Vayner et al.}

\begin{document}

\title{A Spatially-Resolved Survey of Distant Quasar Host Galaxies:\\ I. Dynamics of galactic outflows}

\correspondingauthor{Andrey Vayner}
\email{avayner1@jhu.edu}

\author[0000-0002-0710-3729]{Andrey Vayner}
\affiliation{Department of Physics, University of California San Diego, 
9500 Gilman Drive 
La Jolla, CA 92093 USA}
\affiliation{Center for Astrophysics \& Space Sciences, University of California San Diego, 9500 Gilman Drive La Jolla, CA 92093 USA}

\affiliation{Department of Physics and Astronomy, Johns Hopkins University, Bloomberg Center, 3400 N. Charles St., Baltimore, MD 21218, USA}

\author[0000-0003-1034-8054]{Shelley A. Wright}
\affiliation{Department of Physics, University of California San Diego, 
9500 Gilman Drive 
La Jolla, CA 92093 USA}
\affiliation{Center for Astrophysics \& Space Sciences, University of California San Diego,
9500 Gilman Drive 
La Jolla, CA 92093 USA}

\author{Norman Murray}
\affiliation{Canadian Institute for Theoretical Astrophysics, University of Toronto, 60 St. George Street, Toronto, ON M5S 3H8, Canada}
\affiliation{Canada Research Chair in Theoretical Astrophysics}

\author[0000-0003-3498-2973]{Lee Armus}
\affiliation{Spitzer Science Center, California Institute of Technology, 1200 E. California Blvd., Pasadena, CA 91125 USA}


\author[0000-0003-0439-7634]{Anna Boehle}
\affiliation{ETH Zürich Wolfgang-Pauli-Str. 27 8093 Zürich, Switzerland}

\author[0000-0002-2248-6107]{Maren Cosens}
\affiliation{Department of Physics, University of California San Diego, 
9500 Gilman Drive 
La Jolla, CA 92093 USA}
\affiliation{Center for Astrophysics \& Space Sciences, University of California San Diego, 9500 Gilman Drive La Jolla, CA 92093 USA}

\author[0000-0001-7687-3965]{James E. Larkin}
\affiliation{Department of Physics and Astronomy, University of California, Los Angeles, CA 90095 USA}

\author[0000-0001-7127-5990]{Etsuko Mieda}
\affiliation{National Astronomical Observatory of Japan, Subaru Telescope, National Institutes of Natural Sciences, Hilo, HI 96720, USA}

\author[0000-0002-6313-6808]{Gregory Walth}
\affiliation{Observatories of the Carnegie Institution for Science
813 Santa Barbara Street
Pasadena, CA 91101
USA}

\begin{abstract}
We present observations of ionized gas outflows in eleven z$ =1.39-2.59$ radio-loud quasar host galaxies. Data was taken with the integral field spectrograph (IFS) OSIRIS and the adaptive optics system at the W.M. Keck Observatory targeting nebular emission lines (\hb, \oiii, \ha, \nii, and \sii) redshifted into the near-infrared (1-2.4 \micron). Outflows with velocities of 500 - 1700 \kms are detected in 10 systems on scales ranging from $<1$ kpc to 10 kpc with outflow rates from 8-2400 \myr. For five sources, the outflow momentum rates are 4-80 times $L_{AGN}$/c, consistent with outflows being driven by an energy conserving shock. The five other outflows are either driven by radiation pressure or an isothermal shock. The outflows are the dominant source of gas depletion, and we find no evidence for star formation along the outflow paths. For eight objects, the outflow paths are consistent with the orientation of the jets. Yet, given the calculated pressures, we find no evidence of the jets currently doing work on these galactic-scale ionized outflows. We find that galactic-scale feedback occurs well before galaxies establish a substantial fraction of their stellar mass, as expected from local scaling relationships.

\end{abstract}

\section{Introduction} \label{sec:chap3intro}

In the nearby Universe, at most 10-20$\%$ of the baryonic matter resides in stars \citep{Behroozi10}. In the most massive dark matter halos ($\rm M_{dm} > 10^{12}$ \msun) the lack of baryons inside galaxies has been attributed to negative feedback from active galactic nucleus (AGN)  \citep{Benson03, Kormendy13}. Feedback from AGN is often used to explain the tight correlation between the mass of a supermassive black hole (SMBH) and the velocity dispersion (\msigma) or mass of the bulge/galaxy (\mstellar) \citep{Ferrarese00, Gebhardt00, McConnell13}. In the current theoretical framework, the SMBH and galaxy grow until they reach a mass ratio close to what is observed in the nearby Universe ($\frac{M_{\bullet}}{M_{*}}\sim5\times10^{-3}$). Afterward, the accretion disk surrounding the SMBH is capable of driving a wind that extends deep enough into the host galaxy to drive a shock that may expand adiabatically \citep{Zubovas12, Zubovas14, Faucher12}. The expanding shock produces a galaxy scale outflow that both removes material from the host galaxy and drives turbulence, preventing gas from collapsing and cooling on regular time scales. 

In essence, AGN feedback both prolongs the time necessary for gas to collapse and form stars, while also removing the fuel for future star formation and SMBH growth. The feedback processes start near the vicinity of the SMBH, where radiation pressure from the accretion disk drives powerful winds. These nuclear winds are commonly observed as blueshifted absorption lines in UV spectra and are called Broad Absorption Line (BAL) winds \citep{Chamberlain15} or in X-ray spectra as ``Ultra Fast Outflows" \citep{Tombesi10}. From herein ``winds" will be referred to as ``small" scale (pc to a kpc) structures/events while ``outflows" will refer to galaxy scale ($>$1kpc). Alternatively, the feedback process can start with jets launched from the accretion disk. Similar to BAL and UFO winds, jets can drive outflows that sweep material out of the galaxy \citep{Wagner12, Mukherjee16}. 

To effectively halt star formation or expel gas from a galaxy and establish the observed local relationships (\msigma,\mstellar) the BAL or UFO type wind needs to transfer at least $0.1-5\%$ of the quasar bolometric luminosity into the kinetic luminosity of the galaxy scale outflow \citep{Hopkins12, Zubovas12}. Where $M_{\bullet}$ refers to the SMBH mass, $\sigma$ refers to the stellar velocity dispersion and $M_{*}$ represents the stellar mass of the galactic bulge.  While for jets, the ratio of the power ($P_{jet}$) to the Eddington luminosity of the SMBH accretion disk needs to be higher than $10^{-4}$ \citep{Wagner12}. Another proposed way of clearing gas or inducing turbulence to slow star formation is through radiation pressure on dust grains in the nuclear region of AGN \citep{Thompson15, Costa18}. Although this mode of feedback is not as strong as the other proposed methods, it can still be powerful enough to disrupt star formation within the host galaxy.

Often feedback during luminous AGN phases has been described as transformative, where the galaxy goes from being star-forming to quiescent after episodes of AGN feedback \citep{Hopkins08}. In the past decade, there is growing evidence for feedback from SMBH, yet there appear to be discrepancies between the theoretical predictions and observations of the strength of feedback. These disagreements can stand from observational bias or missing physics within galactic feedback models. Studies have focused on detecting and characterizing nearby \citep{Veilleux03,Morganti05,Holt08,Greene12,Rupke11,LiuG13,Harrison14,Smethurst19} and distant \citep{Nesvadba08,Steinbring11,Cano-Diaz12,Harrison12,Brusa15,Carniani15} outflows through nebular emission lines (e.g, \hb, \oiii, \ha), primarily focusing on the \oiii emission line as a tracer of the ionized gas. There is growing evidence of galaxy-scale outflows; however, it has been challenging to measure accurate outflow rates and energetics for comparison with theoretical predictions \citep{Harrison18}. Detected outflows in quasar host galaxies show a large range in their kinetic luminosity, spanning from 0.001$\%$ to $5\%$ of the quasar's bolometric luminosity. Often the measured energetics fall short of the predicted values from theoretical work, with about half of the detected ionized outflows showing \kineticlum $<0.1\%$ \lbolqso~\citep{Carniani15}. Where \kineticlum refers to the kinetic luminosity of the outflow ($\dot{E}_{outflow} = 1/2\times \dot{M}\times v^2_{outflow}$) and \lbolqso\ is the bolometric luminosity of the AGN/quasar.

Furthermore, for high-redshift ionized outflows, the ratio of momentum flux $\dot{P}_{\rm outflow}=\dot{M}\times v_{outflow}$ to the \momfluxagn have been far lower than what is predicted by theoretical work \citep{Zubovas12}. Where $\dot{M}$ refers to the outflow rate and $v_{outflow}$ to the velocity of the outflowing gas. In contrast, in the nearby Universe, molecular outflows found in systems with AGN have shown energetics consistent with the theoretical predictions \cite{Cicone14}. One of the proposed differences between molecular and ionized outflows is that ionized outflows might constitute only a small fraction of the gas phase in the outflow \citep{Carniani15, Richings18}. The number of observed systems with both ionized and molecular outflows in the distant and nearby Universe is very small. For systems where the multi-phase outflow have been detected they show that the largest fraction of the gas is indeed in a molecular state \citep{Vayner17, Brusa18, Herrera-Camus19}. More extensive studies with ALMA and JWST are necessary to confirm the multi-phase nature of quasar driven outflows.

Following episodes of transformative feedback during a luminous AGN phase it has also been proposed that to keep the galactic halos hot, ``maintenance" feedback takes over in forms of large scale jets formed from a low Eddington accretion AGN. Often, these large scale jets carve bubbles in the hot halo \citep{Fabian94, McNamara00, Hlavacek-Larrondo13}, inducing turbulence, and preventing the halo from cooling to fuel future generations of star formation. Understanding both transformative and maintenance mode feedback is essential to understanding the formation of massive galaxies.

We have conducted the QUART (Quasar hosts Unveiled by high Angular Resolution Techniques) survey to study the host galaxies of z$=1.4-2.6$ radio-loud quasars.  Due to their massive SMBH ($>10^{9}$\msun) the selected objects within our sample are the most likely systems to evolve into massive elliptical galaxies in the nearby Universe. Targets were observed using the OSIRIS near-infrared integral field spectrograph (IFS) behind the laser-guide-star adaptive optics (LGS-AO) system at the W.M. Keck Observatory. The aim of our survey is to understand the gas phase conditions and ionization in high redshift quasar host galaxies, to search for outflows and quantify their feedback on the host galaxies, and weigh the masses of the quasar hosts. We obtained integral field spectroscopy of nebular emission lines (\hb, \oiii 495.9, 500,7 nm, \ha, \nii 654.9, 658.5, \sii 671.7, 673.1 nm) redshifted in the near-infrared bands ($1-2.4$ \micron) at a spatial resolution of $\sim$1.4 kpc. This paper is part one of two, focusing on understanding the driving mechanisms of galaxy scale outflows and their impact on the ISM. Paper-II focuses on resolved line ratio diagnostics, measurement of star formation rates, metallicities, and deriving the host galaxies' dynamical masses to place them on the local scaling relations. We describe our sample selection in \S \ref{sec:sample-selection}, we summarise the observations in \S \ref{sec:observations}, data reduction and analysis is outlined in \S \ref{sec:reduc}. We outline how we identify spatially-resolved outflow regions in \S \ref{sec:regions}, discussion of spatially unresolved outflows can be found in \S \ref{sec:nuclear outflows}, outflow rates, energetics and the driving sources of outflows are presented in \S \ref{sec:out_rate}, we discuss our results in the broader context of massive galaxy evolution in \S \ref{sec:chapter4discussion} and present our conclusions in \S \ref{sec:conclusions}. Kinematic maps for individual sources along with spectra of distinct galactic regions are presented in \S \ref{sec:appendix}. Throughout the paper we assume a $\Lambda$-dominated cosmology \citep{Planck13} with $\Omega_{M}$=0.308, $\Omega_{\Lambda}$=0.692, and H$_{o}$=67.8 \kms~ Mpc$^{-1}$. All magnitudes are on the AB scale unless otherwise stated.

\section{Sample Selection}\label{sec:sample-selection}

Sources were selected from the Sloan Digital Sky Survey (SDSS) data release 10 (\cite{Ahn14}). Quasars with a magnitude brighter than 17.5 at R band or those that have nearby stars with R magnitude $<$ 17.5 within 45\arcsec~were cross-correlated with the FIRST catalog \citep{Becker95} and NASA Extra-galactic Database (NED). Sources with radio flux of 0.3 Jy and greater at 1.4 GHz were selected based on their available archival VLA and single dish observations. The 0.3 Jy criteria was used to ensure  sources are radio-loud in the initial parent sample and to allow sources selected from the 3-7C catalogs, providing a large parent sample. We then selected targets with the most optimal tip/tilt star configuration that have jet sizes $<$ 20 kpc. The selected jet size criteria was imposed so we could study most of the jet-galaxy interaction within the field-of-view of the OSIRIS instrument. These sources fall in the compact steep spectrum (CSS) family except for 3C446, which is a Gigahertz peaked source (GPS). We added 4C09.17 and 4C05.84, which were not observed with SDSS, however, satisfy the radio, tip/tilt star, and archival imaging and spectroscopic data criteria. Our observed sources span a wide range of radio luminosities from the initial parent sample, ranging from the low end (7C1354) to the high end (3C9).

\section{Observations} \label{sec:observations}
Observations taken as part of our survey include OSIRIS adaptive optics (AO) observations at the W.M Keck observatory of nebular emission lines, ALMA observations of the synchrotron emission from the radio jet, and archival VLA.

\subsection{Keck OSIRIS} \label{sec:osiris-reduction}
OSIRIS \citep{Larkin06, Mieda14} observations were performed during semesters 2015A - 2017B\footnote{Program ID: U072OL, U90OL, U121OL, U184OL, U154OL, U110OL, U122OL, PI: S. Wright} on the Keck I telescope behind the laser guide star (LGS) adaptive optics system \citep{Wizinowich06}. For tip/tilt corrections, we used a nearby bright star (R magnitude $<$ 17.5) or the quasar itself if it was sufficiently bright.  For higher order corrections, a laser tuned to the Sodium line at 589.2 nm was used to create an artificial star at an altitude of 90 km centered on our object. The observing sequence was as follows: first, we acquired the tip/tilt star with a short exposure and centered it in the OSIRIS IFS field-of-view (FOV). For off-axis tip/tilt correction, we offset to the quasar using known shifts from archival SDSS or HST imaging with observations after September 2016 utilizing the new Gaia astrometric catalogs \citep{Gaia16, Gaia18}. We took a second short exposure to make sure the quasar was centered and began science observations with 600-second exposures for individual frames. We dithered each source by a few lenslets between science exposures and observed a pure sky region once an hour. Each quasar was observed in multiple filters to cover key nebular emission lines, with the goal to at minimum cover the wavelength range of redshifted \ha, \nii, and \oiii lines. Choices between FOV and wavelength coverage prevented us from obtaining a consistent line coverage for all sources. Table \ref{tab:obs-osiris} includes the nebular lines observed for each quasar. 

Each source was observed at least once in photometric conditions. Data for sources taken in non-photometric conditions were flux scaled to photometric nights using the quasar flux. This done by first constructing a 2D quasar image by taking an average along the spectral axis, and then we extract the flux by performing aperture photometry with an aperture correction from a curve of growth. We use flux conversions from DN/s to cgs from photometric nights. Non-photometric nights had a maximum visual extinction of 1 magnitude as measured from the CFHT sky-probe\footnote{www.cfht.hawaii.edu/Instruments/Elixir/skyprobe/home.html}. 

For some sources, we took observations in both narrow and broadband filter modes. The combined data cubes for these sources have variable noise properties as a function of wavelength and spatial location. The PSF for these sources also varies between overlapping and non-overlapping spectral regions. 

The FOV of our observations ranges from 14$\times$27 kpc$^2$ to 28$\times$55kpc$^{2}$ with a spectral resolution of 80-100 \kms.


\begin{deluxetable*}{llllll}
\tablecaption{OSIRIS-LGS Observational Summary \label{tab:obs-osiris}}
\tablehead{
\colhead{Name} & 
\colhead{Observing Dates} & 
\colhead{Observing} &
\colhead{Integration time} &
\colhead{Nebular} &
\colhead{PSF FWHM} \\
\colhead{} & 
\colhead{UT} &
\colhead{Mode} & 
\colhead{N$\rm_{frames}\times s$} & 
\colhead{Lines} &
\colhead{\arcsec}}

\startdata
3C 9 & 150809 & Kn1 & 6$\times$600 & \ha,\nii &0.14\arcsec$\times$0.14\arcsec  \\
     & 150809 & Hn1 & 6$\times$600 & \oiii & 0.15\arcsec$\times$0.16\arcsec \\
4C 09.17 & 160918  & Hn2 & 10$\times$600 &\oiii & 0.11\arcsec$\times$0.093\arcsec \\
         & 160919  & Kn1 & 12$\times$600 &\ha,\nii & 0.071\arcsec$\times$0.079\arcsec \\
3C 268.4 & 150405 & Hn2 & 6$\times600$ &\ha,\nii,\sii & 0.19$\times$0.17\arcsec \\
         & 150406 & Jn1 & 7$\times600$ &\oiii & 0.3\arcsec$\times$0.27\arcsec \\
         & 150406 & Hbb & 3$\times600$ &\ha,\nii,\sii & 0.19\arcsec$\times$0.17\arcsec \\
7C 1354+2552 & 160619 & Hn1 & 11$\times600$ & \oiii & 0.18\arcsec$\times$0.14\arcsec \\
             & 160620 & Kn1 & 8$\times600$ & \ha,\nii & 0.13\arcsec$\times$0.12\arcsec \\
3C 298 & 140519 & Hn3 & $4\times600$ &\ha,\nii,\sii & 0.113\arcsec$\times$0.113\arcsec\\
      & 140520 & Jn1 & $4\times600$ &\hb,\oiii & 0.127\arcsec$\times$0.127\arcsec \\
3C 318 & 160619 & Jbb & $4\times600$ &\hb,\oiii & 0.34\arcsec$\times$0.26\arcsec \\
      & 160620 & Jn3 & $4\times600$ & \oiii & 0.34\arcsec$\times$0.26\arcsec \\
      & 170717 & Hn4 & $11\times600$ & \ha,\nii\sii & 0.25\arcsec$\times$0.22\arcsec \\
      & 170718 & Hn4 & $9\times600$ &\ha,\nii,\sii & 0.25\arcsec$\times$0.22\arcsec \\
4C 57.29 & 150809 & Hn2 & 7$\times600$ &\oiii,\hb & 0.11\arcsec$\times$0.11\arcsec \\
         & 150809 & Kn2 & 5$\times600$ &\ha,\nii,\sii & 0.13\arcsec$\times$0.13\arcsec \\
4C 22.44 & 160919 & Jn2 & 7$\times600$ & \hb,\oiii & 0.11\arcsec$\times$0.13\arcsec \\
         & 170718 & Hn4 & 10$\times600$ & \ha,\nii,\sii & 0.094\arcsec$\times$0.12\arcsec \\
4C 05.84 & 151009 & Hn3 & 6$\times600$ & \hb,\oiii& 0.11\arcsec$\times$0.10\arcsec \\
         & 160619 & Kn3 & 8$\times600$ & \ha,\nii,\sii& 0.11\arcsec$\times$0.10\arcsec \\
         & 160619 & Jn2 & 4$\times600$ & \oii & -- \\
3C 446 & 160620 & Jn1 & $9\times600$ &\oiii & 0.16\arcsec$\times$0.18\arcsec \\
      & 160918 & Hn2 & 7$\times600$ & \ha,\nii,\sii & 0.14\arcsec$\times$0.14\arcsec\\
      & 160919 & Hn2 & 5$\times600$ & \ha,\nii,\sii & 0.14\arcsec$\times$0.14\arcsec \\
4C 04.81 & 170610 & Hn5 & 5$\times600$ &\hb,\oiii & 0.17\arcsec$\times$0.2\arcsec \\
         & 170717 & Kn5 & 8$\times600$ &\ha,\nii & 0.09\arcsec$\times$0.12\arcsec \\
         & 170901 & Kn5 & 6$\times600$ &\ha,\nii & 0.09\arcsec$\times$0.12\arcsec \\
\enddata
\end{deluxetable*}

\subsection{Archival VLA observations}
All of our sources have been observed with the Very Large Array (VLA) over the past 30 years. We scoured the VLA data archive for readily available high-quality images of the quasar jets in our systems, however not all of the data is readily available. We downloaded data sets that were reduced with the VLA AIPS automated pipeline. We chose data sets that had an angular resolution close to or better than an arc-second with integration long enough to fill the uv-space to produce high fidelity images. Table \ref{tab:VLA-archive} describes the observations. 

\begin{deluxetable*}{ccccr}
\centering
\tablecaption{Archival VLA and ALMA imaging \label{tab:VLA-archive}}
\tablehead{
\colhead{Object} & 
\colhead{Date} & 
\colhead{Project Code} & 
\colhead{Central frequency} &
\colhead{Beam} 
\\
\colhead{} & 
\colhead{} & 
\colhead{} & 
\colhead{(GHz)}& 
\colhead{}}
\startdata
      &               & VLA &         &  \\
\hline
3C318 & 1990 May 5 & AB568 & 8.4399 & 0.24\arcsec$\times$0.24\arcsec\\
4C04.81 & 1985 February 26 & AL93 & 4.86 & 0.43\arcsec$\times$0.40\arcsec\\
3C9 & 1993 January 4 &  AK307 & 8.4399 & 0.22\arcsec$\times$0.21\arcsec\\
3C268.4 & 1991 November 13 & AW249 & 8.2649 & 0.72\arcsec$\times$0.58\arcsec\\
4C57.29 & 1986 May 18 & AG220 & 1.490 & 1.53\arcsec$\times$1.0\arcsec\\
3C298 & 1991 August 6 & AJ206 & 8.4851 & 0.32\arcsec$\times$0.25\arcsec\\
\hline
      &               & ALMA &         &  \\
7C1354 & 2018 January 8-24 & 2017.1.01527.S & 153.364 & 0.48\arcsec$\times$0.36\arcsec\\ 
4C22.44 & 2017 December 17-30 & 2017.1.01527.S & 135.55 & 0.39\arcsec$\times$0.35\arcsec\\
4C05.84 & 2018 January 5-16 & 2017.1.01527.S & 138.87 & 0.41\arcsec$\times$0.30\arcsec\\
4C09.17 & 2017 December 25 - 2018 January 8 & 2017.1.01527.S & 	148.24 & 0.32\arcsec$\times$0.23\arcsec\\
3C446 & 2014 July 6 & 2012.1.00426.S & 609.04 & 0.25\arcsec$\times$0.18\arcsec\\
\enddata
\end{deluxetable*}

\subsection{ALMA}\label{sec:ALMA-reduction}
For objects where there is no readily available VLA imaging data, we have used our observations at shorter radio wavelengths (mm) to construct images of the radio jets. We have a concurrent ALMA program to study the molecular gas content of several quasars within this survey, targeting rotational CO lines redshifted into ALMA band 4 (125–163 GHz, 1.8–2.4 mm). The quasar jet still dominates the continuum emission at these wavelengths and for sources 4C05.84, 3C318, 4C09.17, 4C22.44, and 7C 1354+2552 we can produce high-quality continuum maps of the quasar jets. Data for 3C446 is taken directly from the archive and is not part of our ALMA program.




\section{Data Reduction \& Analysis} \label{sec:reduc}
\subsection{OSIRIS data reduction}

Data reduction was performed using the standard OSIRIS data reduction pipeline version 4.1.0 \cite{OSIRIS_DRP}. First, we constructed a master dark for each observing night by median combining 3 to 5 600s darks taken before/after our observing night. We subtract each master dark from the raw 2D spectra. ``Adjust channel levels", ``Remove Crosstalk" and ``Glitch Identification" routines were only run on data taken before the detector upgrade in early 2016. The new Hawaii-2RG detector does not have the same artifacts which these modules were designed to correct; see \cite{Boehle16} for further discussion. The pipeline performs a Lucy-Richardson deconvolution to extract raw spectra by using unique rectification matrices for each observing mode. The rectification matrices are calibration files that contain the instrumental PSF for each lenslet at each wavelength location. Finally, the ``Assemble Data Cube" routine is run to place the extracted spectra in the correct spatial location and construct a three-dimensional data cube. Sky cubes were subtracted from the science observations using the ``Scaled Sky" module, which scales OH-lines in the sky data cube to that of the science frame before subtraction. The cubes were then combined using the ``Mosaic Frames" routine which registers cubes to the same coordinate system based on AO offset header keywords and uses a 3$\sigma$ clipping algorithm for combining.

An A-type star is observed either preceding or succeeding each science observation for both telluric and flux calibration. We select a star brighter than eighth magnitude from the 2MASS catalog for flux calibrations. The star is selected not to be variable or a known optical/spectroscopic binary. We select stars to have coordinates that when observed, will have an airmass roughly matching the average airmass of our science observations. The observations are taken in NGS mode with a typical exposure time of 1.4-10s depending on stellar brightness and observational mode. For each star, we obtain a pure sky observation by offsetting to an empty sky region a few arcseconds away.

We reduce the standard star in the same manner as regular science observations. We produce a 2D image of the star by averaging the data cube along the spectral axis. We then construct a curve of growth using the 2D stellar image and extract the spectrum using a small radius, approximately at the 34$\%$ of star's spatial width. We apply an aperture correction to each spectrum based on the curve of growth to obtain the total stellar counts. Hydrogen absorption lines in the stellar spectra are fit with a Lorentz profile and removed from each spectrum. We construct a model stellar spectrum using the Planck function normalized to the average flux of the star in the broad J, H, or K bands from 2MASS. We bin the (J, H or K) 1D model spectrum to the OSIRIS spectral sampling and then divide the observed stellar spectrum by the model. This provides a conversion factor between DN/s and \ferga at each channel in the data cube. We estimate the absolute flux accuracy to be at $\sim 10\%$.

\subsection{PSF subtraction}

The broad line region (BLR) of a quasar is spatially unresolved in our observations since the emission occurs on small scales: 10-400 light days \citep{Peterson04}. An IFS is capable of constructing a PSF image from the data cube using only channels confined to the BLR. \cite{Jahnke04a} conducted some of the first IFS observations of low-redshift (z$\sim0.2$) quasar host galaxies to search for extended emission in nebular emission lines. With the deployment of near-infrared IFS and adaptive optics, the search for extended emission from quasar host galaxies shifted to higher redshifts (z$>1$). \cite{Inskip11} presents detection of a quasar host galaxy in \ha for a z = 1.3 quasar using SINFONI on VLT. Our team shortly followed up an observational program to detect nebular emission from quasar host galaxies at z$\sim$2 with OSIRIS at Keck and NIFS on Gemini. We demonstrated that we could detect extended emission on scales $\gtrsim$0.2\arcsec~from the quasar down to flux levels of a few $\times10^{-17}$ \ferg. We present a detailed description of our PSF subtraction routine in \cite{Vayner16}. Herein we provide an overview and some additional improvements that we have made to our PSF subtraction routine.

We select channels that are part of the quasar broad line emission and quasar continuum that do not overlap with strong OH emission from the sky. We avoid regions within $\pm2000$ \kms from the peak of a broad-line \ha, \hb or NLR \oiii 495.9, 500.7 nm, \nii 654.9, 658.5 nm or \sii 671.7,673.1 nm emission to avoid including potentially extended emission from the quasar host galaxy, since we expect the majority of the extended gas to emit close to the redshift of the quasar. Previously we only selected groups of channels between OH lines that are close to the peak of the broad emission line (within $\pm$10,000 \kms from the peak of a line). With further testing and a more extensive data set, we found that selecting more data channels for constructing the PSF image produced less noise in the PSF subtracted data cube. The difference from our 2016 algorithm is that we now select all available channels that do not coincide with OH emission, low atmospheric transparency, spectral edge channels (due to filter transmission artifacts) or near ($<\pm$2000 \kms) the peak of a broad-line Balmer (e.g., \ha, \hb) or forbidden (e.g.,\oiii, \sii) emission lines. Within individual OSIRIS data cubes we find no strong variation of the PSF shape with wavelength. We do not perform any continuum subtraction to remove extended continuum emission from the host galaxy stellar component. This is because the sensitivity of OSIRIS is still below the flux expected from host galaxy continuum emission at these redshifts.  

Our observations are able to achieve exquisite contrast at small separations from the quasar since the PSF is constructed directly from the data cube. The PSF only varies slightly (less than a pixel) as a function of wavelength. Typically, we achieve an inner working angle that matches  the spatial resolution of the observations quoted in Table \ref{tab:obs-osiris}. Only a few (3-10) spaxels within the area of the PSF show noise above a typical sky spaxel. Similar to what we found in our pilot survey \citep{Vayner16}, spaxels at separations $>$0.15-0.2\arcsec\ from the quasar are dominated by the sky background, rather than systematic noise from PSF subtraction. After PSF subtraction, bad spaxels within the inner working angle are identified by comparing their standard deviation to an empty sky spaxel, and their spectra are replaced with an average spectrum of a sky region before performing any analysis. This effectively removes any possible PSF-subtraction artifacts or unresolved emission into the extended quasar host galaxy flux. After subtracting the PSF from the data cube, we apply minor cosmetic smoothing with a 2D Gaussian to improve spaxel-to-spaxel flux variations and to bring the data sets to the same angular resolution. When computing sizes of extended emission we remove in quadrature the size of the Gaussian kernel that was convolved with the data cube. 

\subsection{Emission Line Fitting}\label{sec:line_fitting_maps}
We construct integrated intensity, radial velocity, and dispersion maps from fits to nebular emission lines.

For PSF-subtracted data cubes, we construct a spectrum in a large aperture centered on the quasar. We collapse the cube along the spectral axis using the channels near the peak of the identified emission line, effectively creating a moment zero map for that line. Some data cubes show multiple emission line peaks, from distinct kinematic structures in the galaxy, either from bi-conical outflows, merging galaxies or rotating discs. In such data cubes we create a moment map for each distinct spectral feature. We select a sky region in each map, where we do not see any strong emission, to estimate the background noise. The moment zero maps are divided by the background noise to construct a SNR map for each velocity component. The spectral window for each moment zero map is the FWHM of the spectral feature.

Each emission line with an SNR $>2\sigma$ is fit with a Gaussian model. Channels with strong OH emission are weighted during Least-Squares fitting using a 1 sigma error array constructed in an empty sky region of each data cube. For the \oiii 500.7 nm line, we simultaneously fit the \oiii~495.9 nm line with the position and width held fixed to the redshift and width of the 500.7 nm line. The line ratio between the \oiii~lines is held fixed at 1:2.98 \citep{Storey99}. Spaxels where \ha and \nii SNR maps both show a significant detection have the lines fit simultaneously. All the parameters on the \ha line are free while for the \nii 658.4 lines the width and redshift are held fixed to the \ha line, with only the peak as a free parameter. The 654.8 \nii line has no free parameters with the width and redshift held fixed to the \ha line, and the intensity ratio between the \nii~654.8, 658.4 nm lines are held fixed at 1:2.95.

A fit to an emission line is deemed successful and is considered ``real" if the peak has an SNR of 2 or more above the local noise and the width of the emission line is broader than the width of an OH sky emission line in the sky data cube. We compute the OH sky line width by fitting a Gaussian profile to an isolated sky line. 

Each emission line is integrated from $-3\sigma_{v}$ to +$3\sigma_{v}$ to construct integrated line intensity maps. We compute the error by taking a standard deviation in a sky region at each spectral channel. The error array is added in quadrature over the same spectral range as the integration of the emission line. We construct radial velocity maps by measuring the Doppler shift of the emission line from the redshift of the source, which we assume to be the rest frame of each system. For the case of multi-Gaussian component fit, we use the luminosity-weighted line centroid to compute the Doppler shift. We generate a velocity dispersion map from the $\sigma_{v}$ value of the Gaussian fit. Where $\sigma_{v}$ is velocity dispersion from the Gaussian fit measured in units of \kms. In the case of a multi-Gaussian fit, we use the line width, where we integrated 66$\%$ of the total line intensity as a proxy for $\sigma_{v}$. The errors on the radial velocity and $\sigma_{v}$ values are from the least squares Gaussian fit. While we are fitting emission lines that have a peak value below an SNR of 3 integrated over the line is almost always greater than an SNR of 3.

We create three color composite images from the integrated nebular emission line maps, with green assigned to \oiii 500.7 nm, red to \ha and blue to \nii 658.4 nm.. We utilize the \textit{make Lupton rgb} routine within the \textit{astropy} visualization package \citep{lupton04}. Three color composite images for each source are presented in Figure \ref{fig:color_composite}. The radial velocity and dispersion maps for individual sources are in the Appendix. For each velocity dispersion map, we overlay the radio-maps of the quasars' jets. The two images are aligned by matching the optical/near-IR location of the quasar with the component in the radio maps that shows spectrally flat (S$_{\nu}\sim\nu^{0}$) spatially compact emission \citep{Lonsdale1993,fanti02}.

\begin{figure*}
    \centering
    \includegraphics[width=6.0in]{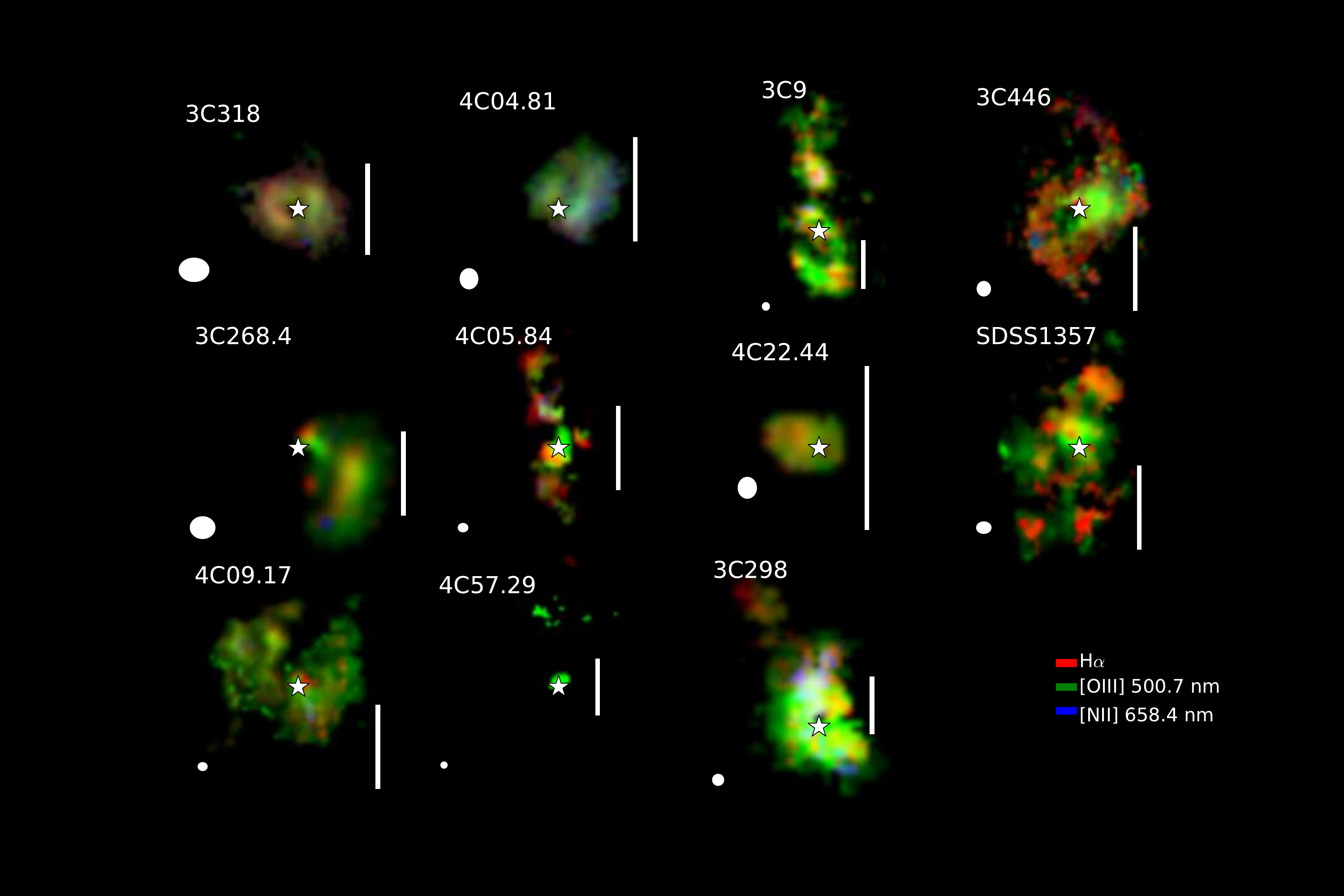}
    
    \caption{OSIRIS - LGS observations of 11 quasar host galaxies in our QUART sample. Each image is a three-color composite showing the distribution of ionized gas in these galaxies detected on scales ranging from 1.6-30 kpc after the quasar emission has been removed. Red is color-coded to \ha, green to \oiii and blue to \nii. The star represents the location of the subtracted quasar, the ellipse in the lower left corner is the measured FWHM of the PSF computed from the quasar-PSF image, and the scale bar to the right of each source represents 1\arcsec~or approximately a projected size of 8.4 kpc at the average distance of the sample.}
    \label{fig:color_composite}
\end{figure*}

\subsection{ALMA data reduction}

Data reduction was performed using CASA (Common Astronomy Software Applications \citep{McMullin07}). There is sufficient signal to noise ratio (SNR) to perform self calibrations directly on 3C318, 4C05.84 and 4C09.17. We used the CASA \textit{CLEAN} function to establish a model for the synchrotron continuum through several interactive runs with clean masks centered on high SNR features. Cleaning is performed with Briggs weighting using a robust value of 0.5 with a pixel scale of 0.05\arcsec. We used the \textit{gaincal} function to perform self-phase calibration. The self-calibrated data was then cleaned again with further phase calibration until we did not see a significant improvement in SNR on the continuum. The final root mean square (rms) improved by a factor of 3-8 in the continuum images. A single round of amplitude self-calibration was successful only for 3C318. The typical spatial resolution of the ALMA observations is 0.4\arcsec. For sources where we could not perform self-calibration, we simply imaged the measurement sets with Briggs weighting using a robust value of 0.5. It should be noted that the ALMA observations trace emission from hot spots due to high energy electrons with in-situ acceleration. Which is likely different from the electron population probed by the VLA observations at longer wavelength. For the objects 7C1354, 4C05.84, 4C22.44 and 4C09.17 we use the ALMA data sets to study the location of the radio jets relative to the ionized gas emission.

\section{Quasar sample properties}
We calculate redshifts from the \ha emission line that originates in the unresolved NLR of these quasars. We extract the spectra of the quasar emission by performing aperture photometry on the unresolved point-source emission in each data cube. Each spectrum has the broad \ha emission from the BLR fit with multiple Gaussian profiles. The number of Gaussian emission lines to fit the BLR is selected to minimize $\chi^{2}$. Most sources only required two broad profiles for the broad-line region emission and one narrow component that signifies the quasar NLR. The centroid of the Gaussian profile associated with the NLR component is used to calculate the redshift quoted in Table \ref{tab:sample}. The average uncertainty for the redshifts derived from the NLR is 10 \kms. In cases where no narrow line component is detected, we select the centroid of the brightest Gaussian component of the broad-line region emission fit for the redshift. The redshift for these sources is slightly more uncertain as we do not know what the intrinsic shape is of the BLR for these quasars. In sources where we can measure both a redshift from the NLR and the BLR we find a maximum offset of $\sim$ 100 \kms, which we take to be the redshift uncertainty for these objects. For 3C9 the broad \ha line is significantly contaminated by telluric absorption, preventing a good fit to the spectrum; therefore, we quote the SDSS redshift for this source.

The bolometric luminosities are computed from monochromatic luminosity at either 1450\AA, 3000\AA~or 5100\AA~for each object, depending on the available spectral coverage. We use the methodology described in \cite{Runnoe12} and include the correction for average orientation angle towards the accretion disk of the quasar. Both rest-frame 1450\AA~and 3000\AA~luminosities are taken from SDSS spectroscopy. For the majority of sources, we use SDSS data release 7 spectra instead of BOSS spectroscopy, as the latter is not properly flux calibrated \citep{Dawson14}. The quoted uncertainties are from the global fit to the linear correlation between monochromatic and bolometric luminosities from \cite{Runnoe12}. Only 4C04.81 and 4C22.44 have their bolometric luminosities computed from BOSS spectra and thus could have larger uncertainties. The bolometric luminosity for 3C446 is from an integrated SED from \cite{Runnoe12}. All quoted luminosity values are reported in Table \ref{tab:sample}. Star formation rates derived from \ha \citep{vayner19b} and when available from total infrared emission \citep{Barthel17,Podigachoski15} are presented in Table \ref{tab:sample}.


\begin{deluxetable*}{lccllllcc@{\extracolsep{-10pt}}}
\tiny
\tablecaption{QUART Sample properties \label{tab:sample}}
\tablehead{
\colhead{Name} & 
\colhead{RA} & 
\colhead{DEC} &
\colhead{z} &
\colhead{L$_{\rm bol}$} &
\colhead{L$_{\rm 178 MHz}$} &
\colhead{M$_{\rm BH}$}&
\colhead{SFR [\ha]}&
\colhead{SFR [Total IR]}\\
\colhead{} & 
\colhead{J2000} &
\colhead{J2000} & 
\colhead{} & 
\colhead{($10^{46}$ \ergs)} & 
\colhead{($10^{44}$ \ergs)} & 
\colhead{\msun}&
\colhead{\myr}&
\colhead{\myr}}
\startdata
3C 9 & 00:20:25.22  & +15:40:54.77 & 2.0199\tablenotemark{a} & 8.17$\pm$0.31 & 9.0 & 9.87 & 160$\pm$16 & $<310$ \\
4C 09.17 & 04:48:21.74  & +09:50:51.46 & 2.1170 & 2.88$\pm$0.14 &2.6 & 9.11 & 9$\pm$1 & 1330  \\
3C 268.4 & 	12:09:13.61 & +43:39:20.92 & 1.3987 & 3.57$\pm$0.14 & 2.3&9.56 & 51$\pm$5 & $<140$  \\
7C 1354+2552 & 	13:57:06.54 & +25:37:24.49 & 2.0068 & 2.75$\pm$0.11 & 1.4& 9.86 & 29$\pm$3 & -- \\
3C 298 & 14:19:08.18 & +06:28:34.76  & 1.439\tablenotemark{b} & 7.80$\pm$0.30 & 12 & 9.51 & 22$\pm$2 & 930$\pm$40 \\
3C 318 & 15:20:05.48  & +20:16:05.49 & 1.5723 & 0.79$\pm$0.04 & 4.0 &9.30 & 88$\pm$9 & 580$\pm$60 \\
4C 57.29 & 16:59:45.85 & +57:31:31.77 & 2.1759 & 2.1$\pm0.1$ & 1.9 & 9.10 & $<9$ & -- \\
4C 22.44 & 17:09:55.01 & +22:36:55.66 & 1.5492 & 0.491$\pm$0.019 & 0.6 &9.64 & 32$\pm$3 & -- \\
4C 05.84 & 22:25:14.70  & +05:27:09.06 & 2.320\tablenotemark{c} & 20.3$\pm$1.00& 4.5 &9.75 & 11$\pm$1 & $<$540 \\
3C 446 & 22:25:47.26 & -04:57:01.39 & 1.4040 & 7.76 & 4.4 & 8.87 & 6 $\pm$1 &-- \\
4C 04.81 & 23:40:57.98 & +04:31:15.59 & 2.5883\tablenotemark{c} & 0.62$\pm$0.02 & 9.3 & 9.58 & $<5$ & 1570 \\
\enddata
\tablenotetext{a}{SDSS redshift}
\tablenotetext{b}{Redshift derived from host galaxy CO (3-2) emission \citep{Vayner17}}
\tablenotetext{c}{Redshift relative to broad-line region emission from \ha}
\end{deluxetable*}

\section{Spatially-Resolved Regions}\label{sec:regions}

In this section, we define how we select distinct regions in our quasar host galaxies and differentiate between gas in different components of a merger system. We define a distinct region as a portion of the data cube that shares similar radial velocity, velocity dispersion, ionized gas morphology, or similar nebular line ratios. These regions may be a part or entirety of a quasar host galaxy or part of the system merging with the quasar host. The number of distinct regions varies depending on how diverse the kinematics and gas morphology are in a given system. This paper will focus on searching for distinct regions that are part of an outflow. Paper II \citep{vayner19b} focuses on distinct regions that are ``dynamically quiescent".

We define ``turbulent outflow" as regions of each galaxy that contain gas with velocity dispersion ($V_{\sigma}$) $>$ 250 \kms. We select this velocity cut-off since massive disk galaxies at z$\sim$1-3 show maximum rotational velocities of 400 \kms \citep{Forster09, Forster18}, close to the maximum rotational velocities in nearby galaxies \citep{Kormendy11}. Using the relationship between rotational velocity and central velocity dispersion of $V_{\sigma}=V_{c}/\sqrt{2}$ \citep{Kormendy11} we can derive a characteristic velocity dispersion associated with gas moving at speeds greater than the escape velocity ($V_{esc}=2\times V_{\sigma}$), corresponding to approximately 250 \kms. This velocity cut-off is likely too small to only encompass gas that will escape the dark matter halo. Star-forming galaxies with a star formation rate (SFR) of $>$1 \myr can drive outflows with velocities $>200$ \kms \cite{murray05}. This is the minimum outflow velocity expected from any energy injecting source; star formation or AGN given our sample and the sensitivity of our observations. OSIRIS is capable of reaching down to a star formation rate of 1 \myr in host galaxies of z$\sim$2 quasars at the exposure length of our observations \citep{Vayner16}. We label the turbulent outflow regions in the following manner: source name + direction + component A/B + outflow.

It should be noted that the selected velocity cutoff of 250 \kms doesn't affect much of our results. The velocity dispersion in all outflow regions is $>$ 300 \kms, and the velocity dispersion in quiescent regions is nearly all below 250 \kms. Furthermore, gas in outflow regions is likely to have a larger systematic velocity offset than ``dynamically quiescent regions".

In some cases, there may be several outflow regions per component per object. In more rare situations there may be outflows associated with multiple components of the merger. For example, in 3C 298 there is an outflow associated with the host galaxy of the quasar and the merging galaxy at 9 kpc from the quasar. We made the distinction by identifying two rotating disks in the system's radial velocity maps \citep{Vayner17}. For each outflow region, we construct a 1D spectrum by integrating over the associated spaxels, as presented in Figures \ref{fig:3C318_all} and for the rest of the sources in the appendix, \S \ref{sec:appendix} (see Figures \ref{fig:3C9_all}, \ref{fig:4C0917_all}, \ref{fig:3C2684_all}, \ref{fig:3C298_all}, \ref{fig:4C0584_all}, \ref{fig:4C0481_all}). The emission lines in each spectrum are fit with multi-Gaussian profiles in the same manner as described in section \ref{sec:line_fitting_maps}. The flux of each emission line along with the uncertainty is presented in Table \ref{tab:fluxes}. 

\begin{deluxetable*}{llccc}
\tablecaption{Fluxes of distinct outflow regions in individual sources, summed over all components with a velocity dispersion $>$ 250 \kms \label{tab:fluxes}}
\tablehead{\colhead{Source}&
\colhead{Region}&
\colhead{$\rm F_{[OIII]}$}&
\colhead{$\rm F_{H\alpha}$}&
\colhead{$\rm F_{[NII]}$}\\
\colhead{}&
\colhead{}&
\colhead{10$^{-17}$ \ferg}&
\colhead{10$^{-17}$ \ferg}&
\colhead{10$^{-17}$ \ferg}
}

\startdata
3C9     & SE component A outflow  & 42$\pm$4   & 10$\pm$1 & 10$\pm$1 \\
4C09.17 & S/E component A outflow & 71$\pm$7 & 6$\pm$1 & --\\
3C268.4 & SW component A outflow & 98$\pm$10&--&--\\
3C 298  & W component A outflow & 1276$\pm$130 & 204 $\pm20$ & 88$\pm$9\\
        & E component A outflow & 178$\pm$20 & 52$\pm$5 & 16$\pm$2\\
        & SE component B outflow & 27$\pm$3 & 14$\pm$2&-- \\
3C318   & E,W component A outflow & 206$\pm$21 & 197$\pm$20 & 98$\pm$10 \\
4C05.84 & S  component A Outflow  &22$\pm$2 & 8.1$\pm$0.8&0.3$\pm$0.03\\
        & NE  component A Outflow  &22$\pm$2 &5.7$\pm$0.6&2.8$\pm$.3\\
4C04.81 & E component A outflow    &453$\pm$45 &127$\pm$13 &51$\pm$5 \\
\enddata
\end{deluxetable*}

\begin{figure*}
    \centering
    \includegraphics[width=7in]{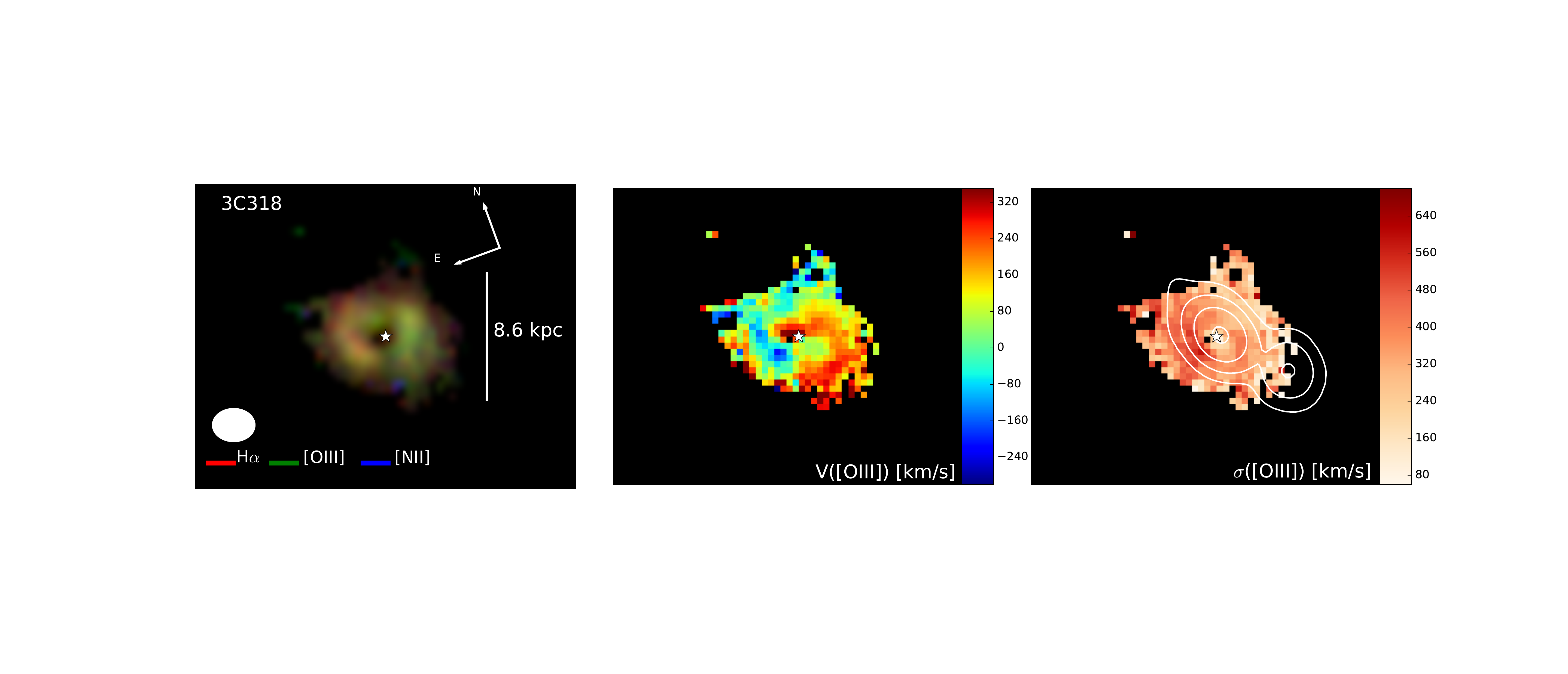}
    \includegraphics[width=7in]{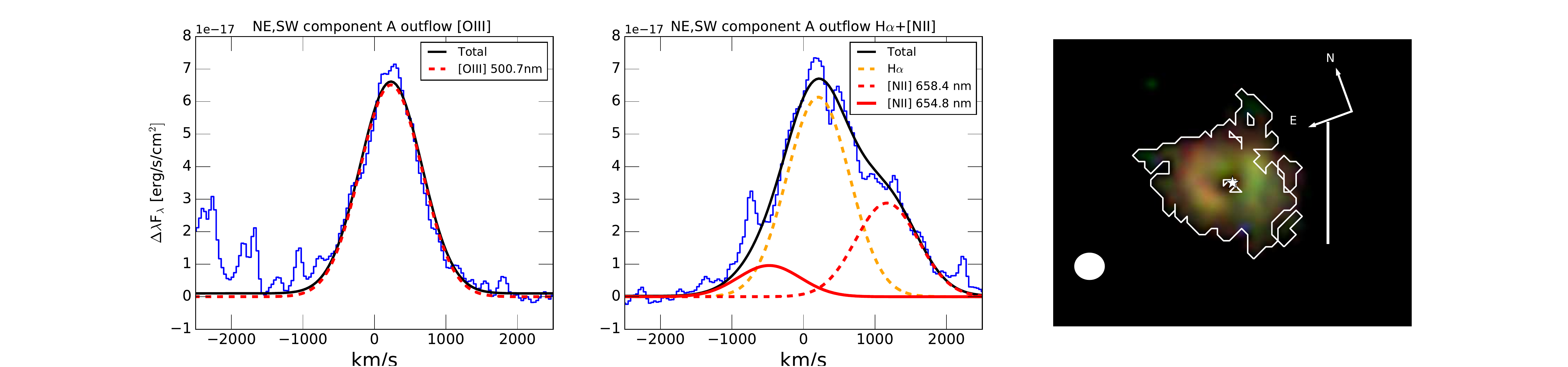}
    \caption{Example of nebular emission line distribution and kinematic maps in the source 3C318 produced from the PSF subtracted data cubes. (Top-left) Three color intensity map of nebular emission lines: \ha (red), \oiii (green), and \nii (blue). Ellipse in the lower left corner showcases the spatial resolution of the observations. (Top-middle) Radial velocity offset (\kms) of the \oiii line relative to the redshift of the quasar. The white star shows the location of the subtracted quasar. (Top-right) Velocity dispersion (\kms) map of \oiii emission. The white contours are VLA observations of radio synchrotron emission from the quasar jet and lobes. (BOTTOM) Spectra of a distinct outflow region along with fits to individual emission lines for the 3C 318 system produced from the PSF subtracted data cubes. On the left we show the fit to the \oiii 500.7 nm emission line, in the middle we present the fit to the \ha and \nii emission lines and on the right we show a three-color composite along with a contour outlining the spatial location of the region. Similar plots for the rest of the sources in our sample are presented in the appendix.}
    \label{fig:3C318_all}
\end{figure*}

\section{Spatially Unresolved Nuclear outflows}\label{sec:nuclear outflows} 
\begin{figure*}
    \centering
    \includegraphics[width=6.5in]{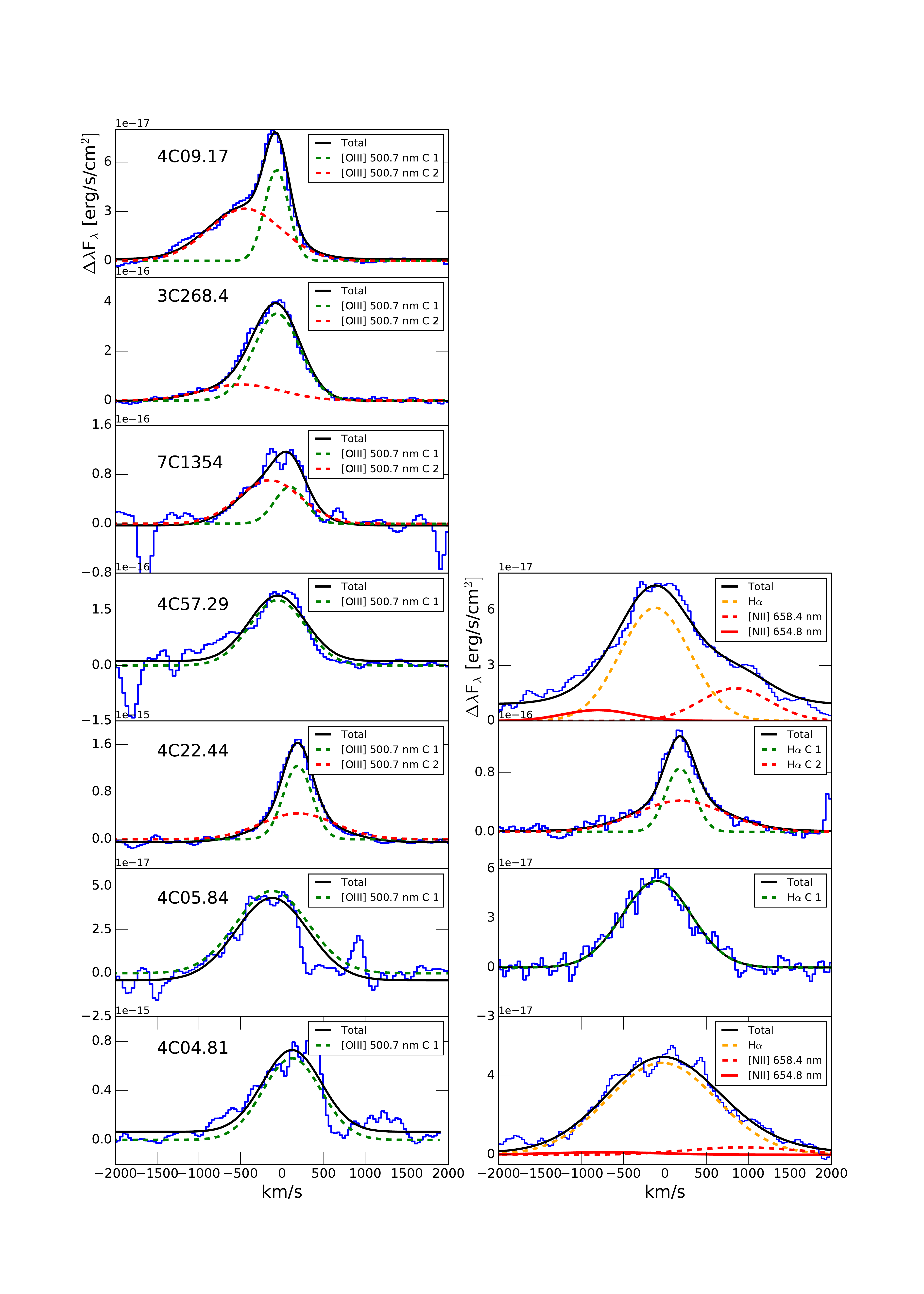}
    \caption{Spectra of unresolved outflows detected within our sample. On the left we present the extracted spectra covering the rest-frame 500.7 nm \oiii emission line, along with the multi-Gaussian fits. On the right, for objects where \ha was detected in the nuclear outflows, we present the spectra along with the fits to both \ha, and when possible to the \nii emission line too. Emission from the broad-line region has been subtracted out from these spectra.}
    \label{fig:nuclear_outflows}
\end{figure*}

For each object, we search for nuclear unresolved outflows by subtracting a model of the extended emission from the PSF un-subtracted data cubes. The model consists of the Gaussian fits at each spaxel to the detected emission lines. We then perform aperture photometry on the point source emission using a curve of growth method. We fit the broad \ha and \hb~emission from the broad-line region with a combination of broad Gaussian profiles (similar to section \ref{sec:sample-selection}) and include intermediate width emission lines ($250<V_{\sigma}<2000$ \kms) in \oiii and \ha when present to account for emission from the narrow-line region and nuclear outflows. For 4C09.17, 4C57.29, 3C268.4, 4C04.81, 4C57.29, 4C22.44. and 7C 1354+2552 we detect broad spatially-unresolved asymmetric emission lines. Integrating along the spectral axis over the velocity range of the intermediate emission lines ($250<V_{\sigma}<2000$ \kms) indeed produces an image consistent with a point source. The spectra along with the fits to the emission lines are presented in Figure \ref{fig:nuclear_outflows}. Due to the detection of \oiii, we believe that these outflows extend far beyond the broad line region for each quasar. The [OIII] 500.7 nm is a forbidden emission line that is suppressed by collisional de-excitation at $n>8\times 10^5$ cm$^{-3}$, hence it must arise on scales of at least a few hundred pc \citep{Hamann11}. In 4C09.17, 4C05.84, 3C268.4, and 4C04.81 we believe the unresolved outflows are connected to the galaxy-wide outflows since they show similar emission line profiles. Most-likely these are the base of the galaxy-scale outflow at small separations ($<1$kpc) from the quasar, within the inner working angle of our LGS-AO observations. For example, note the similarity between the profile of the \oiii emission line for the unresolved outflow in 4C09.17 in Figure \ref{fig:nuclear_outflows} and the extended outflow region in Figure \ref{fig:4C0917_all}. We set a limit on the radius of these regions by measuring the FWHM of the PSF by first integrating the data cube along the spatial axis and fitting a Gaussian plus Moffat profile to the image. For 7C 1354+2552, 4C22.44, and 4C57.29, no extended outflows are detected at separations $>1$kpc.

\section{Outflow rates \& energetics}\label{sec:out_rate}
We derive the outflow rates, energetics, and momentum fluxes for individual outflow regions detected in our sources. We then compare these values to predicted energy and momentum deposition from various origins (e.g., SNe, stellar winds, quasar outflows) to establish the dominant source of the galaxy scale outflows.

\subsection{Ionized gas mass}
First, we estimate the ionized gas mass of individual outflow regions. We assume constant density across each outflow and assume that each cloud inside the outflow has the same density. Under these assumptions, the ionized gas mass can be written as,

\begin{equation}
     M_{gas~ionized}=(n_{p}m_{p} + n_{He}m_{He})Vf . 
     \label{eq:total_ionized_mass}
\end{equation}
Where $V$ is the volume of the emitting region, $f$ is the volume filling factor (the ratio of the volume of emitting clumps to the total volume of the region), and $n_{p}$ is the proton number density. Finally, $n_{He}$ and $m_{He}$ are the number density of Helium and the mass of a helium atom. We assume a solar abundance for Helium and that the gas in the outflow region is fully ionized where Helium is an equal mix of HeII and HeIII. Under these assumptions, we get the following relationships:

\begin{equation}
\begin{aligned}
    n_{He} &= 0.1 n_{p}\\
    n_{e} &= n_{p} + \frac{3}{2}n_{He}\\
    n_{e} &= 1.15 n_{p}
\end{aligned}
\label{eq:relationships}
\end{equation}

Using the formulation of \cite{OsterbrocknFerland06}, the \ha luminosity due to recombination is given by:

\begin{equation}
    L(H_{\alpha}) = n_{e}n_{p}j_{H_{\alpha}}Vf
    \label{eq:halpha_recombination}
\end{equation}

where $j_{H_{\alpha}}$ is the line emissivity assuming Case B recombination. Under case B recombination the Hydrogen gas is optically thick to ionizing radiation, and all photons produced from an excited-to-ground-state transition are immediately reabsorbed. Hence,  we omit downward radiative transitions to the ground state. Isolating $Vf$ in equation \ref{eq:halpha_recombination} and substituting into equation \ref{eq:total_ionized_mass} along with relationships from equation \ref{eq:relationships} we obtain the following equation for the total ionized gas mass:

\begin{equation}\label{equation:ionized_gas_mass}
    M_{gas~ionized} = 1.27 \bigg(\frac{m_{p}L_{H\alpha}}{j_{H_{\alpha}}n_{e}}\bigg)
\end{equation}

where $\rm L_{H\alpha}$ and $\rm n_{e}$ are the integrated \ha luminosity and the average electron density over the outflow region. Since we cannot measure the electron temperature for our objects, we assume a range of 1-2$\times10^{4}$ K, which constrains the \ha line emissivity to 1.8-3.53$\times10^{-25}\rm~erg~cm^{3}~s^{-1}$ \citep{OsterbrocknFerland06} for an electron density of $\sim10^{2-3}$ \eden. Similarly, a lack of information about the electron density can lead to order of magnitude uncertainties on the ionized gas mass, due to a large range of densities in which ionized gas can exist.  Uncertainties on the electron temperature add a factor of a few. 

\begin{figure}
    \centering
    \includegraphics[width=3in]{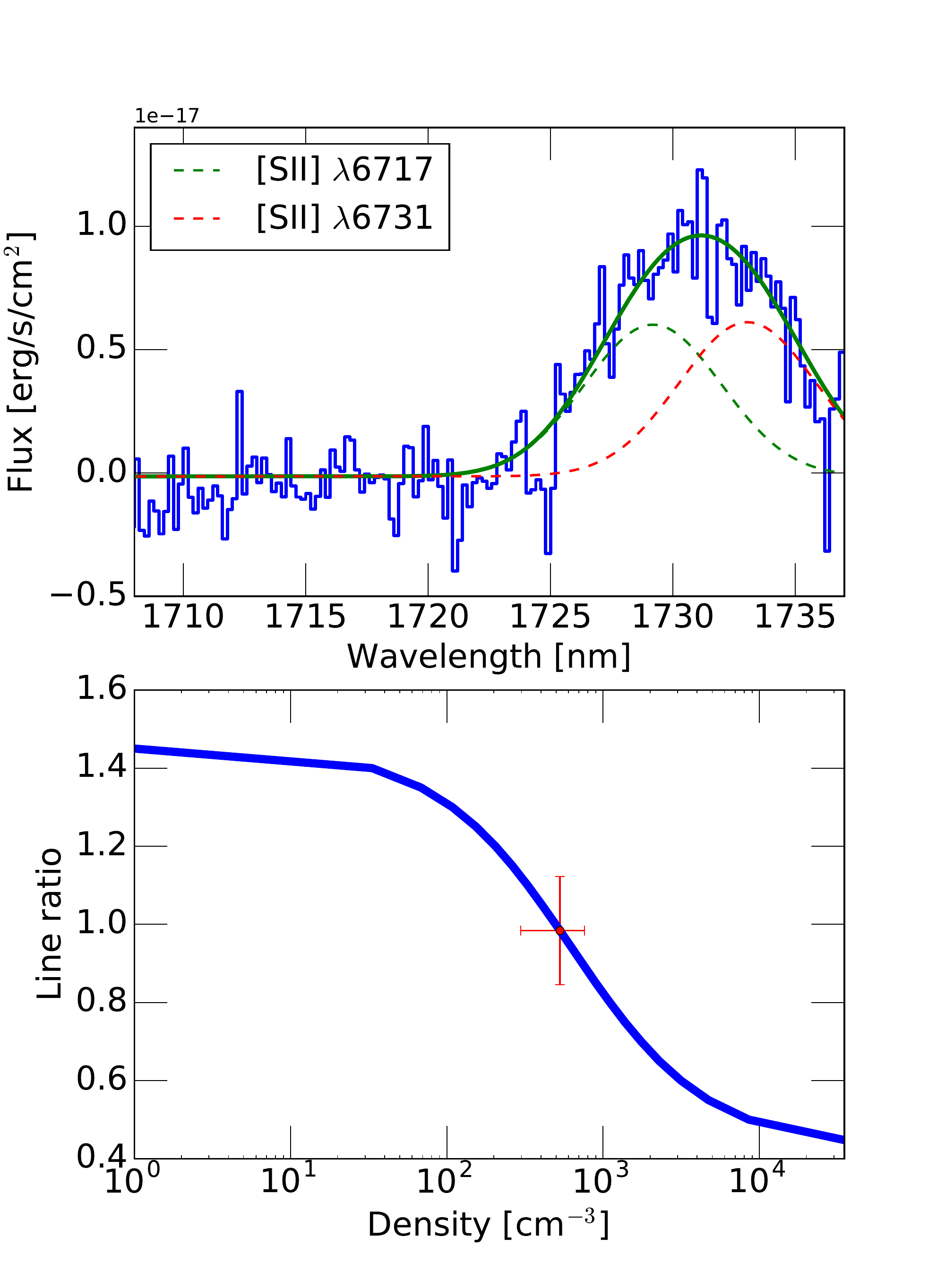}
    \caption{Measuring the electron density for the outflow region in the host galaxy of 3C 318. (TOP): Spectrum in the vicinity of the [SII] doublet along with a double Gaussian fit to each emission line. (BOTTOM): Line ratio vs. density from PyNeb for an electron temperature of 1$\times10^{4}$ K. The red point and associated vertical error bars represents the measured value of the line ratio. The red horizontal error bar shows the uncertainty on the electron density as measured from this line ratio. We obtain an electron density of 529$\pm$232 \eden}
    \label{fig:sii_ne_3C318}
\end{figure}

We were able to measure the electron density directly for two objects (3C318 and 3C298) from the 671.7 nm \& 673.1 nm \sii~lines ratios (Figure \ref{fig:sii_ne_3C318}). We use the \textit{getTempDen} routine from the \textit{PyNeb} \citep{Luridiana15} package to derive the electron density. For 3C318 and 3C298, the largest uncertainty on the electron density comes from measuring the \sii~line ratios with the limited spectral resolving power of OSIRIS (R$\sim$3800). In the rest of the objects where we cover the \sii~doublet, the lines are undetected. For the rest of the targets we assume an electron density of 500 \eden with an uncertainty range of 100-1000 \eden. We assume an electron density in the range of 500-1500 $\rm cm^{3}$ for the nuclear outflows, slightly higher than our assumption on extended outflows since the nuclear outflows might be denser due to their smaller sizes. The selected electron density is within the range of values found in AGN driven outflows in the distant and nearby Universe \citep{Harrison14, Kakkad18, Forster19}. Furthermore, based on gas photoionization models using Cloudy, the assumed electron density is in the range of what the models predict at the extent of the ionized outflow within our observations \citep{Hamann11}. The study by \cite{Carniani15}  also assumes the same electron density for outflows in type-1 quasars at z$\sim 2.4$, allowing for direct comparison. Reducing the uncertainties on outflow rates requires observations at a higher spectral resolving power. Since the emission lines are generally very broad, and the \sii~doublet is hard to separate, future IFS instruments with higher spectral resolving power are crucial for measuring accurate electron densities over outflow regions.

Another way to compute the ionized gas mass is from the \oiii line luminosity and electron density as presented in \cite{Cano-Diaz12}. This method involves an extra uncertainty, as it requires an assumption about the metallicity of the gas in the outflow region. We simply assume solar metallicity. The \oiii based estimate is inversely proportional to gas metallicity, so, if the metallicity is lower than solar, the ionized gas mass and subsequently the outflow rates will be underestimated. A drop in the metallicity from solar to half solar would increase the outflow rate by about a factor of 3. This method also requires an estimate of the fraction of Oxygen atoms in the OIII ionization state. To get a lower limit to the mass loss rate, we assume that all of the oxygen in the outflow is doubly ionized.

\subsection{Outflow rate}
Given a measured gas mass, flow size, and flow velocity, the simplest estimate for the mass loss rate is

\begin{equation}\label{equation:outflow-simple}
    \dot{M}=\frac{Mv}{R},
\end{equation}

\noindent This equation makes the assumption that the majority of the gas is confined to a thin spherical/cone-like shell.

With the assistance of AO and medium resolution spectroscopy, we can resolve each outflow, typically with several resolution elements across each region and many spectral widths across each emission line. We find that the majority of the outflows resemble a cone-like structure with evidence for either a bi-conical shape with blue and redshifted outflows in the same system or one-sided cones. None of the sources appear to show outflow emission covering 4pi steradian on the sky; hence the reason we refer to the outflows as being cones. We refer to an outflow being a bi-conical if we see two distinct components that are spatially symmetric around the quasar. A one-sided conical outflow refers to detecting an outflow spatially asymmetric around the quasar that is predominantly blueshifted or redshifted relative to the quasar. The cones appear to be mostly filled since in all cases the line profile of the outflowing gas spans a large radial velocity range, spanning red and blueshifted velocities continuously along our line of sight, and are mostly best fit with a single Gaussian profile. Furthermore, almost all of our outflows extend towards the quasar. We find bi-conical outflows in 3C318, 3C298 and 4C05.84 while we see one-sided conical outflows in 3C9, 4C04.81, 3C268.4 and 4C 09.17.

The outflow in 3C298 is the only system where one of the cones appear to show emission from both the receding and approaching side of a conical shell. However even in this case the emission lines span a large velocity range and the conical shells appear to be thick enough to assume the majority of the cone is filled with gas and the hollow portion might be relatively thin. 

\noindent A more fitting way to calculate the mass outflow rates for our objects is to assume a conical outflow model with a constant density, where the outflowing cone is filled with gas clouds as presented in \cite{Cano-Diaz12}:
\begin{equation}\label{equation:flow-cone}
    \dot{M}=vR^{2}\Omega \overline{\rho},
\end{equation}
where the density is given by $\overline{\rho}=M/V=\frac{3M}{R^{3}\Omega}$. The velocity of the material in the outflow is $v$, assumed to be constant over the cone, $R$ is the radial extent of the cone, and $\Omega$ is the opening angle of the cone. Substituting the density formula into equation \ref{equation:flow-cone} yields the mass loss rate:
\begin{equation}\label{equation:outflow-cone}
    \dot{M}=3\frac{Mv}{R}.
\end{equation}

\noindent This outflow rate is a factor of 3 larger than the simple outflow rate formula given by equation \ref{equation:outflow-simple}. The wings of the emission lines most likely give the true average velocity of the outflow, as the lower velocities seen in the line profiles are probably due to projection effects of the conical structure \citep{Cano-Diaz12, Greene12}. To measure the velocity in the wings of the emission line we use a non-parametric approach by first constructing a normalized cumulative velocity distribution:
\begin{equation}
    F(v)=\int\limits_{-\infty}^{v}f(v')/\int\limits_{-\infty}^{\infty}f(v')dv'
\end{equation}
on individual Gaussian fits in the spectrum integrated over a distinct outflow region, where $v$=0 is at the peak of the Gaussian profile. For a Gaussian, $F(v)$ is a smooth monotonically increasing function; $v_{10}$ is defined where $F=0.1$ (i.e., the velocity where 10$\%$ of the line is integrated). In Table \ref{tab:outflow-prop} $v_{10}$ is presented under the $\rm V_{out}$ column.  To compute $R$, we construct a curve of growth on the extended \oiii emission line map, integrating from the centroid of the quasar for each outflow region. Individual outflow regions are isolated from the rest of the host galaxy emission using a mask that includes spaxels with broad emission lines (see section \ref{sec:regions} for identifying extended outflow regions). 

The outflow radius is taken to be the radius which contains 90$\%$ of the \oiii flux in the region. This assumes that the outflow began near the quasar and had been expanding outwards. It is possible that the outflow started at some $R_{initial}$ away from the quasar, in which case it would be more appropriate to use $\Delta R = R_{outflow}-R_{initial}$ as the radial extent of the cone. However for all sources but 3C9 we see that the outflow extends all the way to the inner working angle (0.2\arcsec~from the quasar) of our PSF subtracted data cubes, showing that the outflow extends to the innermost resolved regions of the each galactic nucleus. Thus, using $R$ vs. $\Delta R$ would not make a significant difference in the outflow rates. In fact, in 4C05.84, 4C04.81, and 4C09.17 we find that the outflow extends within the inner working angle of our OSIRIS observations, due to the detection of unresolved outflows, after subtracting the extended emission.

We present the outflow rates from extended outflows in Table \ref{tab:outflow-prop}; they range from 23-700 \myr for masses derived from \ha luminosity and 5-460 \myr for masses derived from \oiii. The error bars quoted in Table \ref{tab:outflow-prop} include the photon counting statistics and flux calibration uncertainties, but are dominated by the uncertainty on the electron density over the outflow regions. Outflow rates and energetics for nuclear outflows are recorded in Table \ref{tab:outflow-prop-unreasolved}.


For 3C268.4, we only measure the outflow rate based on ionized gas mass derived from \oiii. This ionized gas mass is most likely a lower limit. It is likely that the ionized outflow rate in 3C268.4 is about 3 times larger if we scale it by the smallest ratio that we find between ionized gas masses derived from \oiii and \ha in extended outflows. For 4C09.17, we measure a higher outflow rate from the \oiii emission line than \ha. Due to the difference in the FOV between the two modes used to observe \oiii and \ha, we were unable to probe the entire extent of the outflow seen in \oiii with \ha, leading to a smaller extended \ha flux.

\subsection{Possible driving source of outflows}\label{sec:drive_outflow}

To constrain the driving mechanism for the outflows, we measure the momentum flux of each outflow region. We compare the measured momentum flux and kinetic luminosity of the outflow to predicted energy and momenta depositions from radiation pressure of the quasar's accretion disk, outflows driven by BAL and/or UFO winds, radiation pressure from stars on dust grains in the host galaxy, or mechanical feedback from supernovae explosions. The total measured momentum flux for the outflow is: 
\begin{equation}
    \dot{P}_{\rm outflow}=\dot{M}\times v
\end{equation}
while the momentum flux of the quasar accretion disk's radiation field is given by:
\begin{equation}\label{eq:mom_deposition_AGN}
    \dot{P}_{\rm quasar}=\frac{L_{\rm bolometric}}{c}    
\end{equation}
where $L_{\rm bolometric}$ is the bolometric luminosity of the quasar from Table \ref{tab:sample} and $c$ is the speed of light. We provide the ratio of these values in Tables \ref{tab:outflow-prop} \& \ref{tab:outflow-prop-unreasolved} which range from 0.004 - 80. We interpret these values in the discussion section below. For each outflow region, we also measure the kinetic luminosity of the outflow, given by 

\begin{equation}
    L_{\rm kinetic} = \frac{1}{2} \dot{M}\times v_{out}^{2}.
\end{equation}

One approach to calculating the bolometric stellar luminosity is simply to use the integrated infrared luminosity from 8-1000 \micron~measured by fitting the mid to far infrared SED. The quasars 3C9, 3C298, 3C318, 4C04.81, 4C05.84, 4C09.17 and 3C268.4 are modeled with AGN and star formation SED from \textit{Herschel} and \textit{Spitzer} photometry presented in \cite{Podigachoski15,Barthel17}. Only the SEDs for 3C298, 4C04.81, 4C09.17 and 3C318 contain points with significant detections at wavelengths from the mid-infrared to sub-mm.

For the rest of the sources, the total infrared luminosity provides and upper limit to the luminosity associated with star formation. In the works of \cite{Podigachoski15,Barthel17} the SED is fit with a model including emission from the dusty torus as well as far infrared emission produced by reprocessed UV emission from young stars. The best fit mid-infrared SED model is subtracted, leaving behind emission from dust associated with reprocessed stellar light. However, these fitting routines do not take into account far-infrared emission associated with cooler emission from the torus and dust on kpc scales heated by the quasar. Several studies have found the quasar itself can heat dust on kpc scales to temperatures similar to heating by young stars. The SFRs derived from total infrared emission should be taken as upper limits in systems with powerful QSOs \citep{Symeonidis16,Symeonidis17}. Works by \cite{Schneider15,Duras17} both find that up to $\sim 70 \%$ of the far-infrared emission in sources with powerful QSOs can be produced by dust heated by the QSO on kpc scales. They find a mild trend between the fraction of dust heated by the QSO and the bolometric luminosity.

A second way to compute the stellar bolometric luminosity is to convert the \ha flux to a SFR. We derive the SFRs from distinct regions in Paper II \citep{vayner19b} using the empirical \ha-SFR relation from \cite{Kennicutt98}. We can then convert this SFR to a total infrared luminosity following \cite{Kennicutt98}. The total luminosity inferred from \ha might be taken as a lower limit because of dust extinction effects, possibly leading to an underestimate of the SFR. In sources with the highest SNR spectra (3C446, 3C298, 4C22.44 and 4C04.81) covering both \ha and \hb, we find a maximum V band extinction 1 magnitude \citep{vayner19b}. We can only compute a spatially integrated extinction measurement as the surface brightness sensitivity at the redshifted location of \hb is lower compared to \ha due to degradation in the AO performance at the shorter wavelength. For this reason measuring dust extinction at high redshift using the Balmer decrement method is very difficult with AO observations. In other sources wavelength coverage prevented us from measuring the ratio between \ha and \hb to compute the dust extinction using the Balmer decrement method. For consistency, we do not apply any extinction correction to the quasars' UV-derived bolometric luminosities to not affect any ratios between the energetics of the outflows and the quasar. It is likely that \ha is not produced by recombination from photoionization by O stars alone; there is also a non-negligible contribution from quasar photoionization. We use the total \ha flux from ``dynamically" quiescent regions when calculating the energy and momentum deposition from stellar feedback, however if we do a cut and only include flux from spaxels whose line ratios are strictly located within the star-formation portion of the BPT diagram the SFR can be lower for some sources. Integrating the flux over all dynamically quiescent regions to calculate the energy and momentum deposition from stellar feedback equates to integrating nearly all the narrow emission at the quasar's systematic redshift. In some sources, we exclude the unresolved narrow emission as the line ratios in these regions falls within the quasar photoionization on the BPT diagram. The contribution from star formation to their flux is minimal. Some of the turbulent outflow regions contain narrow emission gas, however the emission line ratios of the narrow component are consistent with quasar photoionization, hence we do not use their flux when computing the star formation rates. Since we use nearly all narrow \ha emission when computing the energy and momentum deposition, these can be considered strict upper limits. The \ha based star formation rates can be found in Table \ref{tab:sample}, based on the total flux at the systematic redshift of the quasar in the dynamically quiescent regions.

The choice of initial mass function (IMF) will add a factor of $\sim$1.3 uncertainty on the SFRs. We use both the \ha and the SED derived momentum fluxes for comparison with the outflows' momentum flux and momentum deposition. 

The following equations provide the momentum flux from the radiation pressure of stellar systems, given the appropriate observables:

\begin{equation}
    \rm \dot{P}_{SFR} = 5.827\times10^{34}\frac{L_{H\alpha}}{1\times10^{43} erg/s} dynes
\end{equation}

\begin{equation}
    \rm \dot{P}_{SFR} = 3.336\times10^{35}\frac{L_{IR}}{1\times10^{46} erg/s} dynes.
\end{equation}

Using Starburst99 \citep{1999ApJS..123....3L} with a Kroupa initial mass function \citep{2001MNRAS.322..231K}, we find, in terms of the SFR
\begin{equation}
    \rm \dot{P}_{SFR} = 1.5\times10^{33}\frac{\dot{M}_{SFR}}{1M_{\odot}yr^{-1}} dynes.\label{eq:pdot_SFR_SFR}
\end{equation}

These relations assume that the gas surrounding the radiation source is not optically thick to the reprocessed far-infrared radiation emitted after the primary UV or optical radiation is absorbed. If the gas is optically thick to far-infrared radiation, then the possibility arises that the reprocessed radiation may be absorbed and re-emitted multiple times. This can boost the momentum deposited in the outflow up to a factor of $\sim$ 2 on kpc scales \citep[e.g.,][]{Thompson15, Costa18}. The optically thick criteria here refers to the gas condition when the outflows are being driven, not when observed.

\noindent We use recent numerical simulations of supernovae evolution  to compute the momentum and energy available to drive winds from these events. The simulations report either the momentum input per solar mass of new stars $ P_* / M_*$ (typically in kilometers per second) or the momentum per supernova $P_*$ in solar mass-kilometers per second. The results of different studies are very similar. For example, \cite{2015MNRAS.450..504M} report 
\begin{equation}\label{eq:mom_deposition_SNe_1}
\frac{P_*}{ M_*}= 2,663
\left(\frac{Z}{Z_\odot}\right)^{-0.114}
\left(\frac{n}{ {\rm cm}\,{\rm s}^{-1}}\right)^{-0.190}
{\rm km}\, {\rm s}^{-1},
\end{equation}
while \cite{2015ApJ...802...99K} give
\begin{equation}\label{eq:mom_deposition_SNe_2}
\frac{P*}{M_*}= 2,800
\left(\frac{Z}{Z_\odot}\right)^{-0.114}
\left(\frac{n}{ {\rm cm}\,{\rm s}^{-1}}\right)^{-0.170}
{\rm km}\, {\rm s}^{-1},
\end{equation}
where we have assumed that one supernova explodes for every 100 solar masses of stars formed.

Using a value of $2,700\, {\rm km}\, {\rm s}^{-1}$ (average value between the two simulations), and scaling to a mean density of $100\,{\rm cm}^{-3}$, we find that the maximum momentum input rate available to drive a wind is
\begin{equation}
    \dot{P}_{SNe} = 7.5\times 10^{33} \frac{\dot{M}_{SFR}}{1M_{\odot}yr^{-1}} \left(\frac{n}{100\,{\rm cm}^{3}}\right)^{-0.18}
    \left(\frac{Z}{Z_\odot}\right)
    \rm dyne \label{eq:SNe_mom}.
\end{equation} 
Note that this is a factor of five higher than the momentum input rate due to radiation pressure.

Similarly, we use
\begin{equation}
    \dot{E}_{SNe} = \xi E_{SN} \dot{M}_{SFR}f_{SN} \sim 3\times10^{40} \frac{\dot{M}_{SFR}}{1M_{\odot}yr^{-1}} \rm erg s^{-1} \label{eq:SNe_energy}
\end{equation}

As for momentum deposition, equation \ref{eq:SNe_energy} assumes each supernova explosion yields $10^{51}$ erg of energy, an energy coupling fraction to the ISM of $\xi=0.1$, and a supernovae rate per unit rate of star formation, $f_{SN}=10^{-2}$. In Table \ref{tab:AGNvsSFR}, we present the momentum flux and energy deposition values derived from the SFRs in dynamically quiescent regions associated with the quasar host galaxy. When available, we also derive these values from the far-infrared luminosity obtained from SED fitting. As noted above, the quasar itself can contribute to the far-infrared emission; hence, the momentum and energy deposition derived from far-infrared observations can be over-estimated by as much as $\sim 70 \%$ \citep{Schneider15}. Furthermore, the far-infrared observations are taken with the \textit{Herschel space telescope}, where the beam can be as large as 35\arcsec. Likely the dust traced by \textit{Herschel} observations is on larger scale, comes from multiple sources or is mostly heated by the AGN. A recent study of Herschel detected galaxies with AGN have revealed a low detection rate (1/10) with ALMA \citep{Chang20}. This indicates that multiple sources can contribute to the far-infrared emission inside the Herschel beam, further overestimating the momentum deposition from star formation inside the quasar host galaxy. Another possibility is that the dust emits on larger scales and is resolved out by the interferometric observations.

In addition, recombination lines such as \ha probe star formation that happened in the past 6-10 Myr \citep{Calzetti13}. Given the observed dynamical time scales for the galactic outflows in our sample (3-10 Myr, see table \ref{tab:outflow-prop}), \ha probes the starburst event that would have driven these outflows (if it had enough energy and momentum). Therefore, we think that the momentum and energy deposition from stellar feedback is well estimated by the \ha derived SFR. Given the bursty nature of star formation in high redshift galaxies \citep{Muratov15}, recombination lines are most likely the best way to estimate the amount of energy and momentum deposition from stellar feedback in the inner few kpc of a galaxy. Other star formation tracers such as UV or far infrared can trace SFRs averaged over 100 Myr, which does not match the dynamical time scales of our outflows and would provide a SFR averaged over several star formation episodes. This can lead to the computed stellar feedback energy and momentum depositions being under or overestimated. The momentum deposition from SNe and stellar winds can also be over estimated using far-infrared derived SFRs from Herschel observations, and therefore we use the momentum and energy deposition from SFRs derived from recombination lines when deciding if stellar feedback is responsible for driving the galaxy scale outflows. In table \ref{tab:AGNvsSFR} we still list the momentum and energy deposition from all star formation indicators for completeness. 

\subsection{Condition for star formation as a potential driver of outflows}\label{sec:condition_SF_driver} 
During the first 4 Myr after a burst of star formation, radiation pressure and winds from massive stars dominate the energy and momentum deposition from stellar feedback. After about 3-4 Myr the contribution from SNe begin to dominate; with a momentum deposition rate about five times greater than stellar winds driven by radiation pressure at earlier times \citep{2015MNRAS.450..504M,2015ApJ...802...99K}. The SNe rate peaks at about 10 Myr after the burst \citep{Leitherer99}, and after $\sim 40$ Myrs stellar feedback becomes significantly less important. Feedback from supernovae can easily match or surpass the momentum deposition from the quasar accretion disk. Taking a ratio of equation \ref{eq:mom_deposition_AGN} and equation \ref{eq:mom_deposition_SNe_1} or \ref{eq:mom_deposition_SNe_2} leads to the following expression:

\begin{equation}
    \frac{\dot{P}_{SNe}}{\dot{P}_{AGN}} = 10.3 \times \frac{L_{SFR}}{L_{AGN}}
\end{equation}

\noindent assuming an ambient electron density of 100 \eden, solar metallicity, and the \cite{Kennicutt98} relationship between total stellar population luminosity and SFR. This indicates that for the case where the bolometric luminosity of the quasar and the stellar population are equal, feedback from the SNe may dominate. This is especially important in sources such as ULIRGs where a significant fraction of the bolometric luminosity is produced by stars. Understanding the SFRs in sources with AGN is crucial to deciphering the dominant driving source of galactic outflows.

We assume that radiation pressure from stars can drive the outflow if \momfluxsfr $\gtrsim$ \momfluxout. In fact it appears that radiation pressure on dust grains surrounding star-forming regions can drive outflows with \momfluxout up to $\times2$ that of \momfluxsfr on kpc scale \citep{Thompson15} through trapping of far infrared photons in the outflow; at smaller scales, where the optical depth to infrared photons can be much larger than a factor of two, the boost can be larger, since the column densities (and hence optical depths) can be much higher. However, as the gas mass tends to increase only linearly with radius, the column density tends to decrease with radius; on scales larger than $\sim 1$ kpc the available gas is not sufficient to maintain $\tau >> 1$. We are primarily interested in kpc or larger scale outflows, so we will not consider such high IR optical depths further.

It follows that, for the the outflows in our sample with \momfluxout $>2\times$\momfluxsfr, the primary driving source cannot be momentum deposition from young massive stars. 

Similarly, if \energydepsne is higher than or comparable to \kineticlum, then the outflow may be powered by a combination of energy and momentum deposition from SNe feedback and young stars. 

Finally, If $\dot P >$ \momfluxout, then the outflow may be driven by supernovae, but if the inequality is reversed, supernova driven winds cannot be the primary driver of the outflow.

\subsection{Condition for AGN as potential driver of outflow}\label{sec:condition_AGN_driver} 
If both \momfluxsfr and $\dot P_{SN}$ are substantially smaller than $2\times$ \momfluxout, then the outflow cannot be driven by star formation, and we look to the quasar as a primary driving source. There are multiple ways that a quasar can drive a galaxy scale outflow. The first is through radiation pressure on either dust grains or electrons in the host galaxy. If \momfluxagn is greater than \momfluxout, then the outflow can be driven by radiation pressure. In fact, as in the case of stellar luminosity, radiation with a momentum input rate of \momfluxagn can drive an outflow with $2\times$ \momfluxout \citep{Thompson15, Costa18}.  

In our nuclear outflows (with sub-kpc sizes), the optical depth to IR photons may be much larger than unity, resulting in even the driving of outflows with \momfluxout much larger than \momfluxagn. However, in the scenario that the outflow has a very high infrared optical depth, the amount of gas the simulations predict in the outflow is significantly larger than what we observe \citep{Costa18}. 

The momentum available from AGN radiation pressure in our quasar sample is often higher than the momentum deposition by star formation in the same object. It is also larger than the measured outflow momentum rates in most of our objects, the exceptions being 3C 318 and 4C04.81. Thus, most of the outflows in our sample may be driven by radiation pressure from the quasar accretion disk.

The second type of AGN driven outflow occurs when a broad absorption line wind, an ultra-fast outflow, a quasar jet, or a warm absorber-type wind drives a powerful shock in the host galaxy of the quasar. In situations where a fast (v $>$ 30,000 \kms) outflow drives the shock, \cite{Faucher12a} find that the rate-limiting step in cooling the shocked bubble is the coupling of shocked protons to electrons. If the Coulomb heating timescale of the electrons (to the same temperature as the shocked protons) is longer than the dynamical time, the electrons will cool, but the protons will not. Since the protons carry most of the energy, and by themselves cool very inefficiently, under these conditions, the shocked bubble expands adiabatically, leading to a boost in the radial momentum compared to that in the initial wind or jet. The net effect is much like that produced in the Sedov-Taylor phase of supernovae, in which the pressure of the hot gas provides a momentum boost to the outflowing material.

For shocks driven by slower winds, the post-shock gas is cooled by inverse Compton scattering and free-free emission. The condition for an energy conserving shock is then that the cooling time scale ($t_{c}$) is longer than the flow time scale ($t_{flow}=R_{s}/v_{s}$), and much longer than the initial crossing time scale ($t_{cr}=R_{SW}/v_{wind}$) at the radius where the gas is shocked. Here, $v_{s}$ is the velocity of the shock, $R_{s}$ is the radius of the shock, $v_{wind}$ is the velocity of the wind responsible for the shock, and $R_{SW}$ is the radius where the initial shock occurs. As far as the outflow is concerned, $t_{cool}>>t_{flow}>t_{cr}$ implies energy conservation, while $t_{cool}<<t_{cr}<t_{flow}$ implies that the wind cools, and only the initial wind momentum (or wind momentum per unit time) is communicated to the ISM, i.e.,  the wind is momentum conserving \citep{Faucher12a}. 

In the energy conserving or adiabatic scenario, the ratio between \momfluxout and \momfluxagn is expected to be $>$2-10$\times$\momfluxagn \citep{Faucher12a, Zubovas12} on kpc scales. Energy conserving or adiabatic shocks are the most efficient currently known ways to remove gas from a massive galaxy. They can also drive turbulence in the host galaxy's ISM, thereby prolonging the time necessary for gas to cool and form stars. 

In the case of an isothermal shock, \momfluxout will be smaller than \momfluxagn on kpc scales, decreasing as a function of radius \citep{Faucher12a,Zubovas12,King15}. Given that the same condition is true for a radiation pressure wind, with present data, we cannot distinguish between these two driving mechanisms (an isothermal shock vs. radiation pressure). 

\cite{Wagner12} also find a considerable momentum boost for jet driven outflows. In their simulations, a powerful quasar jet slams into the ISM and drives a hot shock. The shocked wind is at a temperature of $10^{7}$ K and expands adiabatically. \cite{Mukherjee16} finds similar results. These simulations do not treat the protons and electrons as having different temperatures past the shock, and they do not have a luminous quasar on while the shocked bubble expands. The temperature of the shocked bubble is about two orders of magnitude lower than that in \cite{Faucher12a}. 

The question is, can the quasar radiation field cool the gas through inverse Compton scatting or free-free emission for jet-driven shocks, in which case the momentum flux would drop as the shocked gas will radiate a portion of the energy provided to it by the jet. Using equation 15 from \cite{Faucher12} inverse Compton scattering can cool the gas in 18 Gyr at a kpc from the quasar for an AGN with a bolometric luminosity of $10^{46}$ \ergs and an electron temperature of $10^{7}$ K. Using equation 24 from \cite{Faucher12a} the estimated cooling time scale from free-free emission is much shorter, about 15-150 Myr for an electron density of  0.1 - 1 \eden \citep{Wagner12}. Even the 15 Myr time scale to cool the gas is much longer than the flow time of the shock. For example, in \cite{Mukherjee16} the shocked bubble propagates from 1 to 3 kpc in a matter of only 1 Myr. These results indicate that even with the presence of an intense radiation field from the quasar or free-free emission, the cooling time scale for the shocked bubble is longer than the flow time. This means that the bubble would expand adiabatically while sweeping material from the inner few kpc of the galaxy, providing a significant momentum boost to the galactic wind. The temperature and phase map in \cite{Mukherjee16} indicate the presence of gas at temperatures $>10^{7}$ with densities 0.1-0.01 \eden. These higher temperatures can prolong the free-free cooling time scale by 1-2 orders of magnitude and decrease the inverse Compton cooling by about an order of magnitude. However, even this time scale will be too long for the flow time of the hot energy bubble at distances greater than 1 kpc.

\subsection{Summary of outflows within our sample}

\begin{figure*}
    \centering
    \includegraphics[width=7in]{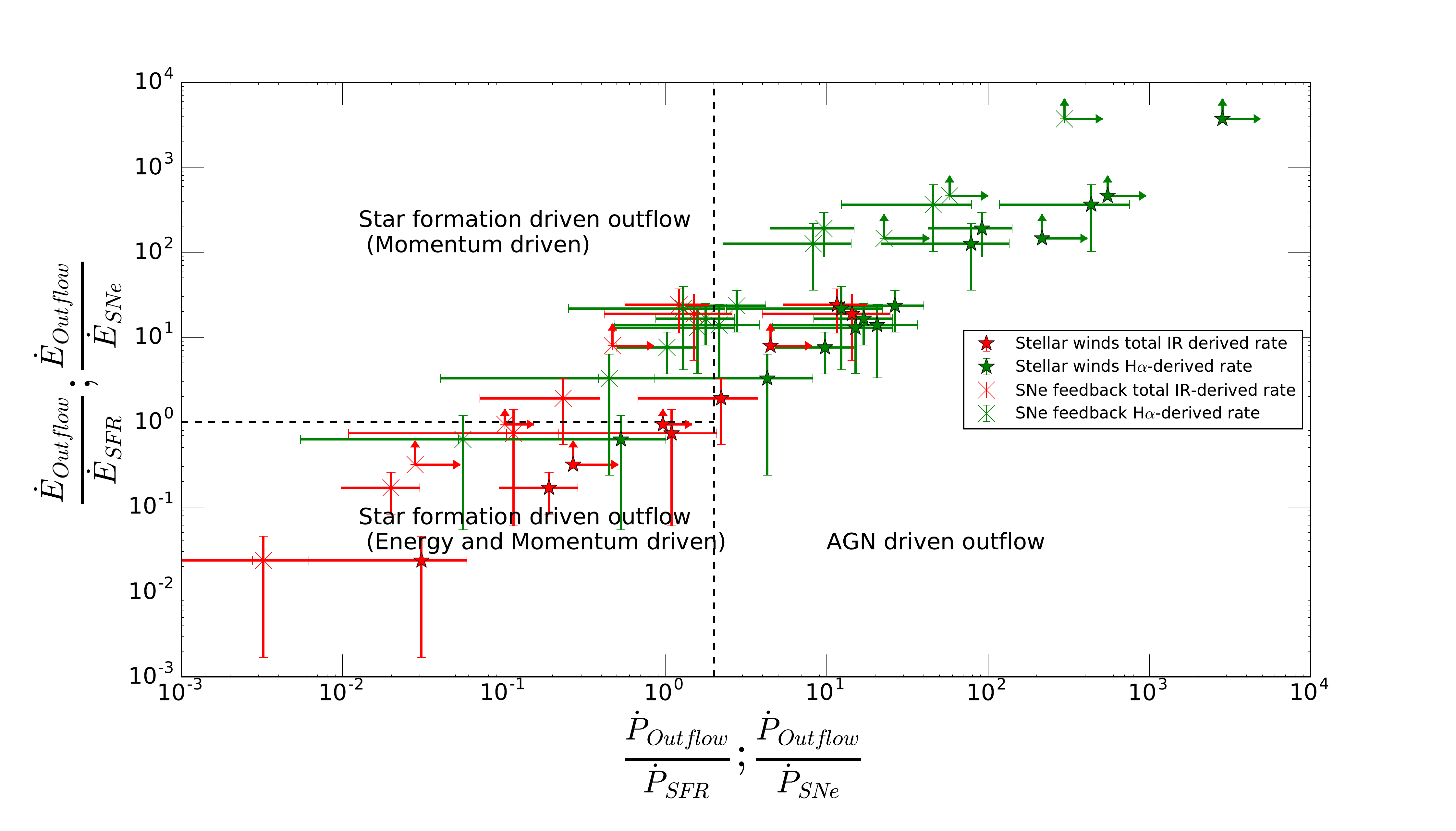}
    \caption{Diagnostic diagram distinguishing between AGN or Star formation as the main driving mechanism of the galaxy outflows. On the Y-axis, we plot the ratio of the kinetic luminosity of the outflow to the energy deposition rate from stellar feedback. On the X-axis, we plot the ratio of the momentum flux of the outflow to the momentum deposition from stellar feedback. Green points depict objects for which energy and momentum deposition are calculated from SFRs derived from the \ha emission line. Red points represent objects for which energy and momentum deposition rates are derived from total infrared luminosity. The stars represent stellar feedback models from stellar winds, while the green Xs represent SNe feedback models. We show the division between AGN vs. Star formation as the main driving mechanism with dashed lines. However some objects that fall the the SF-driven portion of the diagram, radiation pressure from the luminous quasar might still be significant enough to driven the outflow. These criteria are outlined in section \ref{sec:condition_SF_driver},\ref{sec:condition_AGN_driver}. }
    \label{fig:AGN_vs_SFR}
\end{figure*}

\begin{figure*}
    \centering
    \includegraphics[width=6.5in]{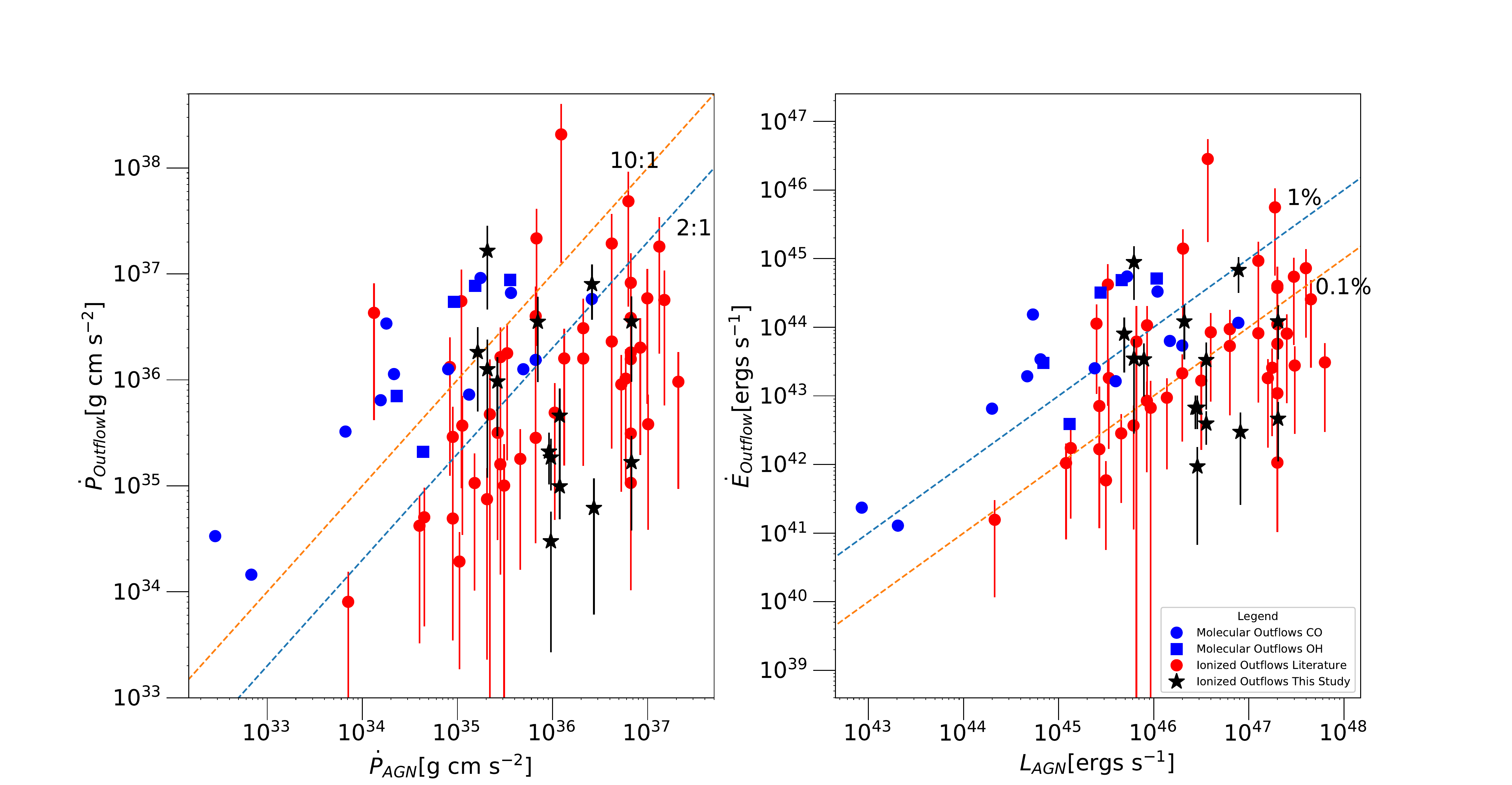}
    \caption{Diagnostic diagram distinguishing between different AGN components that may have driven the outflows. On the left, we plot the momentum flux of the outflow against the radiation momentum flux from the quasar. Red points represent outflows detected in ionized gas emission. Black stars represent points from our sample. Blue squares represent galaxies where a molecular outflow was detected through OH absorption while blue circles represent molecular outflows detected through CO emission. We plot lines of constant ratios of 2:1 and 10:1 between $\dot{P}_{outflow}$ and $\dot{P}_{AGN}$ as these ratios distinguishing between different driving mechanism. Points below the 2:1 line represent outflows that may have been driven either by an isothermal momentum conserving shocks or through radiation pressure on dust grains by the AGN or quasar. Points above the 2:1 line represent galactic outflows that were driven by a hot adiabatic shock if the radius of the outflow is $>1$ kpc. Points above the 2:1 line may also be driven by radiation pressure in very confined $<$1 kpc regions where there is very high opacity to infrared photons. On the right, we plot the kinetic luminosity of the outflow against the bolometric luminosity of the AGN. Green dashed curve represents the minimum coupling efficiency (0.1$\%$) prescribed by theoretical work necessary to clear the galaxy of its gas and establish the observed local \msigma relationship. Points on the right plot are color-coded  to the points on the left plot. About half of our detected outflows are consistent with being energy conserving and show energy coupling above 0.1$\%$. All data points from literature are taken from the following studies: \cite{Carniani15,Nesvadba17A,Nesvadba17B,Genzel14,Drouart14,Leung19,Cresci15,Brusa15,Cicone14,Alatalo11,Aalto12,Feruglio13,Morganti13,Veilleux17,Strum11,Herrera-Camus19,Brusa18}}
    \label{fig:AGN-drive-mechanism}
\end{figure*}

Figure \ref{fig:AGN_vs_SFR} is a diagnostic diagram that distinguishes between the driving mechanism for the outflows (star formation vs. AGN ) based on the criteria defined above in \S \ref{sec:condition_SF_driver} \& \ref{sec:condition_AGN_driver}. We further indicate in Table \ref{tab:AGNvsSFR} whether we think the outflow is driven by AGN, star formation or if we cannot distinguish we then list both sources as the possible driver. As discussed in \S \ref{sec:drive_outflow} the \ha based SFRs are used in determining whether stellar feedback has enough energy and momentum to drive the outflows. However in Figure \ref{fig:AGN_vs_SFR} we still show the comparison between the energy and momenta rates of our outflows compared to energy and momenta deposition from stellar feedback based on far infrared SFRs. Figure \ref{fig:AGN_vs_SFR} is a visual representation of whether the outflows are driven by stellar activity or AGN. In Figure \ref{fig:AGN-drive-mechanism} we distinguish whether radiation pressure from the quasar or an isothermal vs. adiabatic shock is driving the outflow. Points that are below the 2:1 line ratio between \momfluxagn and \momfluxout are outflows that can be driven by either radiation pressure or an isothermal shock, while points that are above the line represent outflows that are most likely driven by an adiabatic shock.

The galaxy-scale outflows in 3C318, 3C 298, and 4C04.81 are consistent within the error bars on the ratio between \momfluxagn and \momfluxout with an energy conserving shock as the driving mechanism. For 3C318 the energy conserving shock can either be driven by AGN or supernovae, since the momentum deposition from stellar feedback is similar to that of the quasar. The nuclear outflows in 4C04.81, 4C57.29 and 4C22.44 are consistent with either being driven by an adiabatic shock or from radiation pressure in a very optically thick environment with a high column density ($\rm N_{H}>10^{24}~cm^{-2}$). 

For 4C09.17, 4C05.84 and 3C268.4 the extended outflow is consistent with either being driven by an isothermal shock or through radiation pressure by the AGN. The nuclear outflow in 4C05.84 is consistent with either being driven by radiation pressure or an isothermal shock. For 3C9, 4C09.17, and 3C268.4 the outflow can either be driven by star formation, by an isothermal shock, or through radiation pressure by the AGN. 

The most likely driving source for 3C9 is star formation, given the geometry of the outflow and the fact that it does not extend to the quasar. The outflow propagates along the minor axis of the rotation disk, similar to star formation driven outflow in M82. Furthermore, the wind in 3C9 is emanating from the location where star formation has recently occurred based on nebular emission line ratios and the presence of extended UV emission in a galactic ring \citep{vayner19b}. For 3C9, if we assume the outflow originates in the galaxy's star-forming region, it would be more appropriate to measure the outflow radius starting from the edge of the turbulent region rather than from the quasar since the outflow itself does not extend down to the quasar. The radius from the edge of the turbulent region measures at 4.3 kpc, so the outflow rate goes up by a factor of 2.5 to 10$\pm$8 \myr and 25$\pm$22 \myr for \oiii and \ha derived ionized gas masses, respectively. In 3C268.4, while SNe feedback does have the required momentum deposition to explain the observed outflow, it appears to be emanating from the location of the quasar while the majority of the star formation is happening in a merging galaxy $\sim$4 kpc away (SW component B).

In 4C05.84, 4C04.81, 3C298, 3C268.4, 3C318 and 3C9 the path of the jet correlates with the direction of the outflow (e.g. see Figures \ref{fig:3C2684_all}, \ref{fig:4C0584_all}). This suggests that the jet could be responsible for driving the outflow. In section \ref{sec:chapter4discussion} we further discuss this scenario.

About half of our detected outflows are consistent with being energy conserving and show energy coupling above 0.1$\%$ between the kinetic luminosity of the outflow and the quasar's bolometric luminosity. The rest of the sources do not show such high coupling efficiency, and an adiabatic shock likely drives those outflows.

\subsection{Sample comparison}

We collate data from the literature on both ionized and molecular outflows in the low and high redshift Universe. We attempt to make a comparison to various AGN host galaxy surveys, both type 1 and type 2 AGN and QSOs that are radio quiet and loud. To perform a direct comparison between the ionized outflows in our sample and the ionized outflows in literature we re-derive the outflow rates and bolometric luminosities (when possible) of the AGN in the same manner as we have in the previous sections. From each paper, we extract the luminosity of a Balmer emission line (either \hb~or \ha), and when possible the electron density to estimate the ionized gas mass with the similar assumption that we made for our sample using Equation \ref{equation:ionized_gas_mass}. In cases where the electron density was not measured, similar to our sample, we assume an electron density in the range of 100-1000 \eden. We obtain the radius and velocity of the outflow, and along with the ionized gas mass, we estimate the ionized gas outflow rate with Equation \ref{equation:outflow-cone}. 

For the type-1 radio-quiet quasar sample from \cite{Carniani15} we use the \hb~emission line luminosity to derive the ionized gas mass, and the radius and velocity from their Table 2 to derive the outflow rates. We assume electron density in the range of 100-1000 \eden. We assume a conical geometry which leads to outflow rates 3 times higher than the original paper. For radio-loud type 2 AGN we extract data from \cite{Nesvadba17A}. We use their Table 4 to extract the velocity of the outflowing gas and the radial extent, which is taken to be half the major axis of the \oiii emitting gas. We use either the extinction corrected (when available in Table 6) \ha or \hb~line luminosity to compute the mass of the ionized gas. For the radio loud type 2 AGN sample we use the bolometric luminosities presented in \citep{Nesvadba17B} computed from mid and far infrared Herschel observations \citep{Drouart14}. Since these are obscured AGN measuring their bolometric luminosities from UV spectroscopy is difficult and would most likely underestimate the true luminosity of the object. For less luminous AGN at redshift $\sim$ 2 we use the \cite{Genzel14} sample of AGN selected from star-forming galaxies observed with near-infrared IFS. We use the \ha line luminosity of the outflow component in their Table 4 along with the velocity and radial extent. The AGN bolometric luminosity for this sample are presented in Table 1 of the paper and are derived from rest-frame 8\micron~luminosity or absorption corrected X-ray luminosity. We only include targets for which both an outflow was detected, and the bolometric luminosity was measured. For one target (GS3-19791) the outflow may not be AGN driven as the system's bolometric luminosity is dominated by star formation. We add all detected outflows in galaxies with AGN from the MOSDEF survey \citep{Leung19}, where we have re-calculated the ionized outflow rates with similar assumptions to our sample using the \ha lines to calculate the ionized gas mass. The bolometric luminosities of AGN in the MOSDEF survey span a range of $10^{44-47}$ \ergs and are calculated from the \oiii line luminosity. The sample includes both type 1 and type 2 AGN and quasars.

We also include outflows from papers on individual sources. From \cite{Cresci15} we include the properties of the outflow detected in the obscured AGN XID2028 at z=1.59. We use an outflow velocity of 1500 \kms and a radius of 13 kpc along with ionized gas mass derived from the \hb~luminosity for a typical range in electron density of 100-1000  \eden. For XID2028, the bolometric luminosity of 2$\times10^{46}$ erg/s is derived from SED modeling. From \cite{Brusa15} we include the ionized outflow detected in XID5395 at z=1.472. We use the \ha emission line to derive the ionized gas mass using an electron density of 780$\pm$300 \eden measured from [SII] line ratios, along with the measured radius of 4.3 kpc and an outflow velocity of 1300 km/s. XID5395 has a bolometric luminosity of 8$\times10^{45}$ erg/s derived from SED fitting similar to XID2028.

We also make a comparison between the ionized outflows detected at z$>$1 to molecular outflows detected in the nearby and high redshift Universe. To date, the majority of the molecular outflows have been detected and studied in detail in nearby systems. With the advent of ALMA, the number of molecular outflows detected at high redshift is growing, but the majority of the sample still consists only of nearby systems. AGN that have bolometric luminosities in the quasar regime are extremely rare in the nearby Universe compared to $z\sim2$ where they are nearly 1000 times more prevalent. As a result of this, the majority of the molecular outflows have been studied in systems with lower bolometric luminosities on average compared to the systems with ionized outflows. From low redshift we include molecular outflows detected and studied in CO from \cite{Cicone14,Alatalo11,Aalto12,Feruglio13,Morganti13,Veilleux17} and in OH from \cite{Strum11}. For comparison we only include objects where the bolometric luminosity of the AGN is greater than 10$\%$ of the total bolometric luminosity of the object, this is selected to avoid including objects where the outflow may primarily be driven by star formation and not AGN activity. The bolometric luminosity of the nearby AGN with molecular outflows spans a range of $10^{43}-10^{46}$ erg/s. For molecular outflows seen at higher redshift we include our detection in the quasar host galaxy of 3C 298 \citep{Vayner17} along with a recent discovery in the star-forming/AGN galaxy zC400528 \citep{Herrera-Camus19} with an AGN bolometric luminosity of $10^{45.5}$ erg/s and in the obscured AGN XID2028 \citep{Brusa18} with an AGN bolometric luminosity of $10^{46.3}$ erg/s.

\begin{deluxetable*}{llcccccccr}
\tablecaption{Extended outflow regions properties \label{tab:outflow-prop}. L$\rm_{[OIII]}$ and L$\rm_{H\alpha}$ are the spatially and line-integrated luminosities of emission solely associated with the outflow. R$\rm_{out}$ is the radial extent of the outflow. V$\rm_{out}$ is the velocity of the outflow. t$_{\rm outflow}$ is the dynamical time scale of the outflow. dM/dt$\rm_{[OIII]}$ and dM/dt$\rm_{H\alpha}$ are the outflow rates using masses derived from the \ha or the \oiii emission lines. $\dot{P}$ is the momentum flux of the outflow, using the dM/dt$\rm_{H\alpha}$ outflow rate. And, $\frac{\dot{P}_{outflow}}{\dot{P}_{AGN}}$ is the ratio of the momentum flux of outflow to the momentum flux of the quasar accretion disk.}

\tablehead{\colhead{Source}&
\colhead{L$\rm_{[OIII]}$}&
\colhead{L$\rm_{H\alpha}$}&
\colhead{R$\rm_{out}$}&
\colhead{V$\rm_{out}$}&
\colhead{t$_{\rm outflow}$}&
\colhead{dM/dt$\rm_{[OIII]}$}&
\colhead{dM/dt$\rm_{H\alpha}$}&
\colhead{$\dot{P}$}&
\colhead{$\frac{\dot{P}_{outflow}}{\dot{P}_{AGN}}$}\\
\colhead{}&
\colhead{10$^{43}$ \ergs}&
\colhead{10$^{43}$ \ergs}&
\colhead{kpc}&
\colhead{\kms}&
\colhead{Myr} &
\colhead{\myr}&
\colhead{\myr}&
\colhead{$10^{35}$dyne}&
\colhead{}
}
\startdata
3C9   &1.3$\pm$0.14 & 0.31$\pm$0.04&10.2$\pm$1&964$\pm$ 50 & 10 & 4$\pm$3&10.2$\pm$9& 0.6$\pm$0.5&0.03$\pm$0.03\\
4C09.17&2.5$\pm$0.2&0.21$\pm$0.02&5.9$\pm$1&623.3$\pm$50 & 9.2 & 9$\pm$7 & 8$\pm$7& 0.3$\pm$0.3&0.08$\pm$0.02\\
3C268.4&1.9$\pm$0.1&--&5.5$\pm$1&1446$\pm$50  & 3.73 & 16.8$\pm$13.5&-&4.6$\pm$4&0.4$\pm$0.3\\
3C298 & & & & & & & & &\\
W component A  &17$\pm$2&2.8$\pm$0.3&4.4$\pm$0.7&1703$\pm$10& 2.5 &401$\pm$164&600$\pm$400&6.5$\pm$4&2.5$\pm$1.6\\
E component A  &2.4& 0.2$\pm$0.1 & 3.94$\pm$0.8 & 1403 &2.75 & 63$\pm$27 & 143$\pm$94 & 1.3$\pm$0.8 &0.5$\pm$0.3\\
3C318 & 3.5$\pm$0.3 & 6.48$\pm$2 & 3.19$\pm $ 0.26 & 703.5$\pm$10 & 4.44 & 20.5$\pm$6.0 & 217$\pm$148 & 9.6$\pm$7 &4$\pm$3\\
4C05.84  & & & &  & & & & & \\
SW component A & 0.95$\pm$0.1&0.36$\pm$0.1&1.6$\pm$1&542.45$\pm$40 & 2.9 & 10.5$\pm$9&41.5$\pm$37&1.4$\pm$1.3&0.03$\pm$0.015\\
NE component A & 0.95$\pm$0.1&0.24$\pm$0.2&8.3$\pm$1&618.68$\pm$40 & 13.2 &  2.3$\pm$1.9 & 6.4$\pm$6 & 0.25$\pm$0.2 &0.004$\pm$0.003\\
4C04.81&26$\pm$3&7.2$\pm$0.7&3.8$\pm$1&552$\pm$20& 6.74 & 122$\pm$100&361$\pm$325& 12.6$\pm$11&6$\pm$5\\
3C268.4&1.95$\pm$0.1&--&5.5$\pm$1&1446$\pm$50  & 3.73 & 16.8$\pm$13.5&-&4.6$\pm$4&0.4$\pm$0.3\\
\enddata
\end{deluxetable*}

\begin{deluxetable*}{lc@{\extracolsep{-10pt}}cccccccr}
\tablecaption{Nuclear outflow regions properties \label{tab:outflow-prop-unreasolved}. L$\rm_{[OIII]}$ and L$\rm_{H\alpha}$ are the spatially and line-integrated luminosities of emission solely associated with the outflow. R$\rm_{out}$ is the radial extent of the outflow. V$\rm_{out}$ is the velocity of the outflow. t$_{\rm outflow}$ is the dynamical time scale of the outflow. dM/dt$\rm_{[OIII]}$ and dM/dt$\rm_{H\alpha}$ are the outflow rates using masses derived from the \ha or the \oiii emission lines. $\dot{P}$ is the momentum flux of the outflow, using the dM/dt$\rm_{H\alpha}$ outflow rate. And, $\frac{\dot{P}_{outflow}}{\dot{P}_{AGN}}$ is the ratio of the momentum flux of outflow to the momentum flux of the quasar accretion disk.}
\tablehead{\colhead{Source}&
\colhead{L$\rm_{[OIII]}$}&
\colhead{L$\rm_{H\alpha}$}&
\colhead{R$\rm_{out}$}&
\colhead{V$\rm_{out}$}&
\colhead{t$_{\rm outflow}$}&
\colhead{dM/dt$\rm_{[OIII]}$}&
\colhead{dM/dt$\rm_{H\alpha}$}&
\colhead{$\dot{P}$}&
\colhead{$\frac{\dot{P}_{outflow}}{\dot{P}_{AGN}}$}\\
\colhead{}&
\colhead{10$^{43}$ \ergs}&
\colhead{10$^{43}$ \ergs}&
\colhead{kpc}&
\colhead{\kms}&
\colhead{Myr} &
\colhead{\myr}&
\colhead{\myr}&
\colhead{$10^{35}$dyne}&
\colhead{}
}
\startdata
4C09.17&3$\pm$0.3&--&$<0.92$&726$\pm$10 & $<$1.3& 40$\pm$20 & -- & 1.8$\pm$0.9 & 0.16$\pm$0.08\\
3C268.4&2.5$\pm$0.25&--&$<$1.7&797$\pm$50 & $<$2.0 & 19$\pm$10&-&1$\pm$0.5&0.08$\pm$0.04\\
7C1354&5.5$\pm$1&--&$<$1.0&636$\pm$10 & $<$1.7 & 52$\pm$27&-&2$\pm$1&0.2$\pm$0.1\\
4C57.29& 15$\pm$2 &7$\pm$1 &$<$1 &686$\pm$20& $<$1.3 & 184$\pm$94&816$\pm$586& 35$\pm$26&5$\pm$4\\
4C22.44& 22$\pm$2 & 2$\pm$0.2 & $<1$ & 878$\pm$ 20 & $<1.1$ & 314$\pm$160 & 330$\pm$240 & 18$\pm$13& 11$\pm$8\\
4C05.84& 5$\pm$0.5 &7$\pm$0.8 &$<$1 &683$\pm$20& $<1.4$ & 60$\pm$30 & 825$\pm$590& 35$\pm$26&0.5$\pm$0.4\\
4C04.81& 10$\pm$1 &14$\pm$1 &$<$1 &1073$\pm$20& $<$1 & 187$\pm$95&2440$\pm$1750& 165$\pm$119&80$\pm$58\\
\enddata
\end{deluxetable*}

\begin{deluxetable*}{lcccccccc}
\tablecaption{Momentum and flux deposition from wind driving mechanisms \label{tab:AGNvsSFR}}
\tablehead{\colhead{Source}&
\colhead{$\dot{P}_{AGN}$}&
\colhead{$\dot{P}_{SFR,rad,T_{ir}}$}&
\colhead{$\dot{P}_{SFR,rad,H\alpha}$}&
\colhead{$\dot{P}_{SNe,T_{ir}}$}&
\colhead{$\dot{P}_{SNe,H\alpha}$}&
\colhead{$\dot{E}_{SNe,T_{ir}}$}&
\colhead{$\dot{E}_{SNe,H\alpha}$}&
\colhead{SF or AGN}\\
\colhead{}&
\colhead{$\times10^{36}$dyne}&
\colhead{$\times10^{36}$dyne}&
\colhead{$\times10^{36}$dyne}&
\colhead{$\times10^{36}$dyne}&
\colhead{$\times10^{36}$dyne}&
\colhead{$\times10^{42}$erg/s}&
\colhead{$\times10^{42}$erg/s}&
\colhead{driven?}
}
\startdata
3C9   &2.7&   $<$0.23&   0.12&   $<$2.2&   1.1&   $<$9.4& 4.8&SF/AGN\\
4C09.17 & 0.96&  0.97&   0.007&   9.28 &   0.066& 40 & 0.29 & SF/AGN\\
3C298 &2.6&   0.69&   0.087&   1.9&   0.24&   28& 3.6& AGN\\
3C318 & 0.26&  0.44&  0.064&   4.1&   0.6 &   18& 0.26 & SF/AGN\\
4C22.44 & 0.16 & -- & 0.023 & 0 & 0.22 & -- & 1.0 & AGN\\
4C05.84 &  6.8& --&   0.0082& --&   0.08& -- &0.33& AGN\\
4C04.81 & 0.21& 1.2 &   $<$0.0058&  11 &   $<$0.05& 47 & $<$0.24& AGN\\
3C268.4& 1.2& $<$0.1&  0.037&  $<$1.0&   0.4&   $<$4.2& 1.5& AGN\\
7C1354& 0.9 &--& $<$0.007 &--& $<$0.07 &--& $<$0.3 & AGN\\
4C57.29 & 0.7 & -- & $<$0.0064 & - & $<$0.06 & -- & 0.3 & AGN\\
\enddata
\end{deluxetable*}

\section{Discussion}\label{sec:chapter4discussion}

\subsection{Driving source of the ionized outflows}

In this section, we discuss the driving mechanisms for the galaxy scale outflows. In section \ref{sec:out_rate}, we derived the outflow rates, momentum fluxes, and the kinetic luminosities of the outflows. We also looked at the expected energy and momentum deposition from stellar feedback to examine if this is sufficient to explain the observed momentum fluxes. These results are summarized in Tables \ref{tab:outflow-prop},\ref{tab:AGNvsSFR} and in Figure \ref{fig:AGN_vs_SFR}. We find that in the majority of the cases stellar feedback either due to supernovae explosions or radiation pressure from stellar winds is insufficient to explain the observed momentum fluxes. In 7/10 of the detected outflows, star formation does not provide the necessary energy and momentum deposition to drive the outflows. In 3/10 systems, star formation can drive the outflows; however, the quasar can still be responsible, and we leave the driving mechanism as ambiguous for these three systems. In one system, 3C446, we find no evidence of an outflow. We therefore believe that the AGN is responsible for driving the majority of the observed outflows within our sample.

\cite{Barthel17, Barthel18, Barthel19} explores a sample of radio-loud quasars that include several objects from our sample (4C09.17, 3C 298, 4C04.81, and 4C05.84). They find a tentative correlation between the SFR and the equivalent width of the CIV $\lambda\lambda$ 1548\AA, 1550 \AA~absorption lines. They do not, however, find a similar correlation between the quasar bolometric luminosity and absorption-line equivalent width. Hence, they consider star formation as the primary driver for the outflows in the quasar host galaxies. In their study, they use far-infrared derived SFRs that could trace a much longer star formation history compared to the dynamical time scale of quasar driven outflows. Furthermore, the derived SFRs can be averaged over several burst episodes, tracing star formation that happened in the past ($>10$ Myr) \citep{Calzetti13}. The velocities of the C IV absorption lines appear significantly lower than the expected outflow velocities for stellar feedback at SFRs of $\sim$ 1000 \myr \citep{murray05, Zhang12}. Additionally, it is hard to measure the extent of the outflows traced in absorption, and therefore difficult to compare their dynamical time scale to the star formation event that may have triggered them. Interestingly, \cite{Talia17} find strong evidence for enhanced outflow velocities in a large sample of galaxies hosting an AGN compared to a control sample of purely star forming galaxies at $1.7<$z$<4.6$. They see a correlation between the X-ray luminosity and the velocity of the outflow using the UV Mg II absorption line. 

To understand which mechanisms drive the galaxy scale outflows, it is important to measure outflow rates, momentum and energy fluxes, and energy and momentum deposition based on tracers of star formation activity on the dynamical time scales of the outflows. It is also important to measure the energy and momentum deposition from the multitude of AGN and quasar mechanisms. Given that we find that star formation activity is typically incapable of driving the galaxy scale outflows, we look towards the several quasar mechanisms that may be able to drive the observed outflows.

Within our sample, we find a mixture of sources with both high and low momentum flux ratios (\momfluxratio). For the extended outflows in 3C318, 3C298 and 4C04.81 we find \momfluxratio of 4-6, but, they can be as low as 0.5 or as high as 11 within the observed uncertainties. However the more likely scenario is that they are all above 2. The momentum fluxes that we measure in our outflows might be lower limits as we are not accounting for the neutral and molecular components of the outflow, except 3C 298 where we have measured both the molecular and ionized gas in the outflow but not the neutral. At the extent of these outflows ($\sim3$ kpc), radiation pressure from the quasar accretion disk is unlikely to drive them. \cite{Thompson15, Costa18} both find that on kpc scales the maximum observed \momfluxratio is about 1.5-2. To invoke radiation pressure as the primary driving mechanism for outflows with momentum flux $>2\times L_{AGN}/c$ and an extent $>1$ kpc requires an extremely obscured ($\rm N_{H}>10^{24}$ cm$^{-2}$) initial environment. However, \cite{Costa18a} argues that driving outflows in much denser ($\rm N_{H}>10^{24}$ cm$^{-2}$) regions through radiation pressure will most-likely not lead to much higher momentum fluxes on kpc scales and instead can produce outflows that barely escape the inner regions of their respective galaxies. Furthermore, the most obscured environments that we find AGN show column densities on the order of $N_{H}\sim10^{24}$ cm$^{-2}$ \citep{Georgantopoulos19}.

The most likely scenario is that an energy conserving shock drives the outflows in 3C318, 3C298 and 4C04.81. Such shocks are produced by either UFO/BAL type winds or through quasar jets. Theoretical work on both of these driving mechanisms predicts \momfluxratio$>$2 on kpc scales \citep{Wagner12, Faucher12, Zubovas12}. In these scenarios the jet/UFO slams into the ISM and drives a hot shock. If this shock cools on a time scale much longer than the flow time then the ``hot shocked bubble" will expand adiabatically sweeping up material along the way. As the material is swept up in a galaxy scale outflow, it is given a significant momentum boost. The outflow that is driven by the bubble also gets shocked, although at far lower temperatures, allowing it to cool through line emission down to perhaps a molecular phase \citep{Richings18}. For both 3C298 and 3C318, we find that the nebular line ratios in the ionized outflow are consistent with being produced by radiative shocks \citep{Vayner17} and both systems are consistent with having an energy conserving shock drive the outflow.

Because the momentum fluxes in the nuclear outflows of 4C04.81, 4C22.44 and 4C57.29 are $>2\times L_{AGN}/c$ an energy conserving shock can drive them. However, due to their smaller sizes ($<1$ kpc), there is the possibility that they are driven through radiation pressure in a high column density environment. Both outflows show high column densities $N_{H}>10^{22}$cm$^{-2}$, and their velocities are consistent with models of radiation pressure on dust grains presented in \cite{Thompson15}. The nuclear outflow in 4C05.84 is also consistent with radiation pressure, however in a less dense environment compared to 4C04.81 and 4C57.29. An isothermal shock \citep{Faucher12} can also drive the nuclear outflow in 4C05.84. 

All of the detected ionized outflows in our sample align with the path of the jet, with the exception of 4C09.17. The jet could be the sole driving source of the galaxy scale outflows. However, in \cite{Vayner17} we found that at present the jet is not doing any work on the extended ionized or molecular outflow in the host galaxy of the quasar 3C 298. This is because the jet pressure is orders of magnitude higher than the ISM pressure measured in both the ionized and molecular gas. Given the similarity in jet luminosity and ISM conditions between 3C 298 and the rest of the sample, we think this likely applies to the rest of the objects. Since we have measured the electron density in the ionized outflow in 3C318, we can also perform the same exercise as we have done for 3C298 in studying whether the jet is currently doing any work on the outflow. To do this, we first extract all available radio fluxes of the object from NED. We construct a radio SED from 22 MHz to 353 GHz and fit it with a double power law. We integrate the fitted power law from rest-frame frequencies of 10 - 10000 MHz and obtain a total radio luminosity of $2.5\pm0.25\times10^{45}$ \ergs. We obtain a jet power for 3C318 on the order of $10^{46}$ \ergs~using the relationship between total radio luminosity and jet power from \cite{Birzan08}. We obtain a similar value using the monochromatic rest-frame 1400 MHz luminosity and jet power from the same study, where the 1400 MHz flux is taken from our best fit SED model. Following \cite{Vayner17} we convert the jet power to jet pressure using the following equation: 

\begin{equation}
    \rm P_{jet}\approx\frac{L_{jet}\times t}{3V},
\end{equation}

\noindent where L$_{jet}$ is the jet luminosity (power), V is the confined volume of the jet for which we simply assume a cylindrical shape with a base radius of 1 kpc and height of 3 kpc, t is the propagation time for the jet for which we simply use the dynamical time scale of the outflow of 4 Myr. We obtain a jet pressure of 1$\times10^{-6}$ dynes/cm$^{2}$. The gas pressure ($P=n_{e}kT_{e}$) in the outflow 3C318 is on the order of $10^{-9}$ dynes/cm$^{2}$, using the measured electron density and assuming an electron temperature of 10$^{4}$K. The gas pressure is much lower than the jet pressure, indicating that the two are not in pressure equilibrium. Similar to 3C298 the jet is venting out of the galaxy along the path of least resistance, and at present is not doing any work in driving the kpc-scale outflow in the system. If the jets were to drive the outflows in our systems, it would have had to be done in a denser environment closer to the quasar in the past. Given the similarities in our detected ionized gas line luminosities and sizes, the ionized gas pressure is likely similar for the rest of our sample sources to 3C318 and 3C298, and likely the jets are venting out of all of our systems. In the case of the galaxy IC 5063, interaction between multiple gas phases and the jet are all found on scales $<$ 1kpc. Likely if the outflows were driven by the jet within our systems they had to happen on similar scales to what is observed in IC 5063 \citep{Morganti07,Morganti15}. Given the strong correlation between the jet's path and the ionized outflows' extent and location, we believe that the jets can be strong candidates for driving the observed galaxy scale outflows. Particularly the jets are excellent candidates for driving the energy-conserving outflows since jets in simulations \citep{Wagner12,Mukherjee16}, can produce hot-shocked gas that does not cool efficiently (see our discussion in section \ref{sec:condition_AGN_driver}) and can sweep the material out of the galaxy. On the other hand, another wind mechanisms such as an Ultra-Fast Outflow (UFO) could have driven the outflow first, and the jet turned on later and is now simply following the path of least resistance and escaping out of the galaxy. Both UFO and BAL type winds can provide a momentum flux of about $L_{AGN}/c$, sufficient to drive the observed galaxy-scale outflows in 3C318, 3C298 and 4C04.81 if shocks produced by the UFO/BAL are energy conserving \citep{Faucher12a}.

The rest of the objects in our sample show ionized outflows with \momfluxratio$<<1$. A significant number of ionized outflows in other studies of distant (z$>1$) quasar host galaxies have found ionized outflows with \momfluxratio$<2$ (Figure \ref{fig:AGN-drive-mechanism}). There are several ways to drive such ionized outflows. First, if the shock produced by a UFO/BAL/Jet radiates its energy efficiently, then only the ram pressure affects the ISM. Such a scenario can produce \momfluxratio$<<$1 \citep{Zubovas12,King15}. Second, if the outflow is driven by radiation pressure in a low-density environment, then the \momfluxratio~can also be very low \citep{Thompson15}. 

Interestingly the majority of the molecular outflows have \momfluxratio$>>$1, different from what is found in ionized outflows that tend to show a much larger range of momentum flux ratios. A significant fraction of molecular outflows have been found and studied in nearby systems. Are there different driving mechanisms between the ionized and molecular outflows? Or are we missing a significant fraction of the gas in the outflows at high redshift by only studying the ionized component \citep{Carniani16}? The latter case would indicate that only a small fraction ($<1\%$) of the total gas in an outflow can be in a warm ionized state for outflows with a radius $>1$ kpc and \momfluxratio$<<1$, the rest of the gas has to be in either neutral, molecular or hot gas phase if all the outflows on these scales ``energy" conserving.

\begin{figure*}
    \centering
    \includegraphics[width=6.5in]{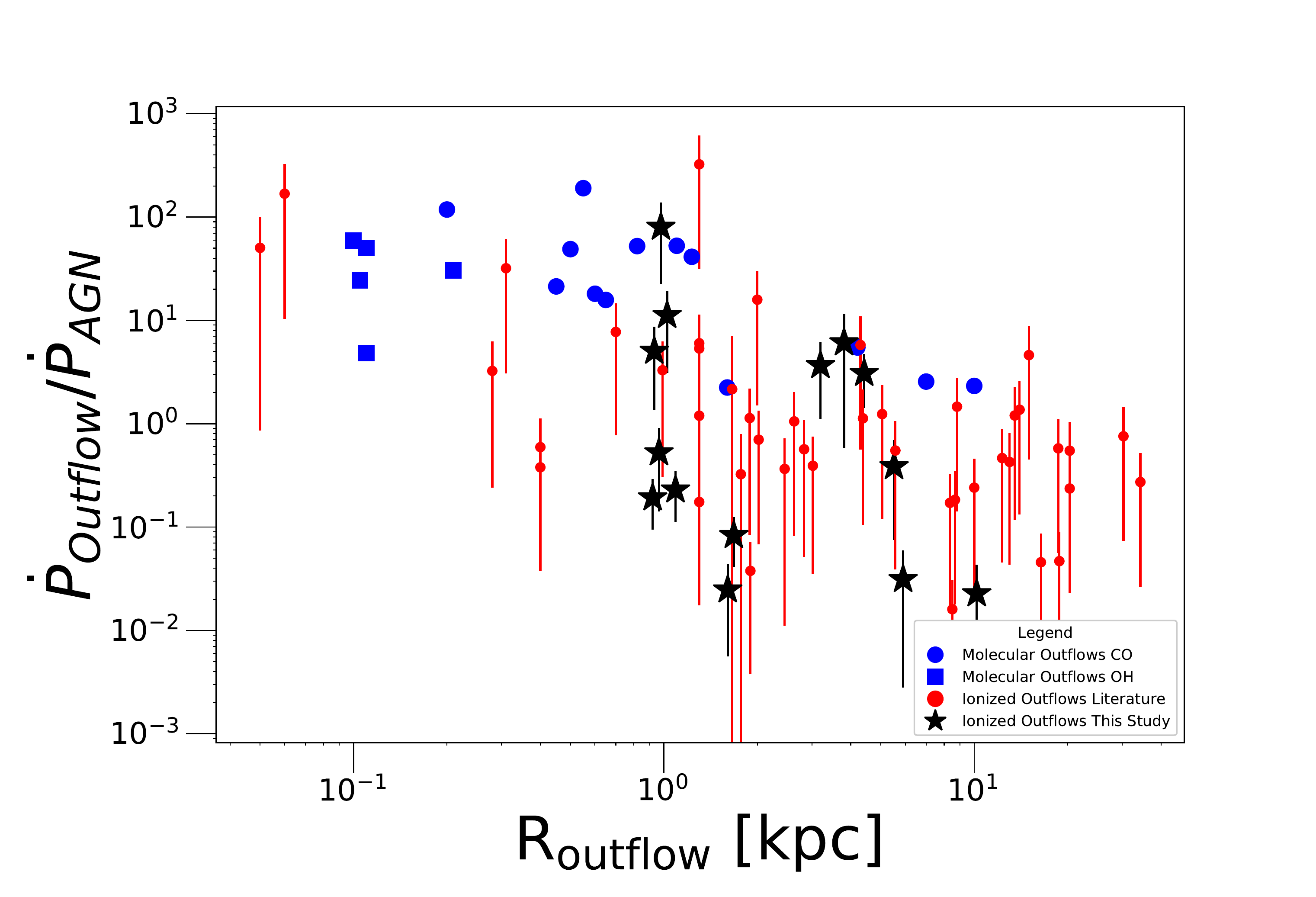}
    \caption{Comparison of momentum flux ratios and sizes of molecular and ionized outflows. We plot the \momfluxratio ratio vs. the radius of the outflow. According to theoretical work on radiation pressure driven outflows the maximum ratio expected at separations $>$1 kpc is 2 and decreases with radius. In contrast, \momfluxratio for outflows driven by an energy conserving shock is expected to increase with radius. Ionized outflows are most consistent with a model where they are driven by radiation pressure or an isothermal shock. About half the points within our sample are consistent with isothermal shock driving, while an adiabatic shock likely drives the others. While the molecular and some ionized outflows are consistent with either being driven by radiation pressure in confined nuclear regions or through adiabatic shocks. Red points represent outflows detected in ionized gas emission, and black stars represent points from our sample. Blue squares represent galaxies where a molecular outflow was detected through OH absorption while blue circles represent molecular outflows detected through CO emission. All data points from literature are taken from the following studies: \cite{Carniani15,Nesvadba17A,Nesvadba17B,Genzel14,Drouart14,Leung19,Cresci15,Brusa15,Cicone14,Alatalo11,Aalto12,Feruglio13,Morganti13,Veilleux17,Strum11,Herrera-Camus19,Brusa18}}
    \label{fig:P_dot_radius}
\end{figure*}

The molecular outflows detected in the nearby Universe are much smaller in size compared to ionize outflows (Figure \ref{fig:P_dot_radius}). The different sizes could indicate that they are probing different evolutionary phases of an AGN driven outflow. According to \cite{Thompson15} for a radiation pressure driven outflow which is highly confined initially, the \momfluxratio and column densities are very high, favoring small scale molecular outflows. It is also important to search for and measure molecular outflows that are of larger extent, similar to that of ionized outflows to make sure we are not missing a significant fraction of the gas in galaxy scale outflows. On the other hand, it could be that nearby AGN are fading quickly, and \momfluxratio~is over-estimated if the AGN bolometric luminosity was smaller when the outflow was driven vs. when it was measured. 

The measured \momfluxratio~in nearby molecular outflows are still consistent with an energy conserving outflow scenario \citep{Faucher12, Zubovas12, Zubovas14}. While uncertainties on ionized outflows rates and dynamics need to be drastically improved, multi-phase studies of outflows will help address the driving source of galactic scale winds and whether there are any evolutionary effects in driving outflows in nearby and high redshift systems during a luminous quasar phase.

\subsection{Gas depletion}
In this section, we discuss the primary source of gas depletion in the quasar host galaxies in our sample. In several systems, we were able to measure upper limits on the SFRs \citep{vayner19b} which we compare to the outflow rates to determine the primary source of gas depletion. 


For 3C298, 3C318, 4C05.84, 4C04.81, 3C268.4, 4C57.29, and 7C1354+2552, the primary source of gas depletion is due to the quasar driven outflows when compared to the \ha derived star formation rate. The coupling efficiency between the kinetic luminosity and the bolometric luminosity of the quasar for 4C04.81, 4C05.84, 4C22.44, 4C57.29, 3C298 and 3C318 are $>$0.1\%; the minimum coupling efficiency necessary for quasar feedback to occur based on theoretical predictions \citep{Hopkins10,Choi12,Costa18}. These results indicate that the outflows can be responsible for removing gas from the inner regions (1-8 kpc) of the quasar host galaxies. Furthermore, in the case of 3C298, the quasar is responsible for removing the molecular gas reservoir, providing direct evidence for negative quasar feedback in multiple gas-phases \citep{Vayner17}. For 3C318 and 4C09.17, the far-infrared derived star formation rates are much greater than the outflow rates. It could be possible that star formation is the main driver behind the outflows and that star formation is the primary source of gas depletion. However, we only measure the ionized component of the outflow, and a large portion of the outflow mass and energetics may be in the molecular \citep{Richings18} or neutral gas-phase \citep{demp18}, so all of the outflow rates that we observe are likely lower limits, and the total multi-phase outflow rate can still match or surpass the far-infrared derived star formation rate.

In a sample of 18 quasar host galaxies studied in dust continuum with ALMA, \cite{Schulze19} find no evidence for regulation of star formation. Their sample matches ours in both bolometric luminosity and SMBH mass. They argue that because they measure SFRs on average that match that of main-sequence galaxies, there is no evidence for regulation of star formation. Interestingly, our galaxies are close to the galaxy main-sequence at z$\sim$ 2 for a stellar mass of $10^{11}$ \msun, however, we see evidence for outflows that are actively removing gas from the galaxy at rates higher than that of star formation. We see the removal of ionized and molecular gas, heating of the ISM by radiative shocks and quasar photoionization, and turbulence in the ISM induced by outflows. This constitutes as negative feedback that will affect future star formation in these quasar host galaxies. Quasars live in galaxies with a very broad range of SFRs \citep{Aird19}. The outflow rates, driven by the quasar, can often surpass the SFRs by factors of a few. This highlights the importance of measuring outflow rates and the dynamics of the gas. Measuring the rate of star formation alone does not give a complete picture of whether feedback is present or not.

\section{Conclusions}\label{sec:conclusions}
We have conducted a near diffraction limited survey of 11 quasar host galaxies, to study the distribution, kinematics, and dynamics of the ionized ISM using the OSIRIS IFS at the W.M. Keck Observatory. The aim of this survey paper was to detect ionized outflows on galactic scales and to quantify their effects on the ISM.

\begin{itemize}

\item Outflows are detected in 10/11 objects on spatial scales from $<$1 kpc to 10 kpc. Outflow rates range from 8 - 2400 \myr, with momentum flux ranging from 0.03-80 $L_{AGN}$/c, and energy rates of 0.01-1\% $L_{AGN}$. 

\item For 5/11 sources the momentum fluxes are $4-80 \times L_{AGN}$/c and therefore the outflows are consistent with being driven by an energy conserving shock. The energetics of these outflows are consistent with the theoretical predictions of energy coupling necessary to clear gas out of the galaxy and establish some of the local scaling relations \citep{Faucher12, Hopkins12, Zubovas14}.

\item In the rest of the objects, the outflows are either driven by star formation, radiation pressure on dust grains, or an isothermal shock. Their energetics fall short of the predicted energy and momentum rates necessary to establish the \msigma relationship and the clearing of galaxies' gas reservoirs predicted by theoretical work \citep{Faucher12, Hopkins12, Zubovas14}. Another possibility is that a large fraction of the gas inside the outflows is not in an ionized phase, but rather in either molecular or neutral, hence, we are drastically underestimating the gas outflow rates and energetics.

\item For the majority of the systems, the outflow rates are higher than that of star formation. Along the path of the outflows, we see no strong evidence of active star formation. 

\item For 7/8 extended galactic outflows, the path of the jet is consistent with the direction of the outflow. However, we find that the jet pressure is orders of magnitude higher than that of the ionized galactic outflows. Because the jets and the ionized outflows are not in pressure equilibrium, the jets are not doing any work on the outflows. If the jet were to drive the outflows, it had to happen in dense environments close to the SMBH.

\end{itemize}

\acknowledgments
The authors wish to thanks Jim Lyke, Randy Campbell, and other SAs with their assistance at the telescope to acquire the Keck OSIRIS data sets. We want to thank the anonymous referee for their constructive comments that helped improve the manuscript. The data presented herein were obtained at the W.M. Keck Observatory, which is operated as a scientific partnership among the California Institute of Technology, the University of California and the National Aeronautics and Space Administration. The Observatory was made possible by the generous financial support of the W.M. Keck Foundation. The authors wish to recognize and acknowledge the very significant cultural role and reverence that the summit of Maunakea has always had within the indigenous Hawaiian community. We are most fortunate to have the opportunity to conduct observations from this mountain. This research has made use of the NASA/IPAC Extragalactic Database (NED) which is operated by the Jet Propulsion Laboratory, California Institute of Technology, under contract with the National Aeronautics and Space Administration.

\software{OSIRIS DRP \citep{OSIRIS_DRP}, CASA\citep{McMullin07}, PyNeb \citep{Luridiana15}, Matplotlib \citep{Hunter07}, SciPy \citep{2020SciPy-NMeth}, NumPy \citep{2020NumPy-Array}, Astropy \citep{Astropy18}}

\appendix \label{sec:appendix}

\begin{figure*}[!th]
    \centering
    \includegraphics[width=6.5in]{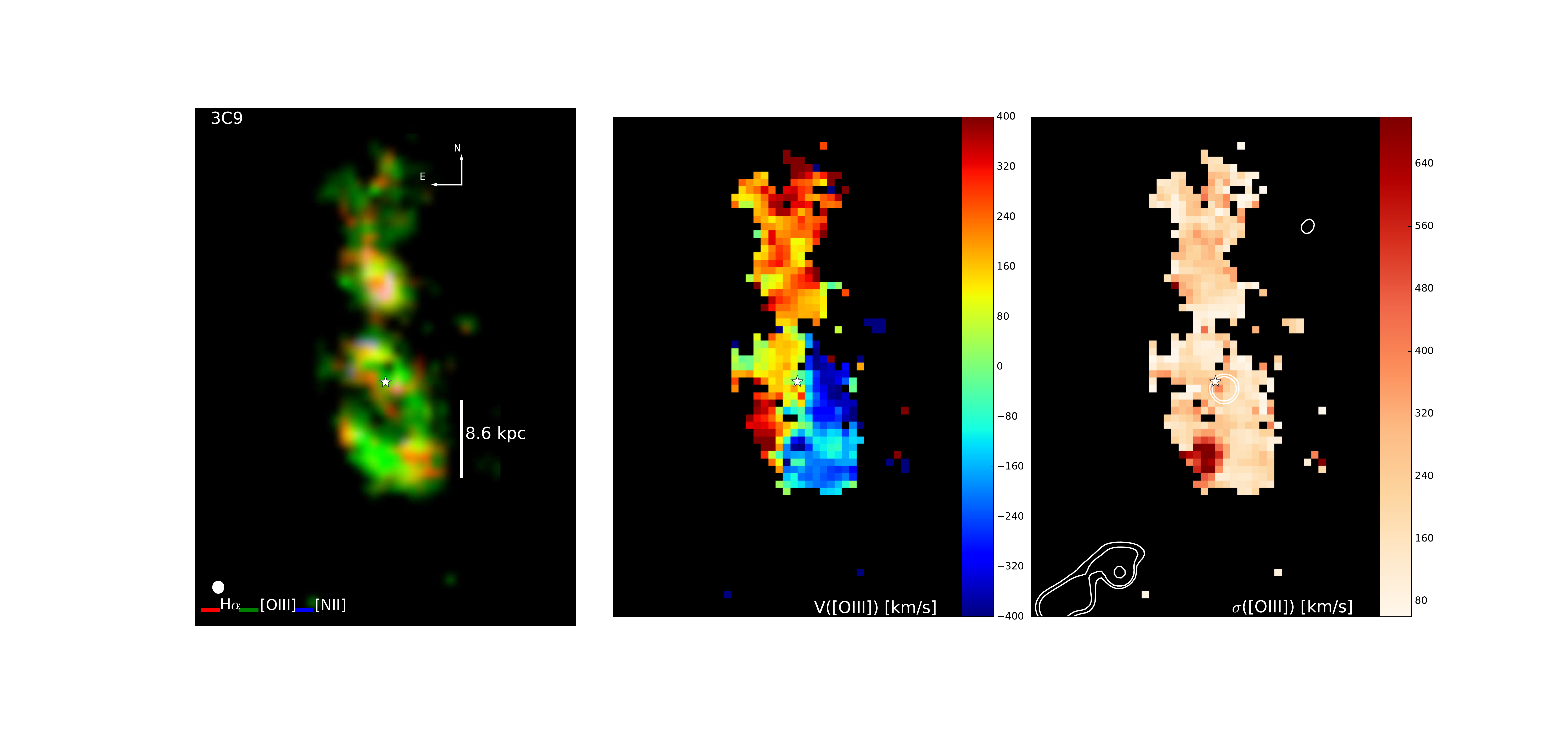}
    \includegraphics[width=6.5in]{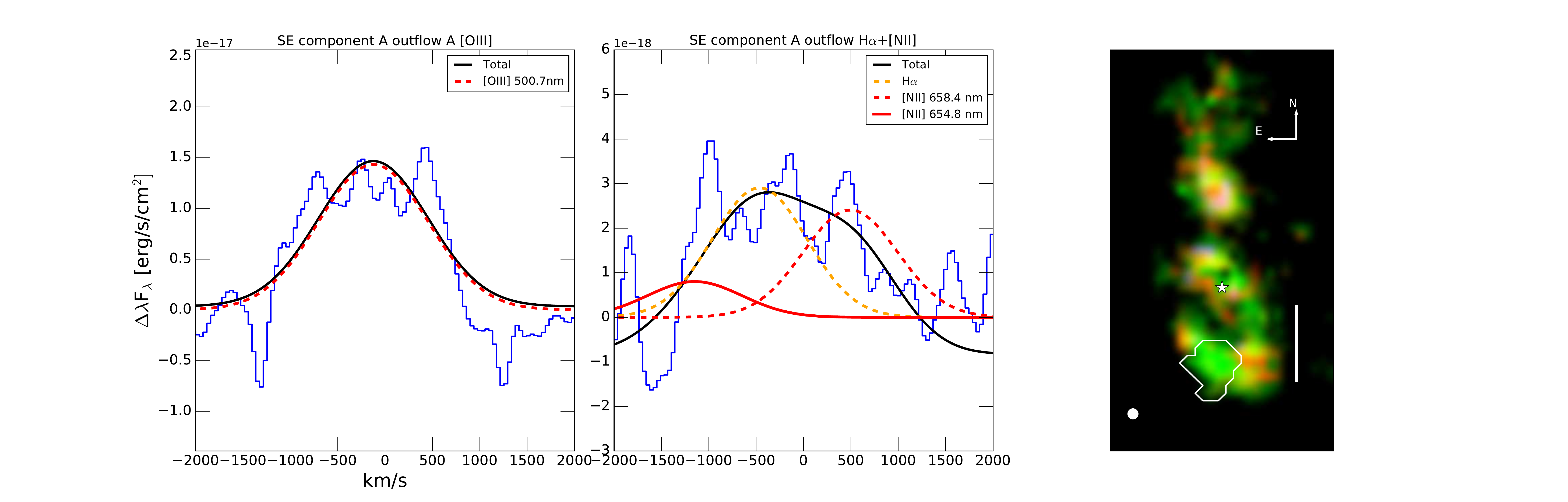}
    \caption{OSIRIS observations of the 3C9 nebular emission line distribution and kinematics produced from the PSF subtracted data cubes.(Top-left) Three color intensity map of nebular emission lines: \ha (red), \oiii (green), and \nii (blue). Ellipse in the lower left corner showcases the spatial resolution of the observations. (Top-middle) Radial velocity offset (\kms) of the \oiii line relative to the redshift of the quasar. The white star shows the location of the subtracted quasar. (Top-right) Velocity dispersion (\kms) map of \oiii emission. The white contours are VLA observations at 8439.900 MHz of radio synchrotron emission from the quasar jet and lobes. (BOTTOM) Spectra of a distinct outflow region along with fits to individual emission lines On the left we show the fit to the \oiii 500.7 nm emission line, in the middle we present the fit to the \ha and \nii emission lines and on the right we show a three-color composite along with a contour outlining the spatial location of the region.}
    \label{fig:3C9_all}
\end{figure*}

\begin{figure*}[!th]
    \centering
    
    \includegraphics[width=6.5in]{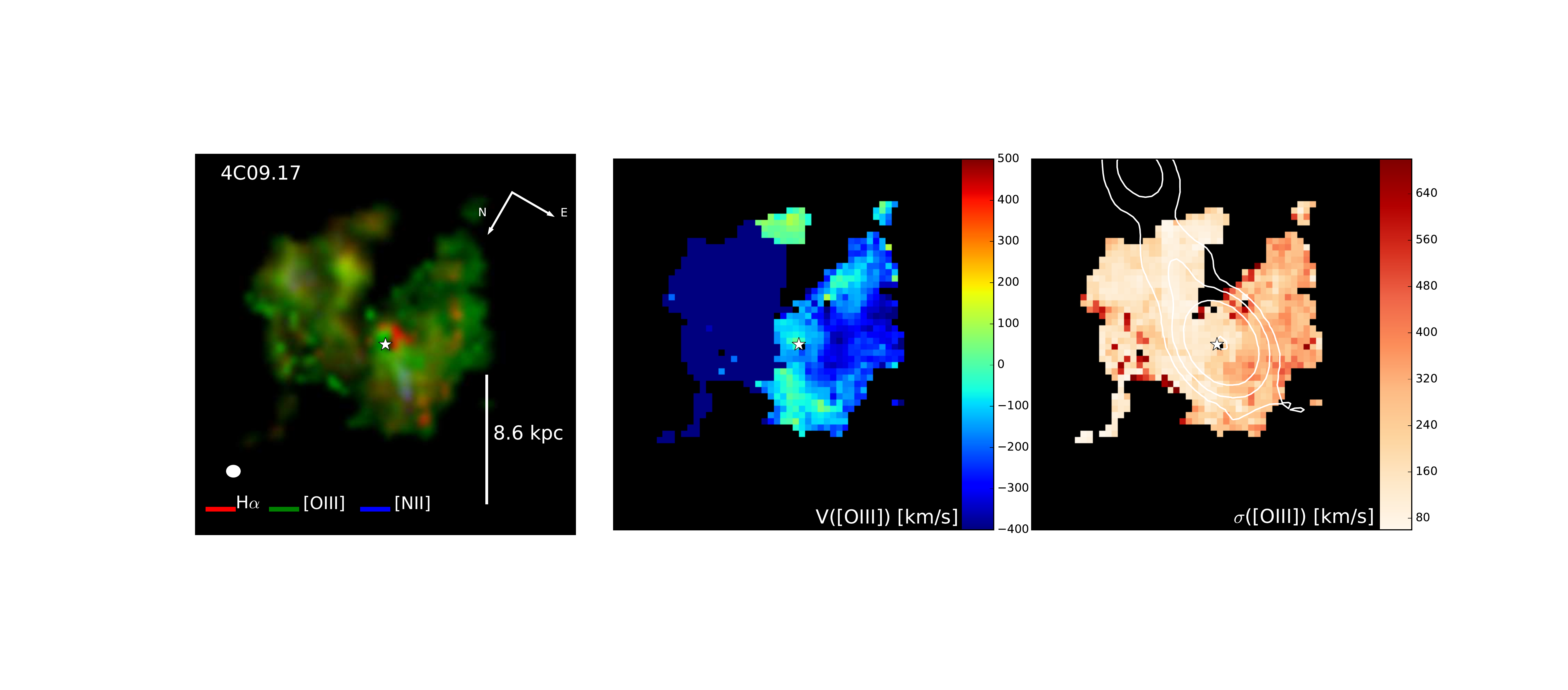}
    \includegraphics[width=7in]{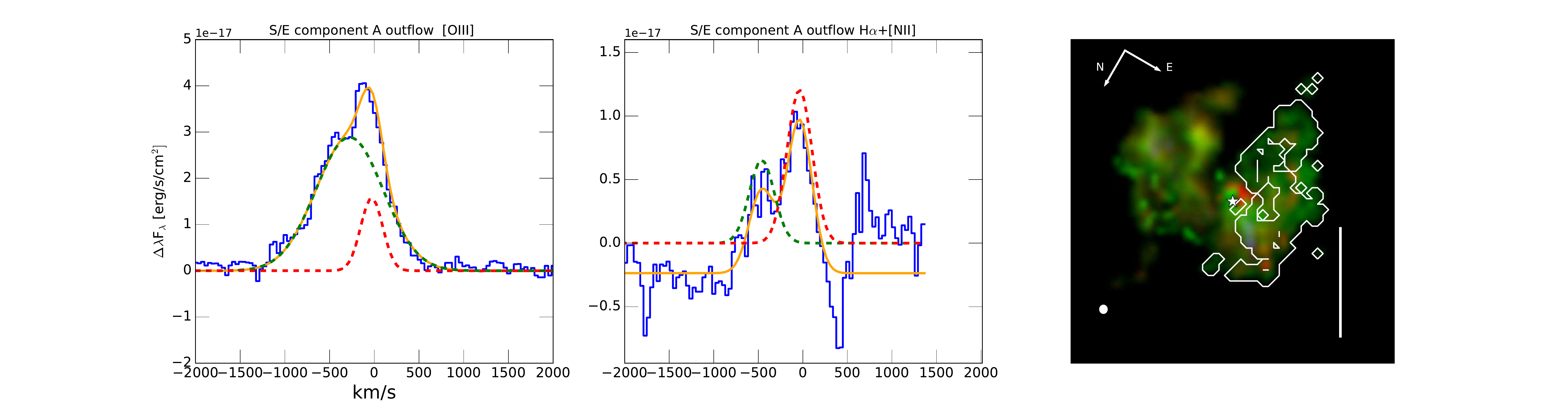}
    \caption{OSIRIS observations of the 4C09.17 nebular emission line distribution and kinematics produced from the PSF subtracted data cubes.(Top-left) Three color intensity map of nebular emission lines: \ha (red), \oiii (green), and \nii (blue). Ellipse in the lower left corner showcases the spatial resolution of the observations. (Top-middle) Radial velocity offset (\kms) of the \oiii line relative to the redshift of the quasar. The white star shows the location of the subtracted quasar. (Top-right) Velocity dispersion (\kms) map of \oiii emission. The white contours are ALMA observations in band 4 of radio synchrotron emission from the quasar jet and lobes. (BOTTOM) Spectra of a distinct outflow region along with fits to individual emission lines On the left we show the fit to the \oiii 500.7 nm emission line, in the middle we present the fit to the \ha and \nii emission lines and on the right we show a three-color composite along with a contour outlining the spatial location of the region. The blueshifted and low-velocity dispersion emission towards the west is from a merging system.}
    \label{fig:4C0917_all}
\end{figure*}

\begin{figure*}[!th]
    \centering
    \includegraphics[width=6.5in]{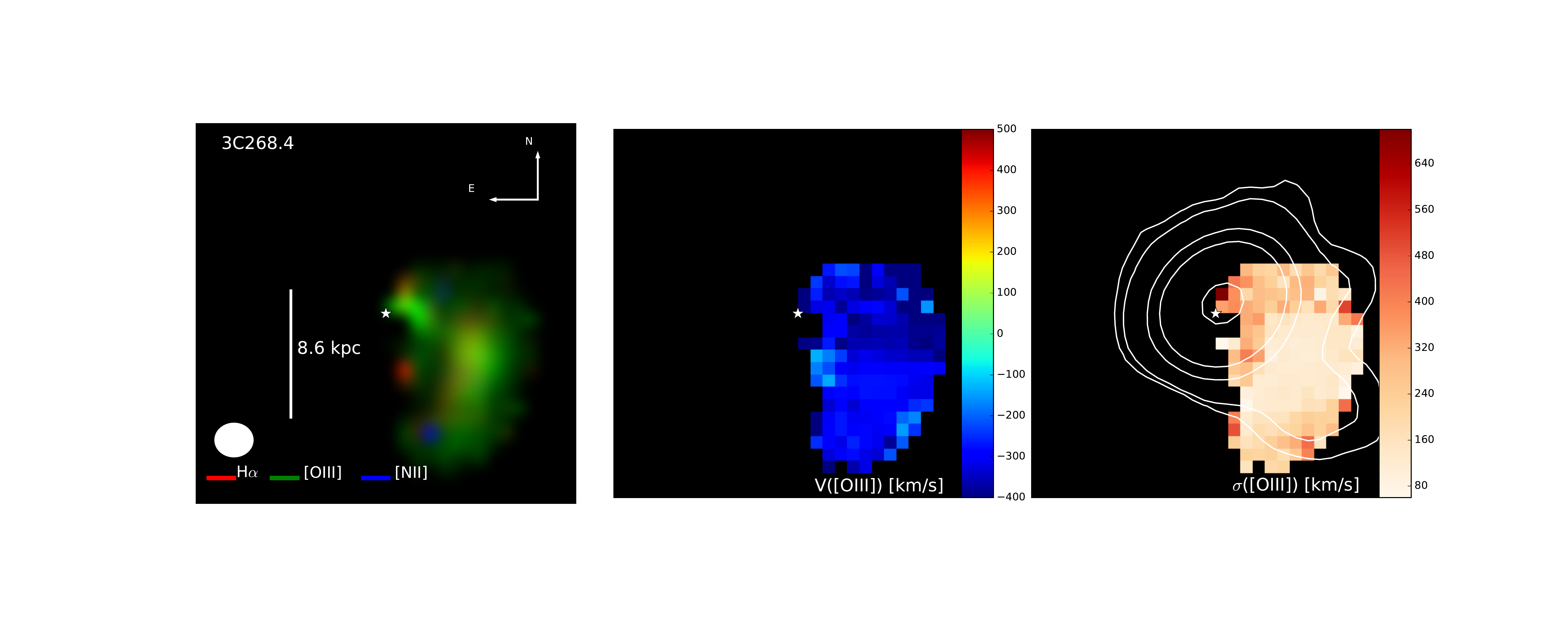}
    \includegraphics[width=7in]{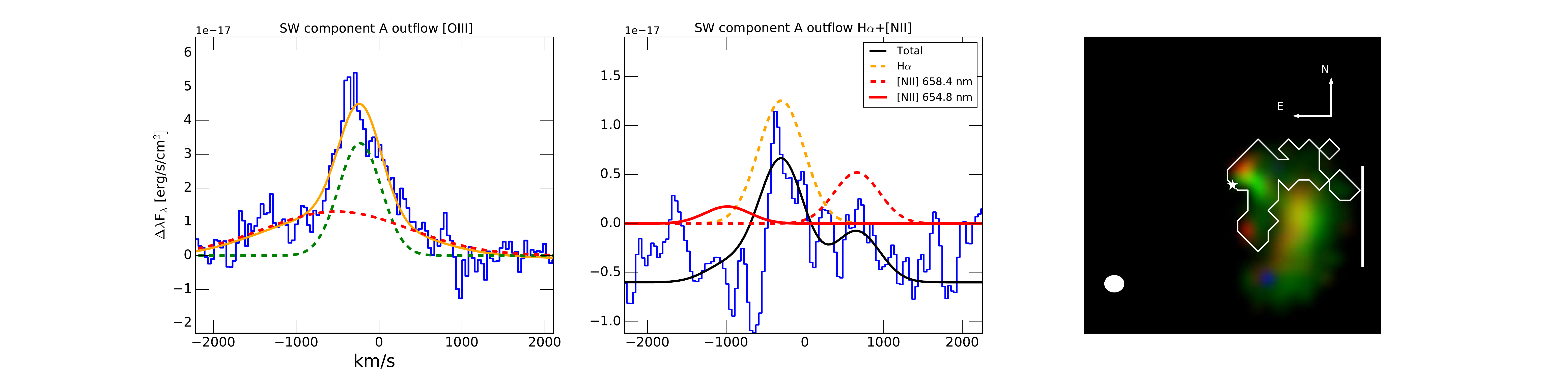}
    \caption{OSIRIS observations of the 3C268.4 nebular emission line distribution and kinematics produced from the PSF subtracted data cubes.(Top-left) Three color intensity map of nebular emission lines: \ha (red), \oiii (green), and \nii (blue). Ellipse in the lower left corner showcases the spatial resolution of the observations. (Top-middle) Radial velocity offset (\kms) of the \oiii line relative to the redshift of the quasar. The white star shows the location of the subtracted quasar. (Top-right) Velocity dispersion (\kms) map of \oiii emission. The white contours are VLA observations at 8264.9 MHz of radio synchrotron emission from the quasar jet and lobes. (BOTTOM) Spectra of a distinct outflow region along with fits to individual emission lines On the left we show the fit to the \oiii 500.7 nm emission line, in the middle we present the fit to the \ha and \nii emission lines and on the right we show a three-color composite along with a contour outlining the spatial location of the region. The blueshifted and low-velocity dispersion emission towards the south-west is from a merging system.}
    \label{fig:3C2684_all}
\end{figure*}

\begin{figure*}
    \centering
    \includegraphics[width=6.5in]{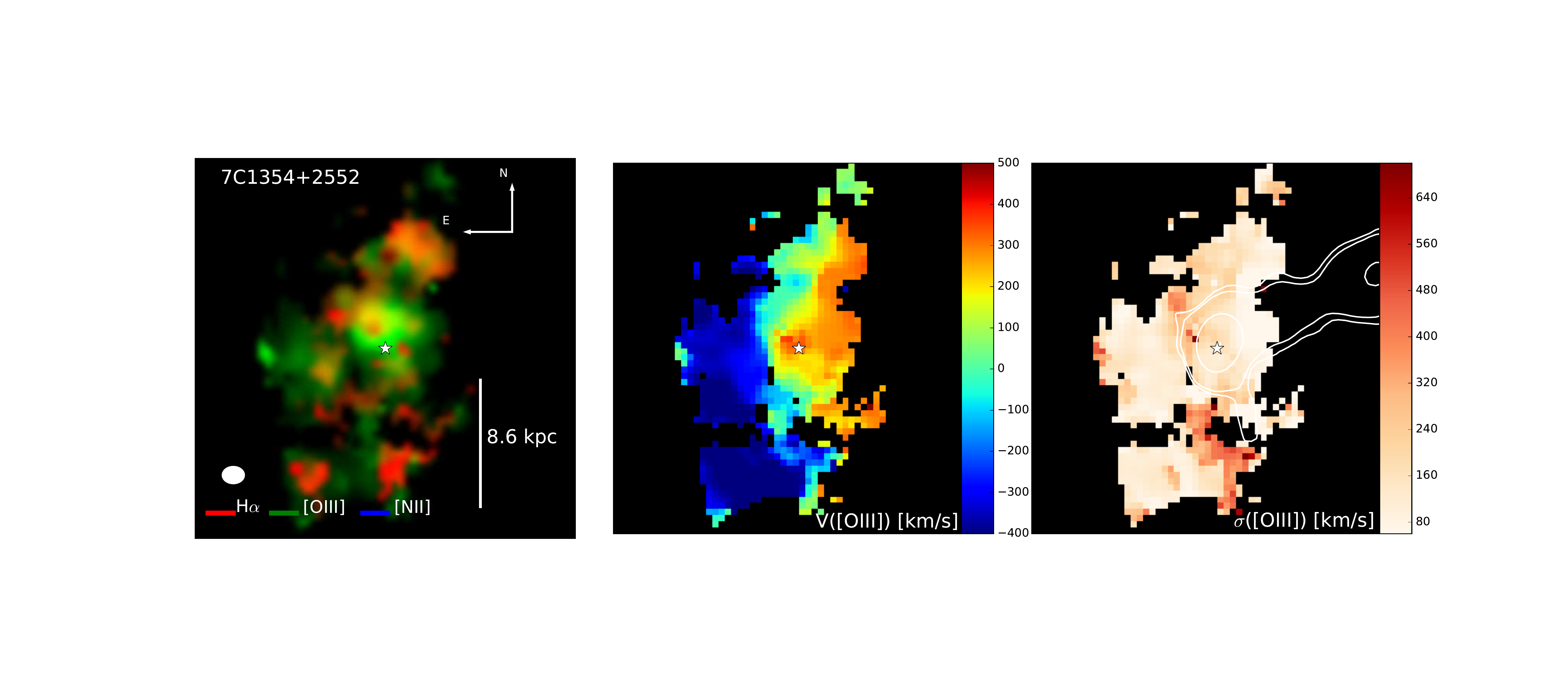}
    \caption{OSIRIS observations of the 7C1354 nebular emission line distribution and kinematics produced from the PSF subtracted data cubes.(Left) Three color intensity map of nebular emission lines: \ha (red), \oiii (green), and \nii (blue). Ellipse in the lower left corner showcases the spatial resolution of the observations. (Middle) Radial velocity offset (\kms) of the \oiii line relative to the redshift of the quasar. The white star shows the location of the subtracted quasar. (Right) Velocity dispersion (\kms) map of \oiii emission. The white contours are ALMA band 4 observations of radio synchrotron emission from the quasar jet and lobes.}
    \label{fig:7C1354_all}
\end{figure*}

\begin{figure*}
    \centering
    \includegraphics[width=7in]{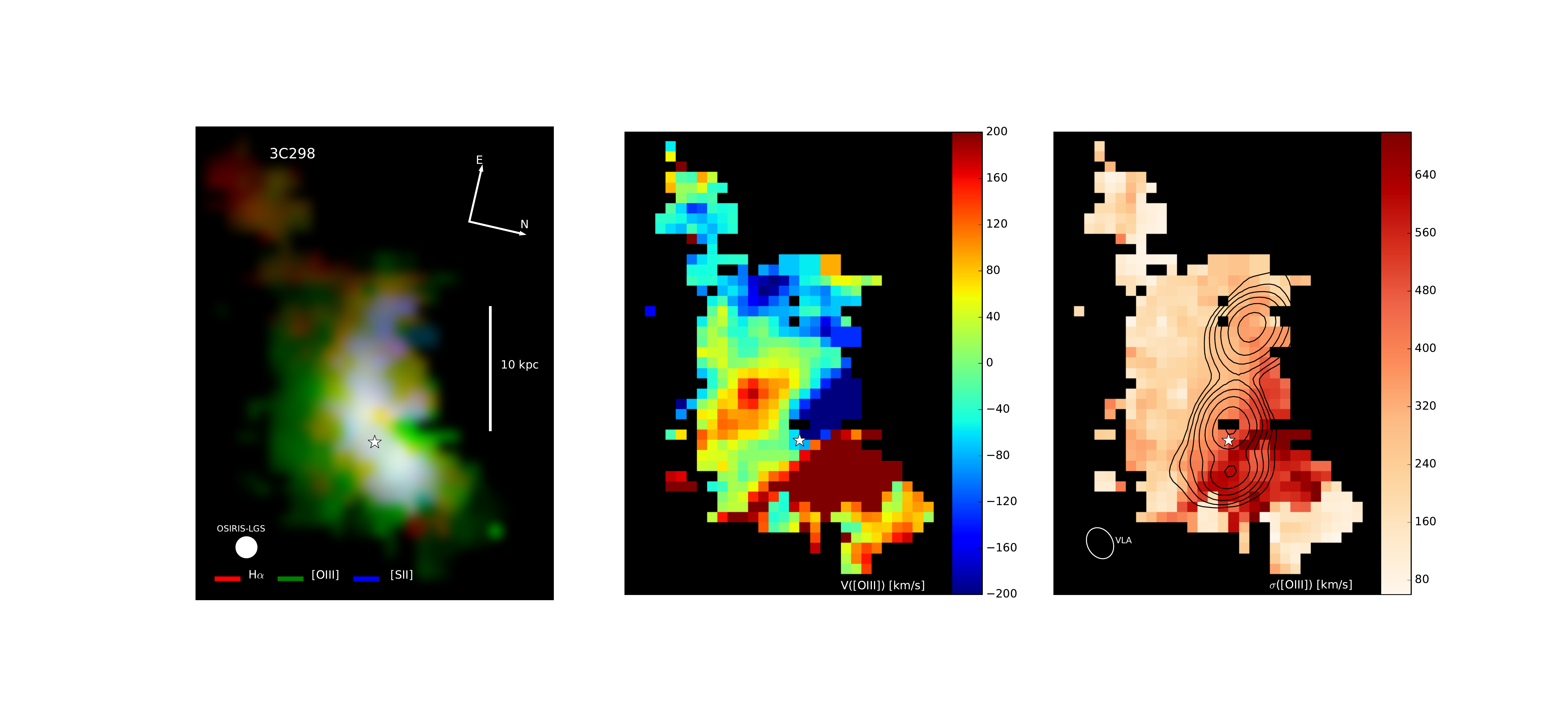}
    \includegraphics[width=7in]{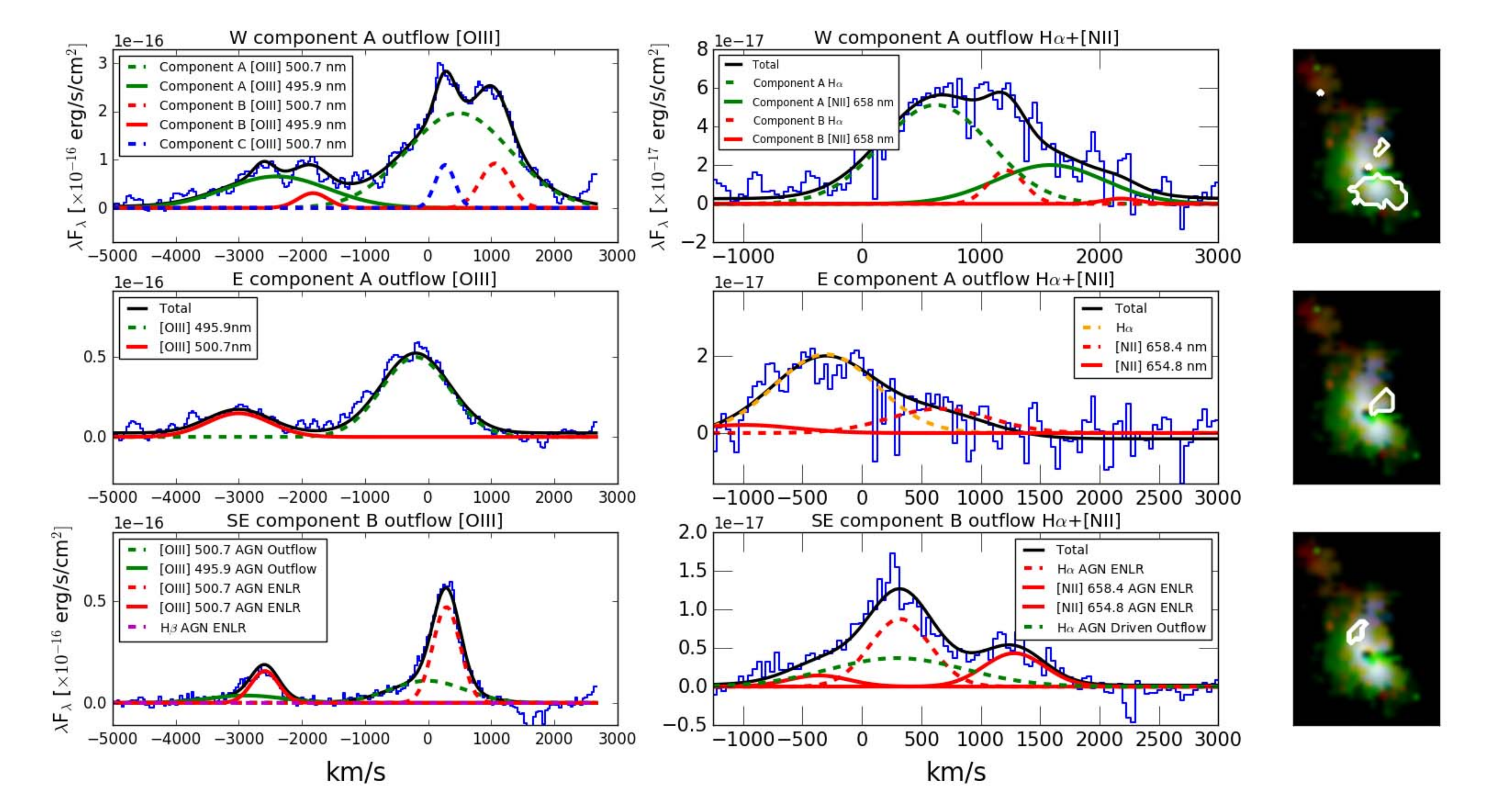}
    \caption{OSIRIS observations of the 3C298 nebular emission line distribution and kinematics produced from the PSF subtracted data cubes.(Top-left) Three color intensity map of nebular emission lines: \ha (red), \oiii (green), and \nii (blue). Ellipse in the lower left corner showcases the spatial resolution of the observations. (Top-middle) Radial velocity offset (\kms) of the \oiii line relative to the redshift of the quasar. The white star shows the location of the subtracted quasar. (Top-right) Velocity dispersion (\kms) map of \oiii emission. The white contours are VLA observations at 8485.1 MHz of radio synchrotron emission from the quasar jet and lobes. (BOTTOM) Spectra of a distinct outflow region along with fits to individual emission lines On the left we show the fit to the \oiii 500.7 nm emission line, in the middle we present the fit to the \ha and \nii emission lines and on the right we show a three-color composite along with a contour outlining the spatial location of the region.}
    \label{fig:3C298_all}
\end{figure*}

\begin{figure*}[!th]
    \centering
    \includegraphics[width=6.5in]{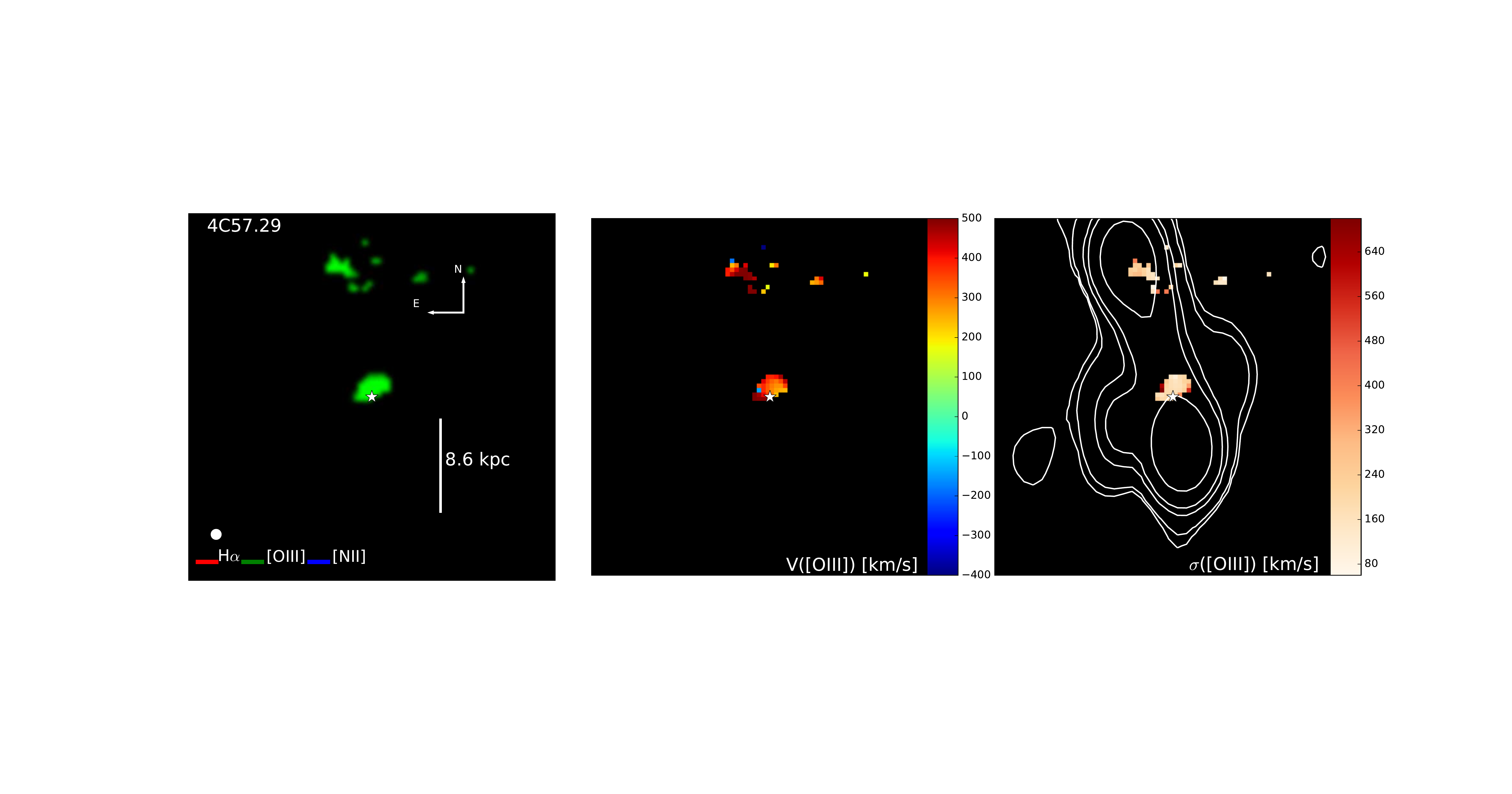}
    \caption{OSIRIS observations of the 4C57.29 nebular emission line distribution and kinematics produced from the PSF subtracted data cubes. (Left) Three color intensity map of nebular emission lines: \ha (red), \oiii (green), and \nii (blue). Ellipse in the lower left corner showcases the spatial resolution of the observations. (Middle) Radial velocity offset (\kms) of the \oiii line relative to the redshift of the quasar. The white star shows the location of the subtracted quasar. (Right) Velocity dispersion (\kms) map of \oiii emission. The white contours are VLA observations at 1490.0 MHz of radio synchrotron emission from the quasar jet and lobes. The redshifted and low-velocity dispersion emission towards the north is likely from a merging system.}
    \label{fig:4C5729_all}
\end{figure*}

\begin{figure*}[!th]
    \centering
    \includegraphics[width=6.5in]{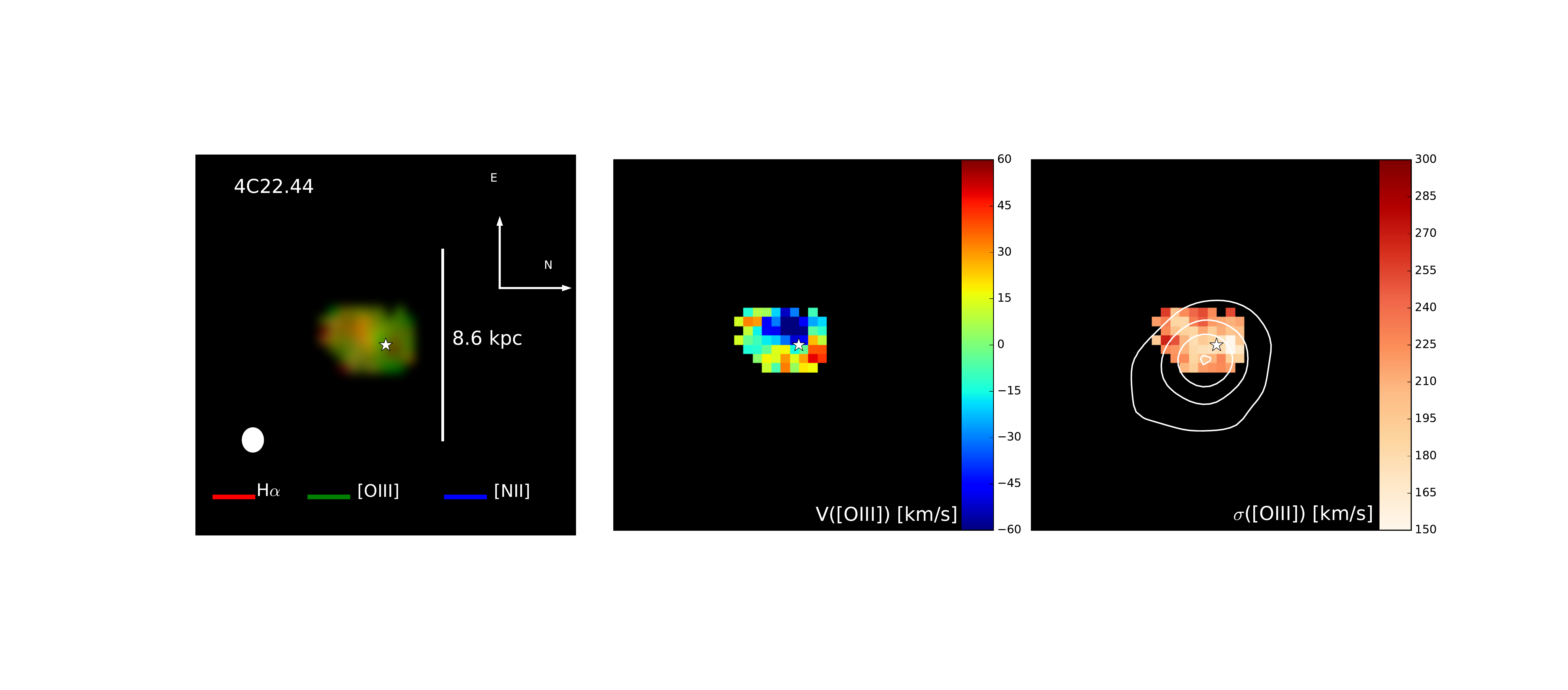}
    \caption{OSIRIS observations of the 4C22.44 nebular emission line distribution and kinematics produced from the PSF subtracted data cubes.(Left) Three color intensity map of nebular emission lines: \ha (red), \oiii (green), and \nii (blue). Ellipse in the lower left corner showcases the spatial resolution of the observations. (Middle) Radial velocity offset (\kms) of the \oiii line relative to the redshift of the quasar. The white star shows the location of the subtracted quasar. (Right) Velocity dispersion (\kms) map of \oiii emission. The white contours are ALMA band 4 observations of radio synchrotron emission from the quasar jet and lobes.}
    \label{fig:4C2244_all}
\end{figure*}

\begin{figure*}[!th]
    \centering
    \includegraphics[width=6.5in]{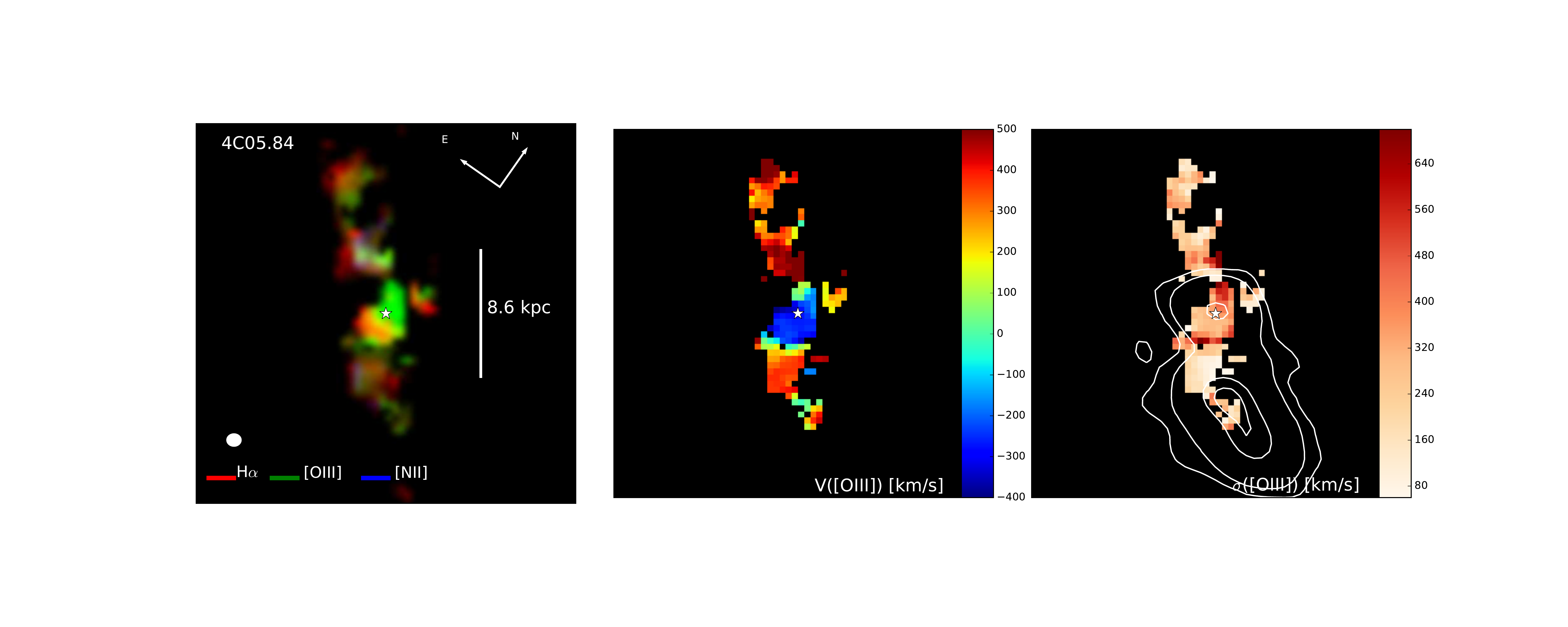}
    \includegraphics[width=7in]{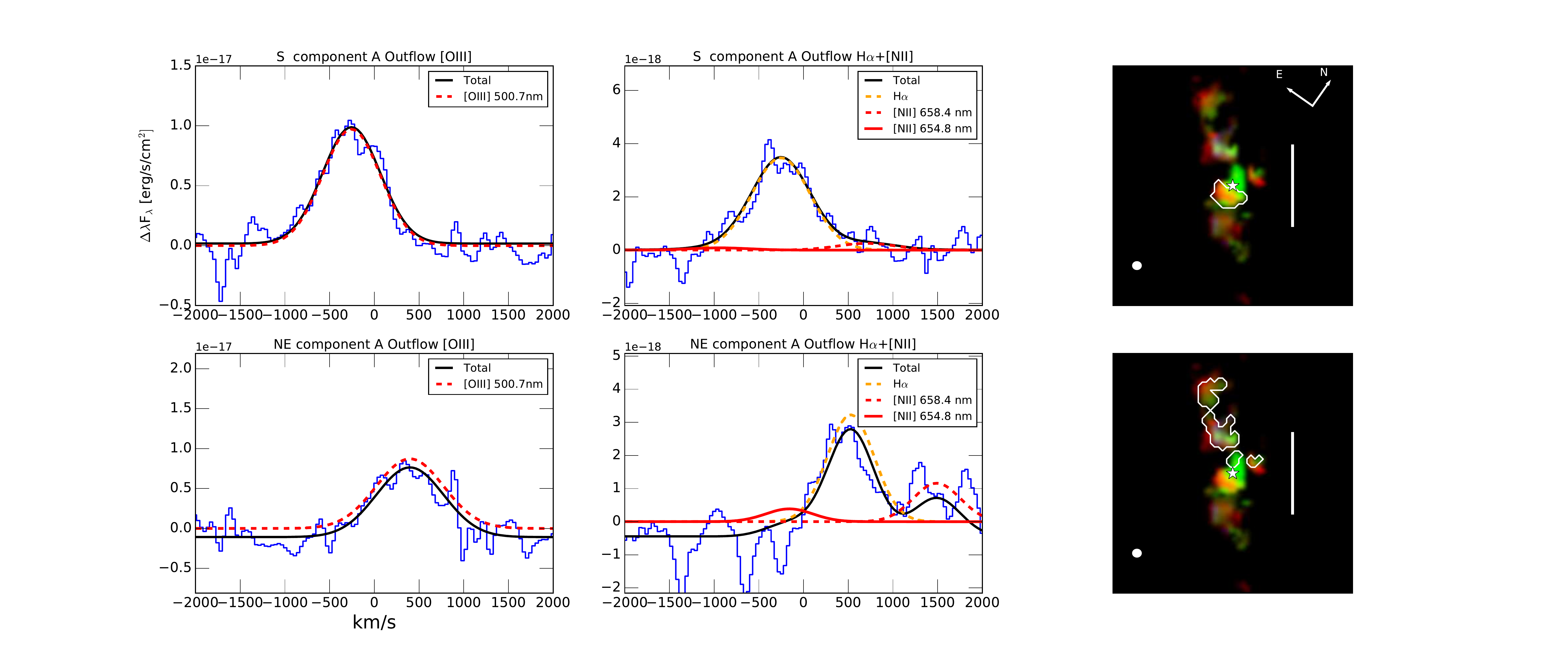}
    \caption{OSIRIS observations of the 4C05.84 nebular emission line distribution and kinematics produced from the PSF subtracted data cubes.(Top-left) Three color intensity map of nebular emission lines: \ha (red), \oiii (green), and \nii (blue). Ellipse in the lower left corner showcases the spatial resolution of the observations. (Top-middle) Radial velocity offset (\kms) of the \oiii line relative to the redshift of the quasar. The white star shows the location of the subtracted quasar. (Top-right) Velocity dispersion (\kms) map of \oiii emission. The white contours are ALMA band 4 observations of radio synchrotron emission from the quasar jet and lobes. (BOTTOM) Spectra of a distinct outflow region along with fits to individual emission lines On the left we show the fit to the \oiii 500.7 nm emission line, in the middle we present the fit to the \ha and \nii emission lines and on the right we show a three-color composite along with a contour outlining the spatial location of the region.}
    \label{fig:4C0584_all}
\end{figure*}

\begin{figure*}[!th]
    \centering
    \includegraphics[width=6.5in]{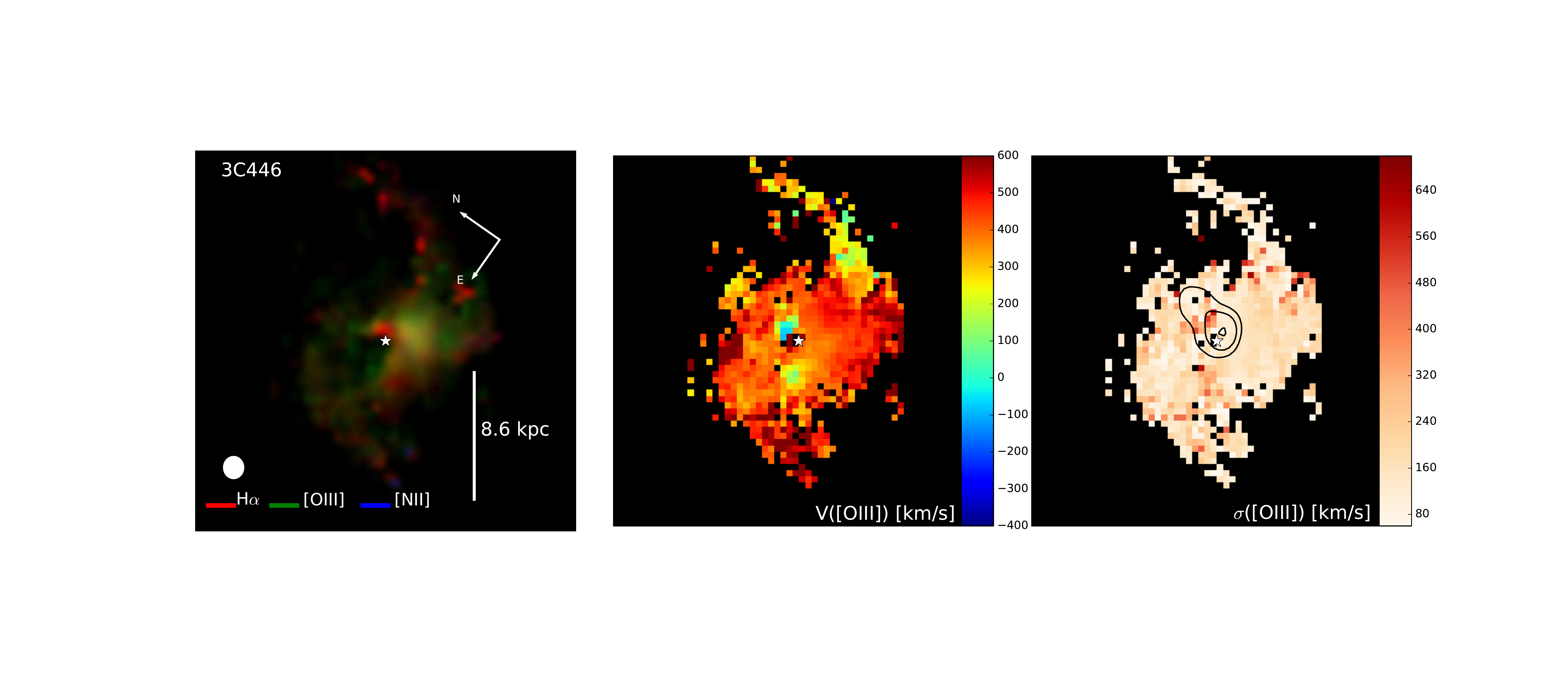}
    \caption{OSIRIS observations of the 3C446 nebular emission line distribution and kinematics produced from the PSF subtracted data cubes. (Left) Three color intensity map of nebular emission lines: \ha (red), \oiii (green), and \nii (blue). Ellipse in the lower left corner showcases the spatial resolution of the observations. (Middle) Radial velocity offset (\kms) of the \oiii line relative to the redshift of the quasar. The white star shows the location of the subtracted quasar. (Right) Velocity dispersion (\kms) map of \oiii emission. The white contours are radio synchrotron emission from the quasar jet and lobes.}
    \label{fig:3C446_all}
\end{figure*}

\begin{figure*}[!th]
    \centering
    \includegraphics[width=6.5in]{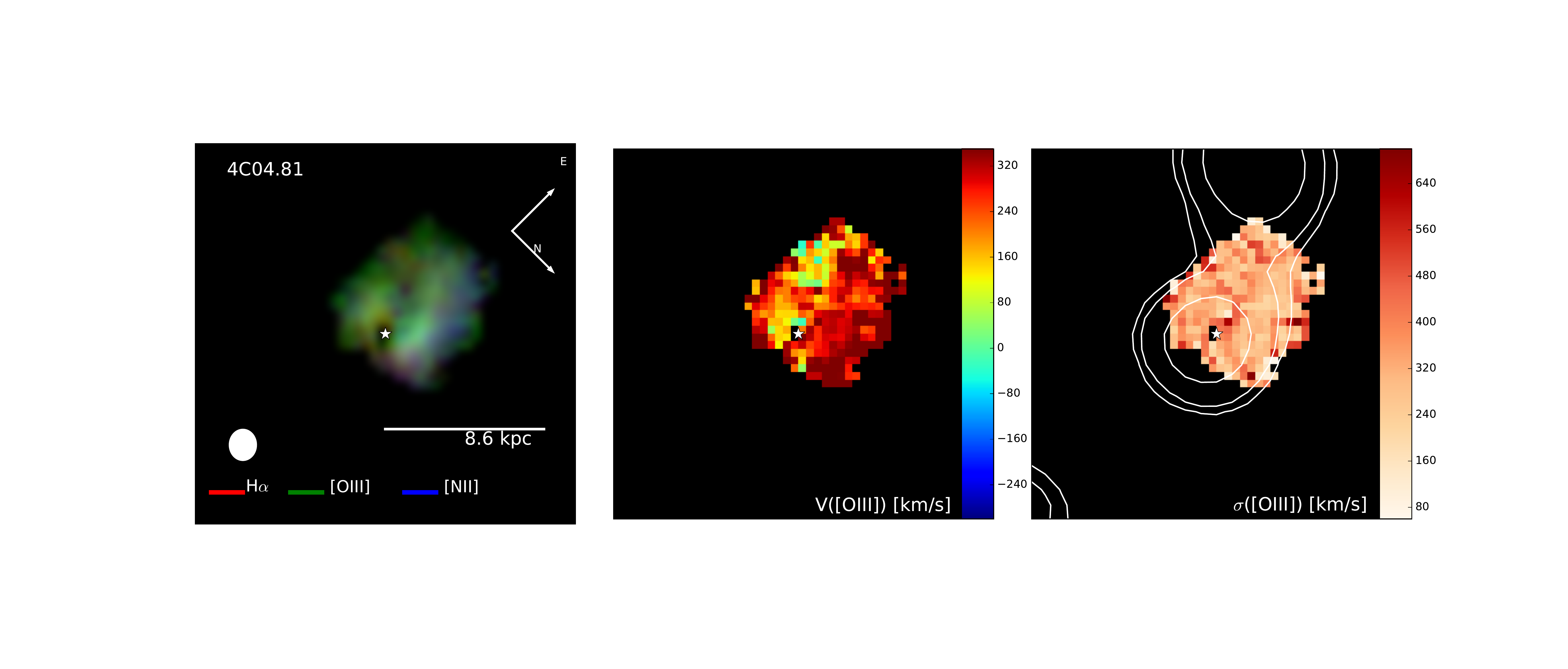}
    \includegraphics[width=7in]{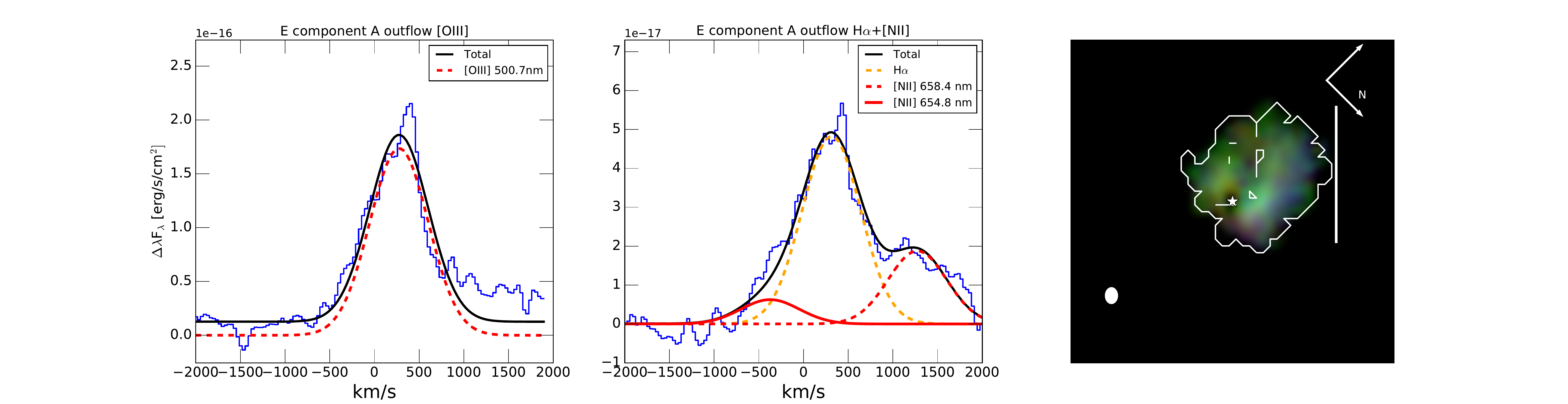}
    \caption{OSIRIS observations of the 4C04.81 nebular emission line distribution and kinematics produced from the PSF subtracted data cubes.(Top-left) Three color intensity map of nebular emission lines: \ha (red), \oiii (green), and \nii (blue). Ellipse in the lower left corner showcases the spatial resolution of the observations. (Top-middle) Radial velocity offset (\kms) of the \oiii line relative to the redshift of the quasar. The white star shows the location of the subtracted quasar. (Top-right) Velocity dispersion (\kms) map of \oiii emission. The white contours are VLA observations at 4860.0 MHz of radio synchrotron emission from the quasar jet and lobes. (BOTTOM) Spectra of a distinct outflow region along with fits to individual emission lines On the left we show the fit to the \oiii 500.7 nm emission line, in the middle we present the fit to the \ha and \nii emission lines and on the right we show a three-color composite along with a contour outlining the spatial location of the region.}
    \label{fig:4C0481_all}
\end{figure*}

\begin{figure}
    \centering
    \includegraphics[width=6.0in]{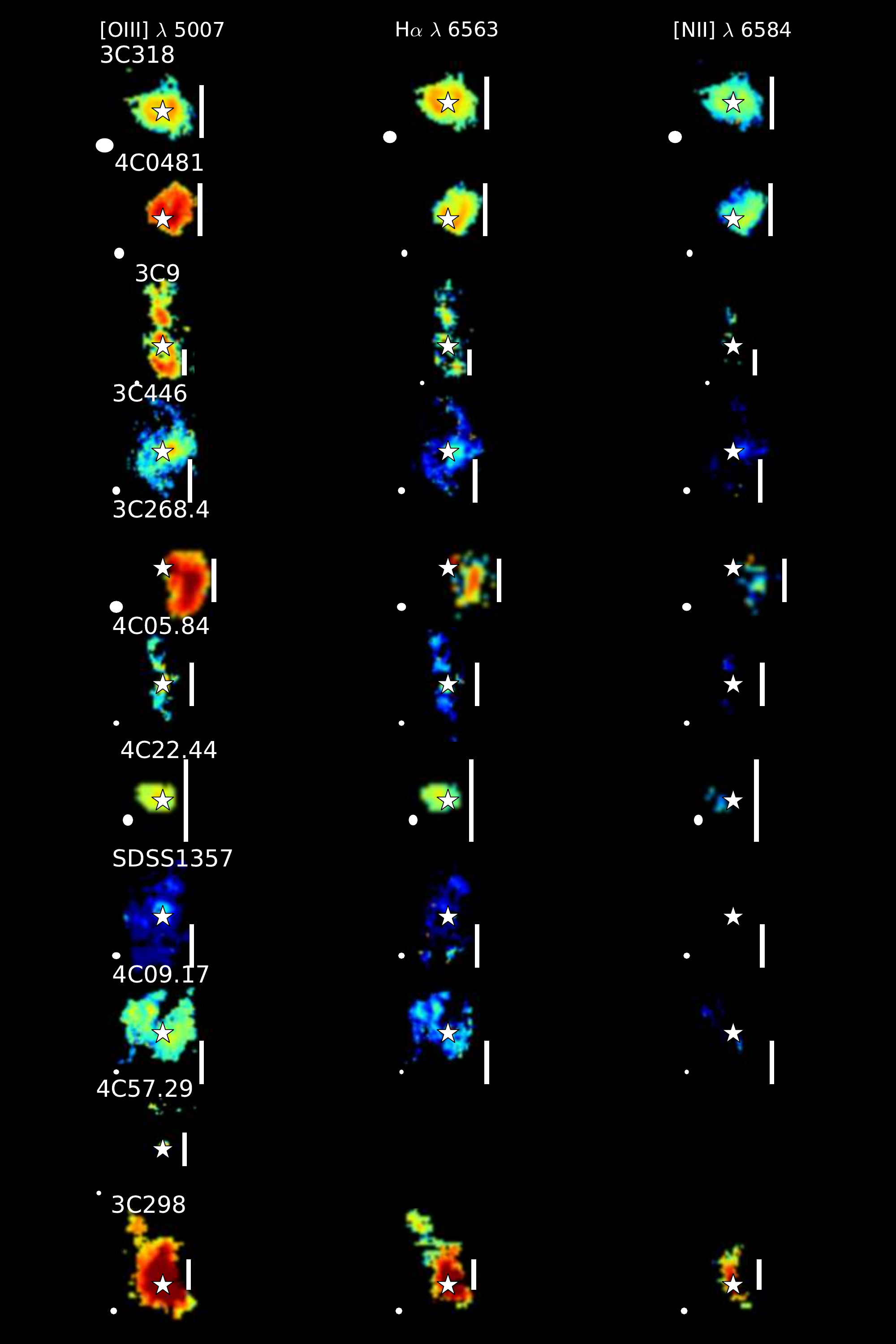}
    \caption{A figure illustrating the individual line integrated maps for \oiii (left) \ha (middle) and \nii (right) produced from the PSF subtracted data cubes. The bar in each stamp represents 1\arcsec, while the ellipse in the lower-left corner represents the PSF's FWHM for that filter.}
    \label{fig:all_maps}
\end{figure}

\bibliography{bib}
\end{document}